\documentclass[namedreferences]{solarphysics}
\usepackage[optionalrh]{spr-sola-addons} 
\usepackage{graphicx}        
\usepackage{color}           
\usepackage{url}

\usepackage{pdflscape}

\newcommand{\adv}{    {\it Adv. Space Res.}} 
 
\newcommand{\aap}{    {\it Astron. Astrophys.}}

\newcommand{\apj}{    {\it Astrophys. J.}}
\newcommand{\apjl}{   {\it Astrophys. J. Lett.}}

\newcommand{\nat}{    {\it Nature}}

\newcommand{\solphys}{{\it Solar Phys.}}
 
\newcommand{\ssr}{    {\it Space Sci. Rev.}} 
\newcommand{\NewA}{   {\it New Astron.}} 

\begin{document}

\begin{article}

\begin{opening}

\title{Observations of Photospheric Vortical Motions During the Early Stage of Filament Eruption}

\author{Sajal Kumar~\surname{Dhara}$^{1}$\sep
        B.~\surname{Ravindra}$^{1}$\sep
        Ravinder Kumar~\surname{Banyal}$^{1,2}$      
       }
\runningauthor{S. K. Dhara \textit{et al.}}
\runningtitle{Photospheric Vortical Motions During Filament Eruption}

   \institute{$^{1}$ Indian Institute of Astrophysics, Bangalore 560034, India
                     email: \url{sajal@iiap.res.in} email: \url{ravindra@iiap.res.in}\\ 
              $^{2}$ Institut f\"{u}r Astrophysik, Georg-August-Universitat G\"{o}ttingen, Friedrich-Hund-Platz
              1, 37077 G\"{o}ttingen, Germany \\  
             }

\begin{abstract}
Solar filaments/prominences exhibit rotational motion during different phases of their
evolution from their formation to eruption. We have observed the rotational/vortical 
motion in the photosphere near the ends of ten filaments during their initial phase of eruption, 
at the onset of the fast rise phase. All the filaments were associated with active regions. 
The photospheric vortical motions we observed lasted for 4--20 minutes.
In the vicinity of the conjugate ends of the filament the direction of rotation was opposite, 
except for two cases, where rotational motion was observed at only one end point.
The sudden onset of a large photospheric vortex motion could have played a role 
in destabilizing the filament by transporting axial flux into the activated filament 
thereby increasing the outward magnetic pressure in it. The outward magnetic pressure may 
have pushed the filament/flux rope to the height where the torus instability 
criterion was satisfied, and hence it could have caused the filament instability and eruption.

\end{abstract}
\end{opening}

\section{Introduction}
     \label{S-Introduction} 

Solar filaments are large structures which contain dense and cool ($\sim$~10$^{4}$~K) plasma 
embedded in the tenuous and hot corona. The filaments form above the neutral line of the photospheric magnetic 
field and survive for days to weeks \cite{Martin98}. During their formation and evolution
and prior to eruption they exhibit a variety of dynamical processes. Bidirectional flows of plasma 
were observed within quiet region filaments \cite{Zirker98} and in active region 
filaments \cite{Alexander13}. Oscillations were observed in erupting filaments 
\cite{Bocchialini12}. Rotational motion in erupting filaments/prominences have been observed \cite{Panesar13}.

Normally, the filament is considered to be situated in the lower part of a flux rope, where the dense matter 
is suspended in the magnetic cavity.  In the CME structure, the dark cavity, below the 
bright leading edge, is considered to correspond to the twisted flux rope \cite{Gibson06} and the filament is 
located below the dark cavity. Identification of cavities is difficult 
in active region filaments, as the cavity or flux ropes are low lying, compact structures in active regions. Hence, 
it is not trivial to identify the flux ropes in active regions prior to eruption.  On many 
occasions the ends of filaments are considered as the ends of the lower part of a flux rope 
that has roots in the photosphere (\opencite{vanBallegooijen04}; \opencite{Kliem13}).

During the rising phase of eruptions, filaments are sometimes observed to undergo a rotation about the vertical axis
(see, {\it e.g.,} \opencite{Zhou06}; \opencite{Green07}; \opencite{Liu09}; \opencite{Thompson11}). 
This kind of filament rotation is interpreted as a conversion of twist 
into writhe in a kink-unstable flux rope. The flux rope axis rotation is clockwise (as viewed 
from above) if it has right-handed twist and counter-clockwise if it 
has left-handed twist (see, {\it e.g.,} \opencite{Rust05}; \opencite{Green07}; \opencite{Wang09}).
Erupting prominences which are considered to be contained in magnetic flux ropes, very often develop into a 
helical-like structure (see, {\it e.g.,} \opencite{Rust03}; \opencite{Rust05}; \opencite{Williams05}; 
\opencite{Liu07}), which indicates the signature of a magnetohydrodynamics (MHD) kink instability of a twisted magnetic
flux rope \cite{Rust05}. In the twisted flux rope, the upward directed kink instability leads 
to the upward motion of the flux rope, which eventually turns into helical deformation and 
forms current sheets. The helical kink instability of a pre-existing flux rope can trigger 
 solar eruptions \cite{Torok05}. In the MHD instability, half the magnitude of the twist will be 
converted into writhe having the same sign as twist \cite{Torok14}. The `S' shaped stable filament reverses its
shape to inverse `S' while it is rising. During the rising phase, the shape of the erupting filament 
straightens out and later acquires the deformation of the axis in opposite direction suggesting
the transition from `S' to inverse `S'. The possibility of this mechanism is confirmed in 
the numerical simulation of a kink--unstable flux rope by \inlinecite{Torok10}. During the observation 
of such events the deformation of the shape can be seen as a rotation of the axis.

 The filament/flux rope can face an another type of instability called 
``torus instability''(see, \opencite{Kliem06}; \opencite{Zuccarello14}). The toroidal 
current carrying flux rope experiences an outward directed self Lorentz force 
(also called a hoop force) which counteracts the external poloidal fields of 
the background magnetic fields (\opencite{Chen89}; \opencite{Titov99}).
However, if the external poloidal field decreases faster with the radial distance  
than the self Lorentz force, then the flux rope can become unstable to the lateral expansion.  
A loss of equilibrium of the flux rope due to this effect can lead to its eruption. 

\inlinecite{Liggett84} studied rotational motion in five non-eruptive prominences. 
In some events they reported only a part of the 
prominence rotating, while in the other the entire body was in rotation. 
They interpreted the rotation in terms of a twisting of the magnetic structure resulting from
the reconnection. \textit{Atmospheric Imaging Assembly} ({AIA;} \opencite{Lemen12}) observations show 
that the feet of solar prominences exhibit coherent rotation for over three hours in the 
lower corona as the material is flowing along helical structures~\cite{Li12}.

\inlinecite{Su12} reported two solar tornadoes during the formation of a 
quiet prominence. This tornado was observed for about two days. A systematic analysis of 
giant tornadoes by \inlinecite{Wedemeyer13} using AIA 171~\AA~ images and high-resolution 
SST observations shows that giant tornadoes are an integral part of solar prominences. 
The tornadoes inject mass and twist into the filament spine until it becomes unstable and 
erupts. \inlinecite{Su13} also reported a quiescent filament eruption in which a part of the 
quiescent filament exhibits a strong clockwise rolling motion.

When a filament erupts non-radially, the top of its axis bends first to one side and
propagates into sideways rolling motion, known as the {\it roll effect}, which results in a 
large scale twists in both legs of erupting filaments~(see, {\it e.g.,} \opencite{Bangert03}; \opencite{Filippov01};
\opencite{Panasenco11}; \opencite{Panasenco13}). This rotational mass motion
is observed to flow down as the structure of the filament rises to higher heights.
If the observed sense of rotations was generated by the external forces at the legs of the 
erupting filament, then the sense of the twists in both legs is opposite to each other (\opencite{Panasenco13}).
In the case of asymmetric eruption with respect to the polarity inversion line, the spine of the 
filament bends one side, which does not necessarily result in rotation in both legs of the filament, 
depending upon the instability
(see, {\it e.g.,} \opencite{Martin03}; \opencite{Panasenco08}). 

In the aforementioned cases, either the whole prominence was rotating around its vertical 
axis or photospheric vortex motions were observed at individual footpoints. Only in some cases a
part of the prominence was observed to be rotating. \inlinecite{Dhara13} 
have reported a photospheric vortex-like rotational 
motion at the ends of a filament during its eruption. 
In that observation they have reported only a single event in which the rotational motion 
started soon after the filament eruption was initiated and lasted for a few minutes.

In this paper, we extended this study to ten erupting active region filaments and
searched for such rotational motions in and around the end points of these filaments.
The rotational motions were studied by applying a local correlation tracking technique 
(LCT; \opencite{November88}) to the photospheric dopplergram
data. We looked for rotational (photospheric vortex) motion during the onset of 
the filament eruption. This paper is organized as
follows. In Section \ref{section2} we describe the data used. 
The results are described in Section \ref{section3}, and 
a brief summary and discussion are presented in Section \ref{section4}.

\section{Data Set} 
      \label{section2}  
      
The AIA instrument onboard the \textit{Solar Dynamics Observatory}
({SDO;}~\opencite{Pesnell12}) provides imaging of the solar atmosphere in seven extreme 
ultraviolet (EUV) wavelength channels centered on specific lines. The temperature diagnostics of
the EUV emissions cover the range from 6$\times$10$^{4}$ to 2$\times$10$^{7}$ K.
It has high image cadence (12~sec) and spatial resolution (0.6$^{\prime\prime}$/pixel).
We obtained the chromospheric data observed in He~II~304~\AA~wavelength for eight active regions 
(NOAA 11226, 11283, 11515, 11560, 11451, 11936, 12027 and 12035) associated with ten events. 
The full-disk EUV images of the Sun observed in 304~\AA~wavelength are obtained 
on 07 June 2011, 07 September 2011, 02 July 2012, 02 September 2012, 07 April 2012, 31 December 2013, 01 January 2014,
04 April 2014, and 15 April 2014, covering the ten events of filament eruption. The obtained data is 
corrected for the flat fielding, the bad pixels, and the spikes due to high energy particles.
From these data sets the filament regions are extracted and tracked over time. This process
provided data cubes showing the filament eruption for each of the events.      
      
We also acquired the dopplergram data set from \textit{Helioseismic and Magnetic Imager} 
({HMI;} \opencite{Scherrer12}; \opencite{Schou12}) at a cadence of 45~sec.  We obtained the dopplergram 
data for each event for about six hours covering the whole event of filament eruption. The 
obtained dopplergrams show several velocity patterns, and while looking for the small scale 
feature motions one has to remove the long term and very short term velocity patterns in sequence. To start with, 
we first removed the overall solar rotation from the dopplergram data. This has been done 
by using the following method. We first averaged the dopplergram data set without correcting 
for solar rotation. The gradient part of the rotational velocity
pattern was extracted and subtracted from each of the dopplergram data. This procedure 
removed the overall solar rotational velocity patterns in the full-disk dopplergrams.
Later, we interpolated these dopplergram data set to the AIA pixel resolution to match the 
spatial size of each pixel in the data set. Further, 
the dopplergrams were differentially rotated to the central meridian passing time of the 
active region.  The usual 5-minute oscillations in the dopplergrams were suppressed by applying 
a subsonic filter with an upper cutoff velocity magnitude of 4~km~s$^{-1}$. This corrected data 
set is used to determine the horizontal velocity of features  
near the footpoints of the filament and its surrounding regions at the photospheric level. 
This has been done by applying the LCT technique to
the processed dopplergram data set as has been done in \inlinecite{Derosa04} and \inlinecite{Ravindra06}.   

Along with the aforesaid data sets, we also obtained the line-of-sight magnetograms from HMI 
for each events at a cadence of 45~sec for about a few hours before, during and after the 
eruption. We corrected the data set for the solar rotation in a similar way to the 
dopplergrams. The correction for the line-of-sight effect is made by multiplying 
1/cos$\theta$, where $\theta$ is the heliocentric angle. We used these magnetograms to 
locate the filament position in the active region (AR).

\section{Observations and Results} 
      \label{section3}         
\subsection{Events description}
 \label{fcorona}
\begin{figure}    
   \centerline{\hspace*{0.015\textwidth}
               \includegraphics[width=0.8\textwidth,clip=]{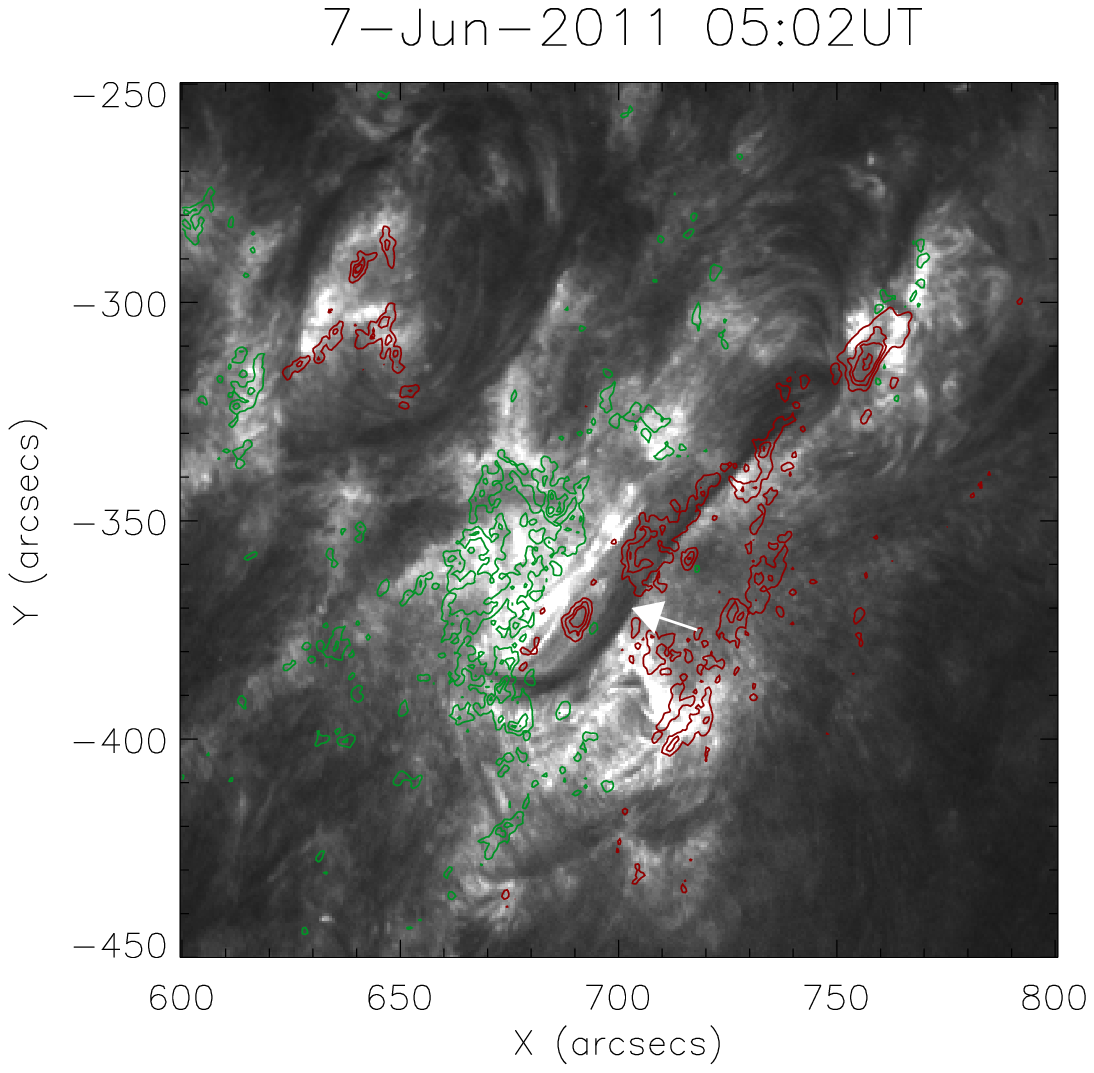}
               \hspace*{-0.32\textwidth}
               \includegraphics[width=0.8\textwidth,clip=]{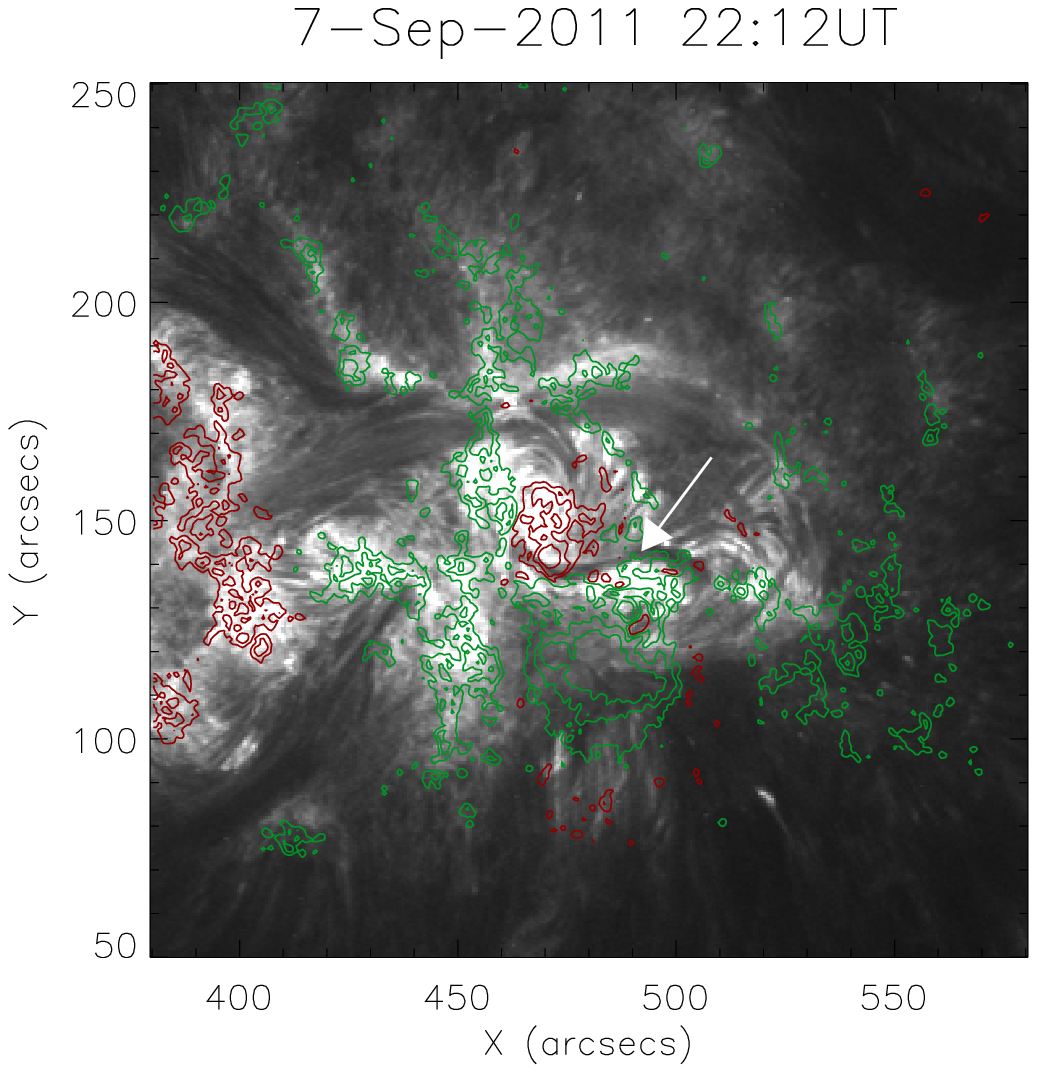}
              }
     \vspace{-0.15\textwidth}   
     
     \centerline{\Large \bf     
    \hspace{0.38 \textwidth}  \color{white}{(a)}
      \hspace{0.41\textwidth}  \color{white}{(b)}
         \hfill}
         \vspace{0.03\textwidth}
         
   \centerline{\hspace*{0.015\textwidth}
               \includegraphics[width=0.8\textwidth,clip=]{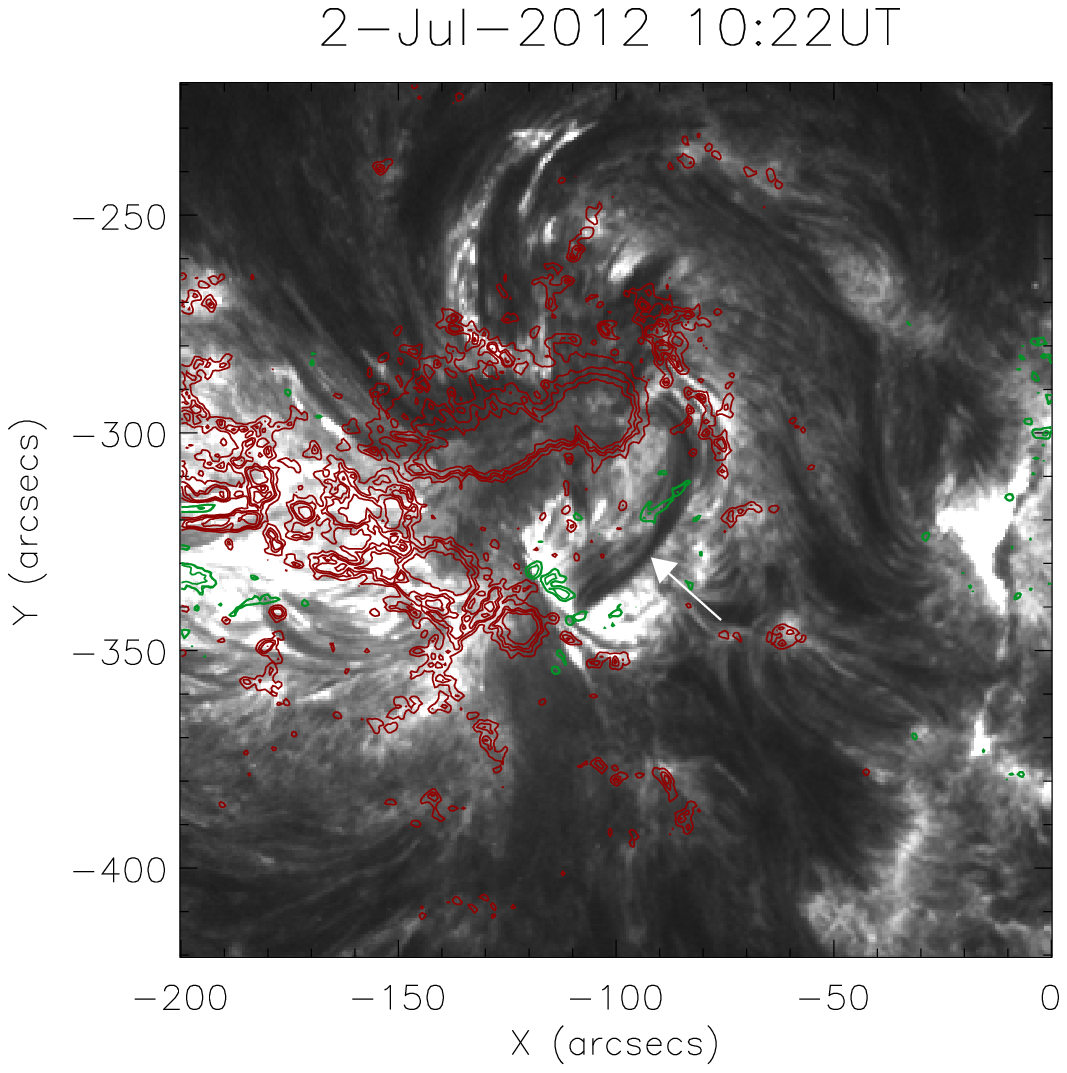}
               \hspace*{-0.32\textwidth}
               \includegraphics[width=0.8\textwidth,clip=]{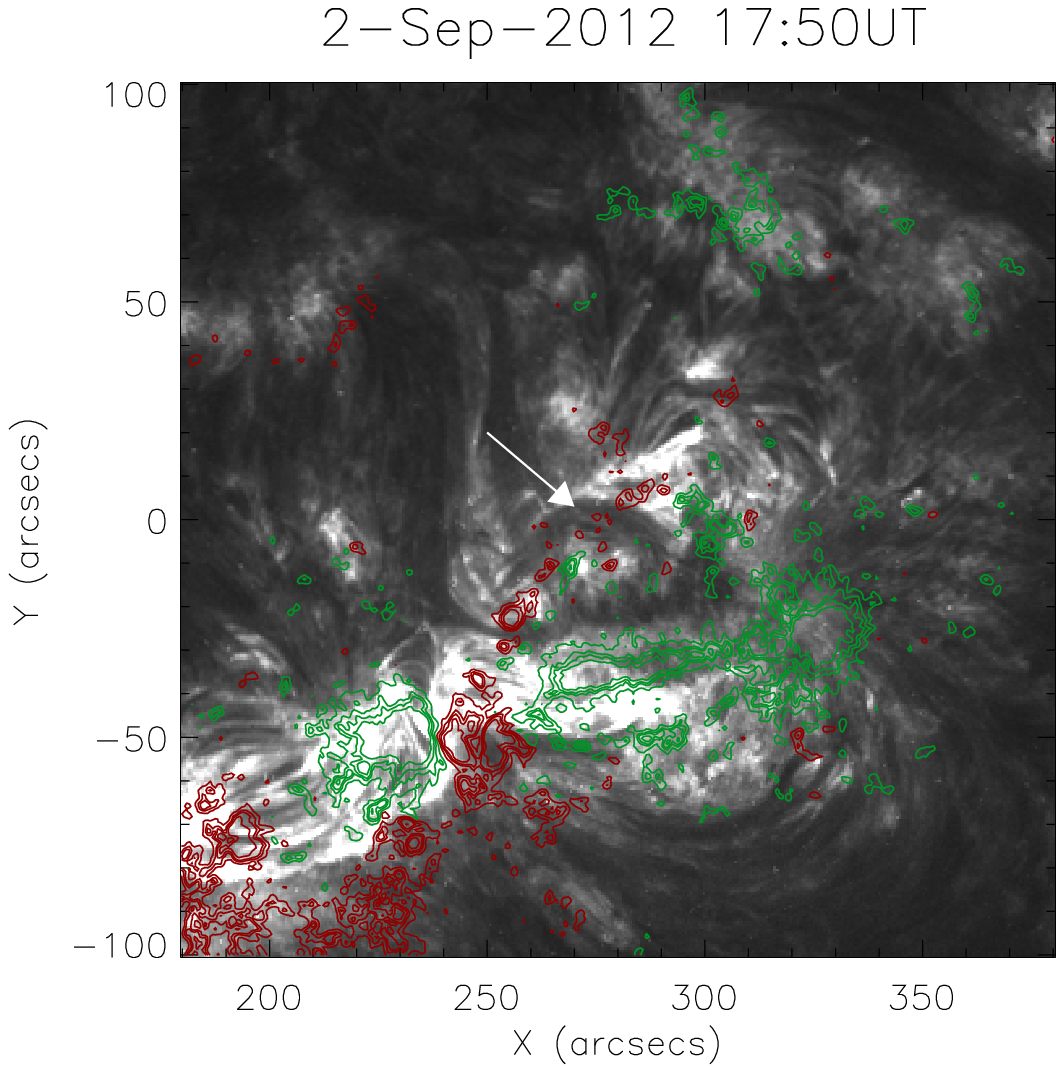}
              }
               \vspace{-0.15\textwidth}   
     
     \centerline{\Large \bf     
    \hspace{0.38 \textwidth}  \color{white}{(c)}
      \hspace{0.41\textwidth}  \color{white}{(d)}
         \hfill}
      \vspace{0.1\textwidth}    
      
\caption{Filaments observed in SDO/AIA 304~\AA~channel in different active 
regions. The filament positions are shown by  white arrows. Isocontours of the SDO/HMI magnetic 
strength values are overlaid on the images. The red and green contours represent the 
positive and negative polarities with magnetic field strength values of $\pm$ 150, 400, 600, 
and 900~G, respectively.}
   \label{fig:1}
   \end{figure}

In this section we show the images only for four events. The remaining six events are shown 
in the appendix.
The general appearance and location of filaments are depicted in Fig.~\ref{fig:1} for 
four different events observed in He~II 304~\AA. In Fig.~\ref{fig:1} their location is marked by an arrow.
Isocontours of the magnetic field are overlaid upon the 304~\AA~images 
indicating  that the filaments are associated with active regions.
The location of these filaments on the Sun is given in Table~\ref{Table:1}. We used 
the events whose apparent central meridian longitude is less than 65$^{\circ}$ to avoid 
large projection effects on the observations. 
The filaments in EUV wavelengths appeared as dark features. The shape of the 
filament in Fig.~\ref{fig:1}b appears as `S' in 304~\AA~wavelength
before the activation and the other two filaments (Fig.~\ref{fig:1}c, d) appeared 
as small arcade shaped structures. While erupting, the filaments became dark over large 
areas and a large amount of material was ejected.

A time sequence of images showed that for events 1 (Fig.~\ref{fig:1}a) and 
2 (Fig.~\ref{fig:1}b), both the eastern and the western
footpoints of the filament were detached near simultaneously from the solar surface after 
activation, the filament reached some height, and eventually fell back to the solar surface. The movies 
for these failed eruption events are available on our ftp site\footnotemark[1].
In events 3 (Fig.~\ref{fig:1}c) and 4 (Fig.~\ref{fig:1}d), the eruptions 
took place in a different manner. The eastern footpoints of both filaments were
detached first from the solar surface in association with brightening at the same
location and later the bulk of filament material was seen escaping from the solar surface. 
As soon as the eastern footpoint of the filament detached, the filament expanded in its 
structure and started to rotate. The rotational motion was observed 
starting from $\sim$10:50~UT to $\sim$11:45~UT for event~3 and it was counter-clockwise 
as seen from above.  A similar kind of rotational motion was observed for event 
4, which lasted for about 9 minutes starting from $\sim$18:11~UT to $\sim$18:20~UT. The direction 
of rotation was observed to be clockwise. These movies are available on our 
ftp site\footnotemark[1]. The zoomed in versions of the erupting filaments are also shown
in the same movies. This kind of rotation of the filament was not observed for events 1 and 2.
The movies for events 5--10 are also available on the same ftp site\footnotemark[1].
    
By examining the time series of 304~\AA~images for each event we determined the filament 
activation time (see Table~\ref{Table:1}). The projected average velocities of 
the erupting filaments also shown in Table~\ref{Table:1}, are computed by tracking the 
features of the erupting filaments within the field-of-view during their eruptions.

\footnotetext[1]{Movies generated from SDO/AIA 304~\AA~images of the filament 
eruptions associated with active region NOAA 11226, 11283, 11515, and 11560 discussed in 
this paper are available on our ftp site (ftp://ftp.iiap.res.in/sajal/). The movies are 
named according to the date of observations of the events. The movies for events 5--10
discussed in this paper are also available on the same ftp site.}

\subsection{Flows in and around the filament footpoints} 
\begin{figure}    
   \centerline{\hspace*{0.015\textwidth}
               \includegraphics[width=0.53\textwidth,clip=]{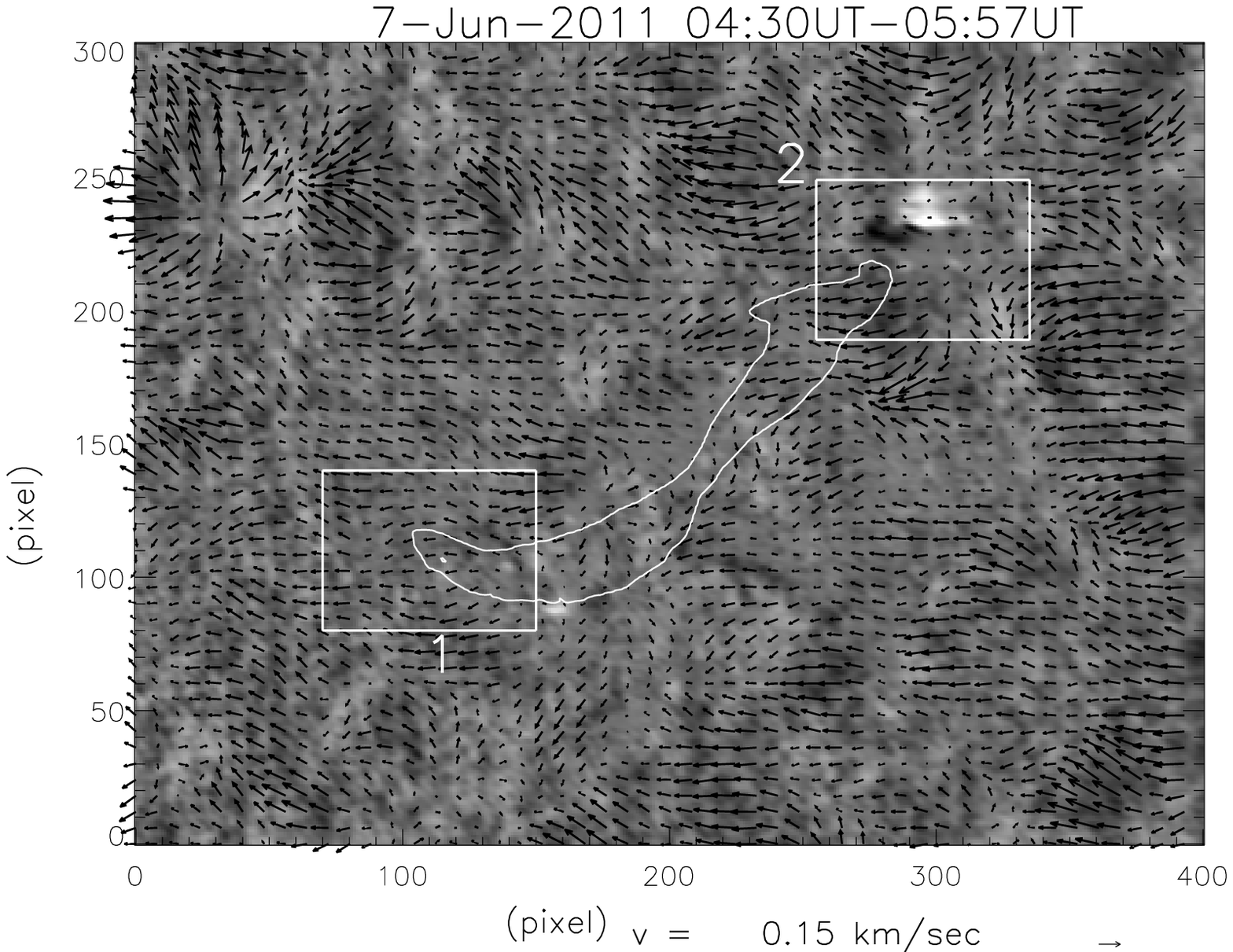}
               \hspace*{-0.06\textwidth}
               \includegraphics[width=0.53\textwidth,clip=]{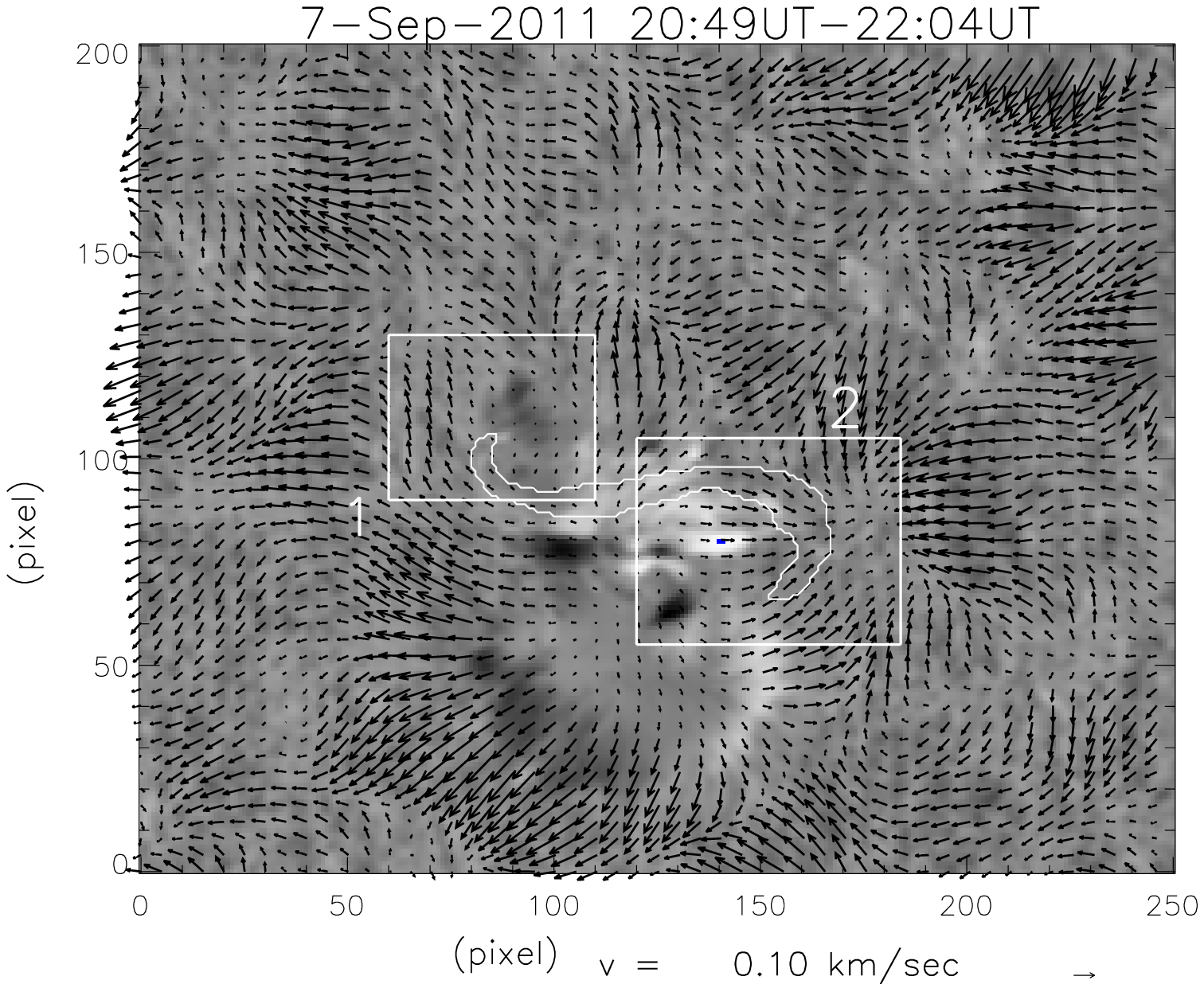}
              }
           \vspace{-0.1\textwidth}   
     
     \centerline{\Large \bf     
    \hspace{0.4 \textwidth}  \color{white}{(a)}
      \hspace{0.38\textwidth}  \color{white}{(b)}
         \hfill}
         \vspace{0.07\textwidth}

   \centerline{\hspace*{0.015\textwidth}
               \includegraphics[width=0.53\textwidth,clip=]{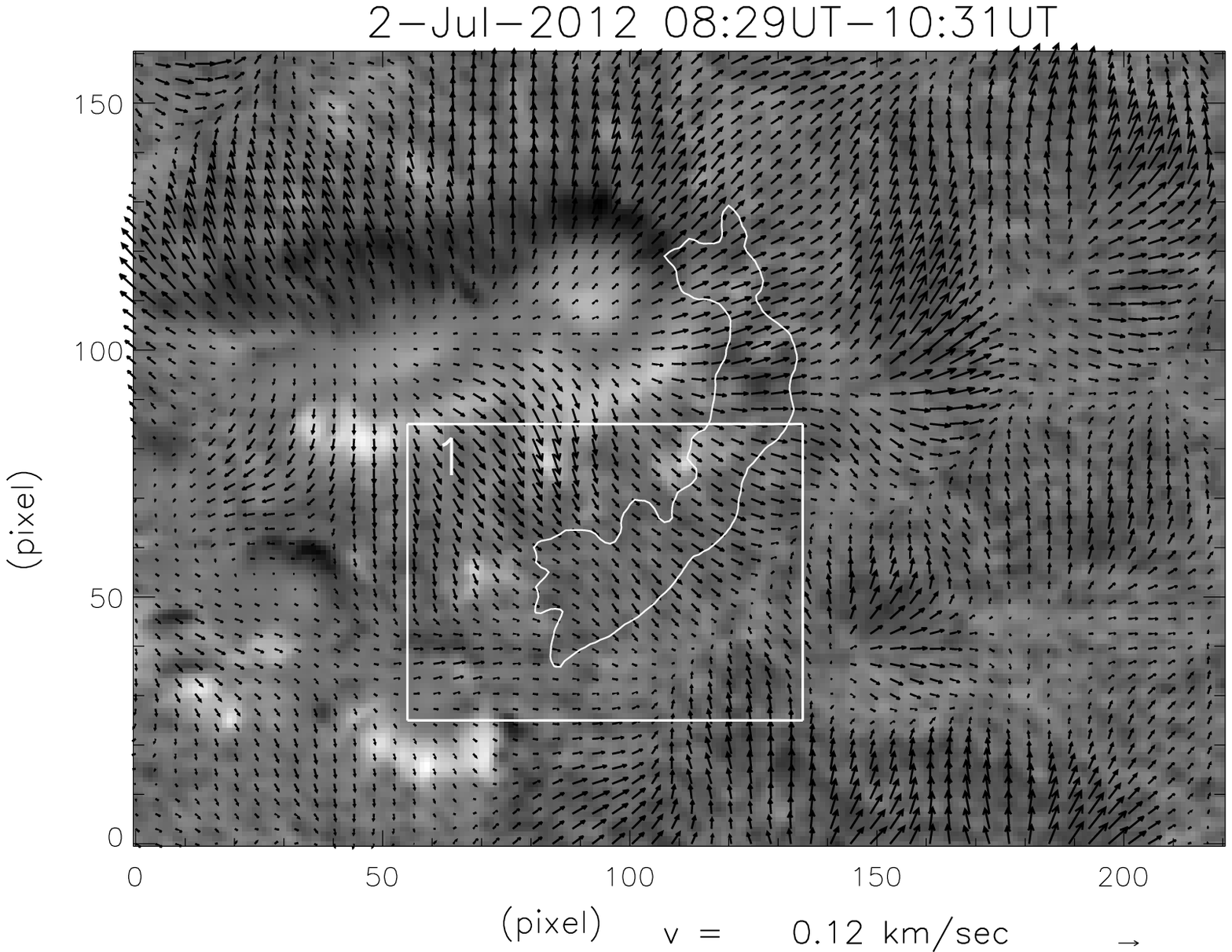}
               \hspace*{-0.06\textwidth}
               \includegraphics[width=0.53\textwidth,clip=]{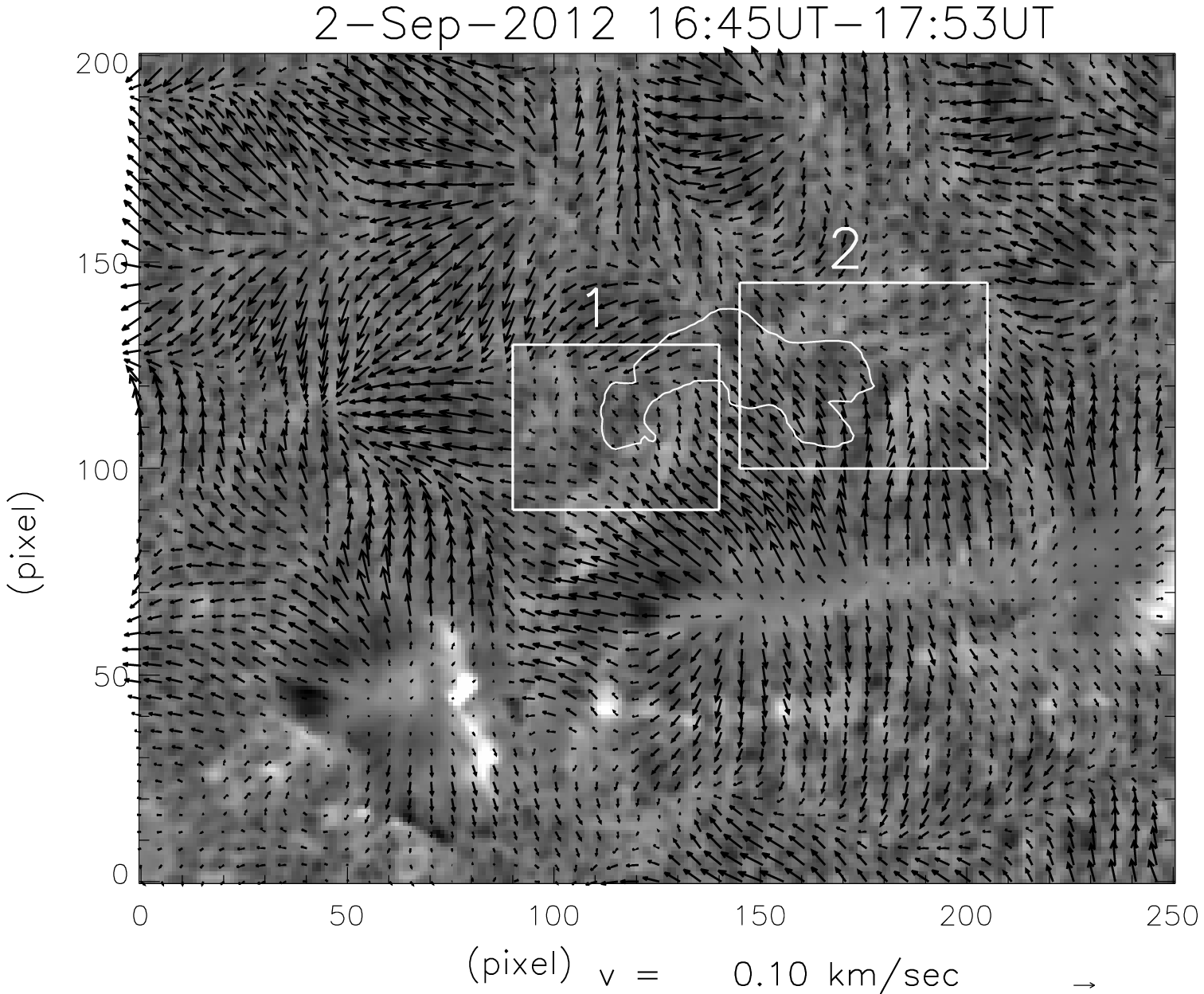}
              }
      
       \vspace{-0.1\textwidth}   
     
     \centerline{\Large \bf     
    \hspace{0.42 \textwidth}  \color{white}{(c)}
      \hspace{0.35\textwidth}  \color{white}{(d)}
         \hfill}
      \vspace{0.10\textwidth}    
      
\caption{Averaged horizontal velocity vectors (arrows) are overlaid upon the 
averaged dopplergram for the first to fourth events. The black, white, and gray  regions in 
the averaged dopplergram represent, respectively, the line-of-sight doppler velocities 
with upflow, downflow, and close to zero velocity. The averaging time interval is given above each figure.
The white contours of the filaments, extracted from the 304 \AA~SDO/AIA images are 
overlaid upon the averaged dopplergram. The boxed regions 1 and 2 show the location of the eastern 
and western footpoints of the filament, respectively.}
   \label{fig:2}
   \end{figure}

In order to measure photospheric flows in and around sunspot regions and filaments, we applied
the Fourier local correlation tracking ({FLCT;} \opencite{Welsch04}) technique to dopplergrams.
Selecting the two parameters such as the time difference between the two images and the size of 
the Gaussian apodizing window functions are crucial in finding the velocity vectors.
We used the dopplergram images which are 3 minutes apart and the apodizing window width 
of 9$^{\prime\prime}$ as a window function. The obtained velocity vectors are averaged over $\sim$1.5 hours to
examine the long term flows in and around the active regions. Figure~\ref{fig:2} shows
the long lived flows for the first four events listed in Table~\ref{Table:1}. The contours of the 
filament are overlaid upon the velocity map to mark the location of the filament and its end
points. Filament contours are extracted from the 304~\AA~ images for each event. The boxed 
regions 1 and 2 show the eastern and western footpoints of the filament regions, respectively. 
Around the sunspots outward flows are observed. At filament end points either converging motions or large 
scale outflows were observed.

\begin{landscape}

\begin{table}[h!]

\tiny
\vspace{0.2in}
\caption{Event number is shown in the first column of the table. The date of observations, filament 
location on the Sun, associated active region number, filament activation time (AT), GOES class of the associated flare,
average velocities and fast rise time (FRT) of the erupting filaments are given 
in the next seven columns. The duration and type of rotational motions observed for each events 
are listed in the final four columns. In the tenth and twelfth columns of the table, the abbreviation AC corresponds to 
anti-clockwise and C corresponds to clockwise direction of rotational motions.}

\begin{tabular}{lcccccccccccccl}

\hline

Event &Date & Filament & AR     & AT     & GOES      & Velocity &  FRT &\multicolumn{3}{c}{Rotational Motion Observed in }\\
 No.  &     & Location & NOAA   & (UT)   & Flares    & (km s$^{-1}$)   & (UT) &\multicolumn{2}{c}{Eastern footpoint}  &  \multicolumn{2}{c}{Western footpoint}\\
      &     &          &        &        &           &          &      & Duration (UT)   &  Type           &  Duration (UT) & Type\\
 
\hline
1  & 07--Jun--2011      & $\sim$S22W64 &  11226 & $\sim$05:56UT & M2.5  & 181.7$\pm$7.1& $\sim$06:14 & 05:55:26--06:02:11 & AC & 06:08:56--06:23:11 & C  &\\
2  & 07--Sep--2011      & $\sim$N14W30 &  11283 & $\sim$22:08UT & X1.8  & 193.9$\pm$1.9& $\sim$22:30 & 22:13:22--22:17:07 & C  & 22:17:52--22:24:37 & AC & \\
3  & 02--Jul--2012      & $\sim$S17E03 &  11515 & $\sim$10:31UT & M5.6  & 90.9$\pm$0.8 & $\sim$10:43 & 10:34:25--10:46:25 & AC & --                 & -- &\\
4  & 02--Sep--2012      & $\sim$N03W18 &  11560 & $\sim$17:52UT & C5.5  & 42.9$\pm$0.4 & $\sim$18:03 & 18:08:51--18:14:51 &  C & 17:56:51--18:04:21 & AC &\\
5  & 07--Apr--2012      & $\sim$N18W30 &  11451 & $\sim$17:45UT & C2.4  & 39.4$\pm$1.6 & $\sim$17:58 & 17:45:34--17:52:19 & AC & 17:45:34--17:56:04 &  C &\\
6  & 31--Dec--2013      & $\sim$S16W32 &  11936 & $\sim$21:26UT & M6.4  & 149.0$\pm$2.7& $\sim$21:46 & 21:37:54--21:43:54 & AC & 21:16:54--21:36:24 &  C & \\
7  & 01--Jan--2014      & $\sim$S16W46 &  11936 & $\sim$16:18UT & --    & 11.4$\pm$0.3 & $\sim$16:23 & 16:09:24--16:18:24 & AC & 16:19:09--16:23:39 &  C  &\\
8  & 01--Jan--2014      & $\sim$S16W48 &  11936 & $\sim$18:21UT & M9.9  & 73.4$\pm$2.9 & $\sim$18:40 & 18:14:39--18:23:39 & AC & 18:10:09--18:25:09 &  C  &\\
9  & 04--Apr--2014      & $\sim$N13E15 &  12027 & $\sim$13:28UT & C8.3  & 109.5$\pm$3.0& $\sim$13:33 & 13:28:17--13:39:32 &  C & 13:46:17--13:58:17 & AC &\\
10 & 15--Apr--2014      & $\sim$S18E23 &  12035 & $\sim$17:33UT & C7.3  & 44.6$\pm$1.4 & $\sim$17:51 & 17:31:19--17:44:49 & AC & --                 & -- &\\

\hline 
\end{tabular}
\label{Table:1}
\end{table}
\end{landscape}

\begin{figure}    
   \centerline{\hspace*{0.015\textwidth}
               \includegraphics[width=0.53\textwidth,clip=]{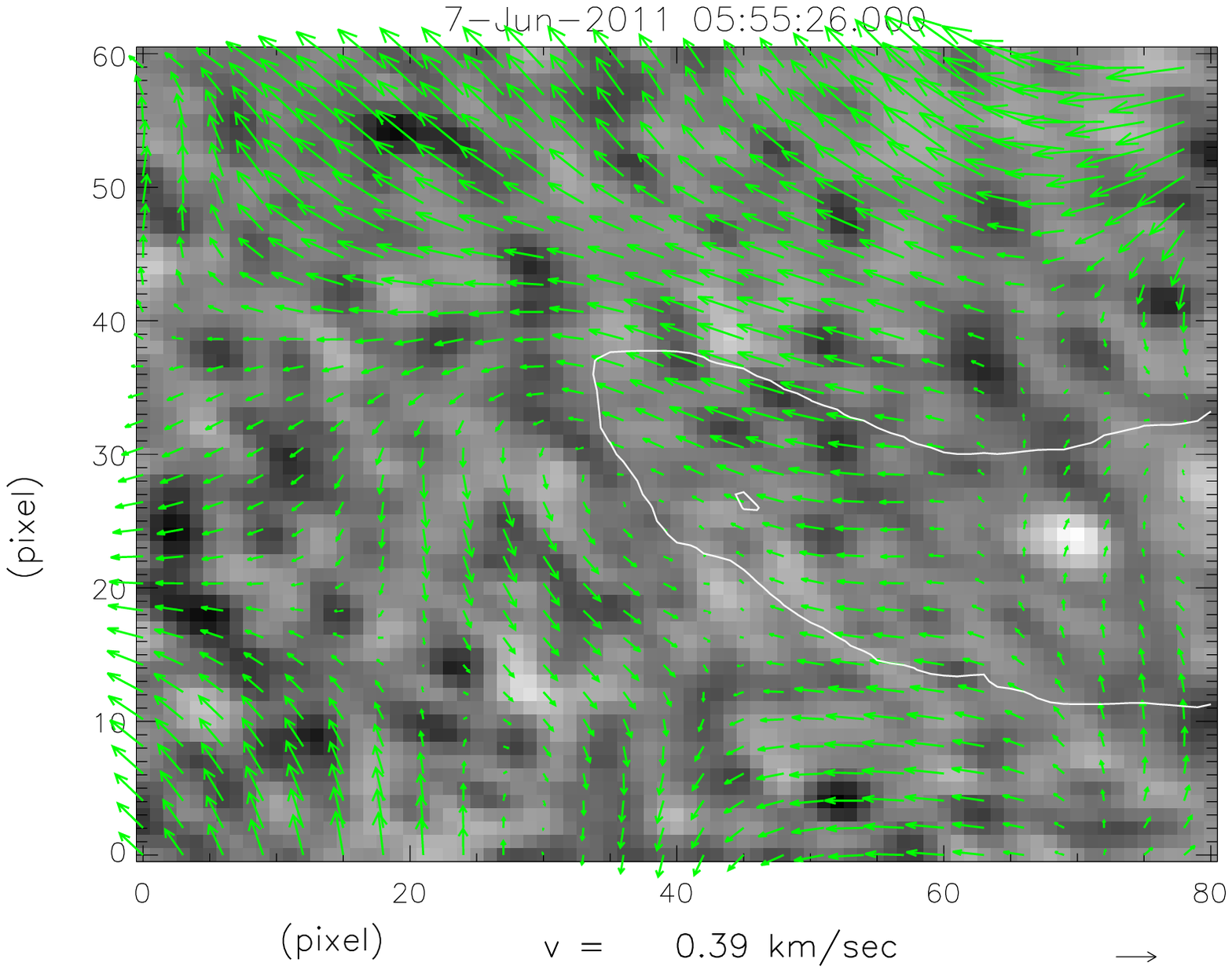}
               \hspace*{-0.06\textwidth}
               \includegraphics[width=0.535\textwidth,clip=]{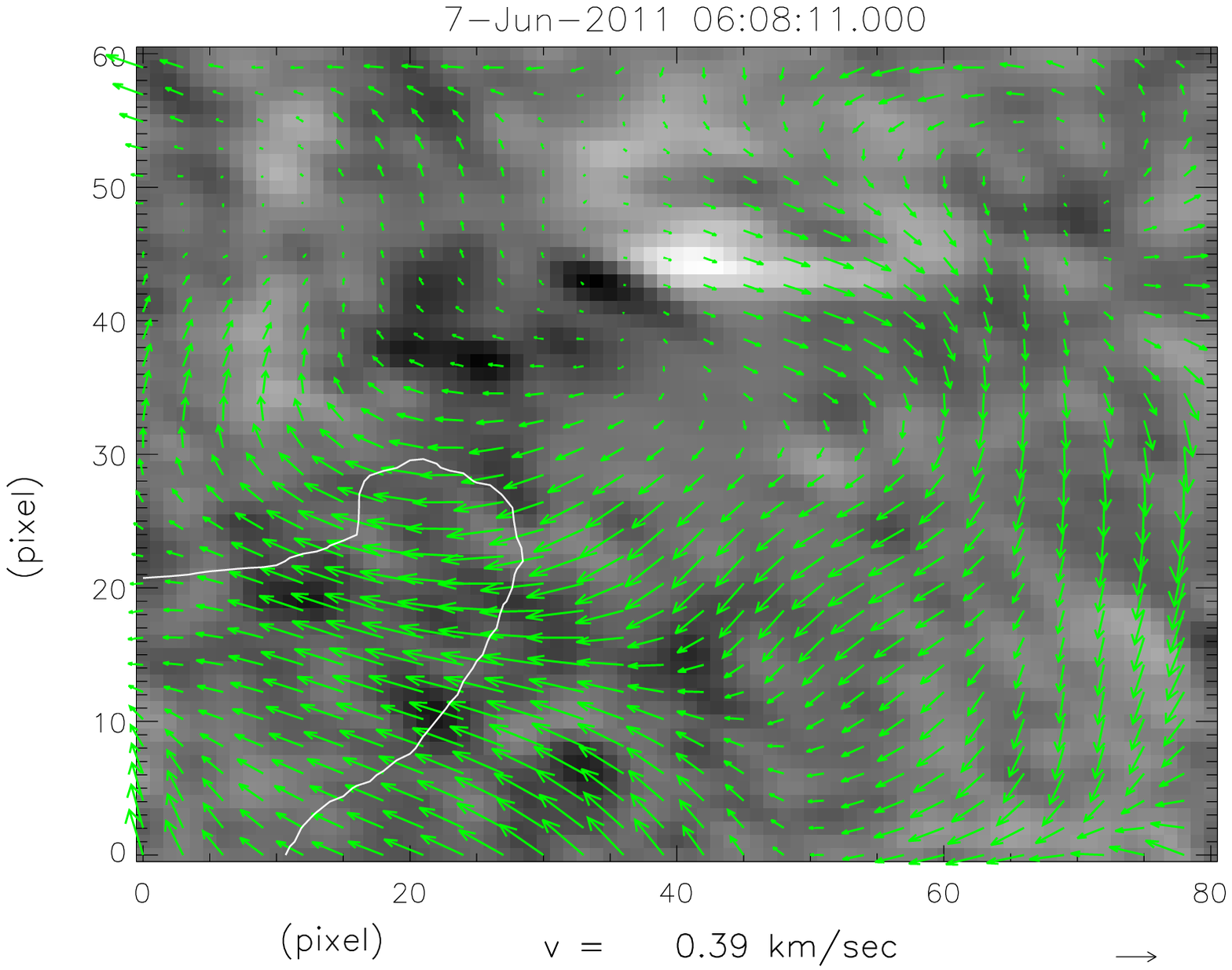}
              }
     \vspace{0.01\textwidth}   
   \centerline{\hspace*{0.015\textwidth}
               \includegraphics[width=0.53\textwidth,clip=]{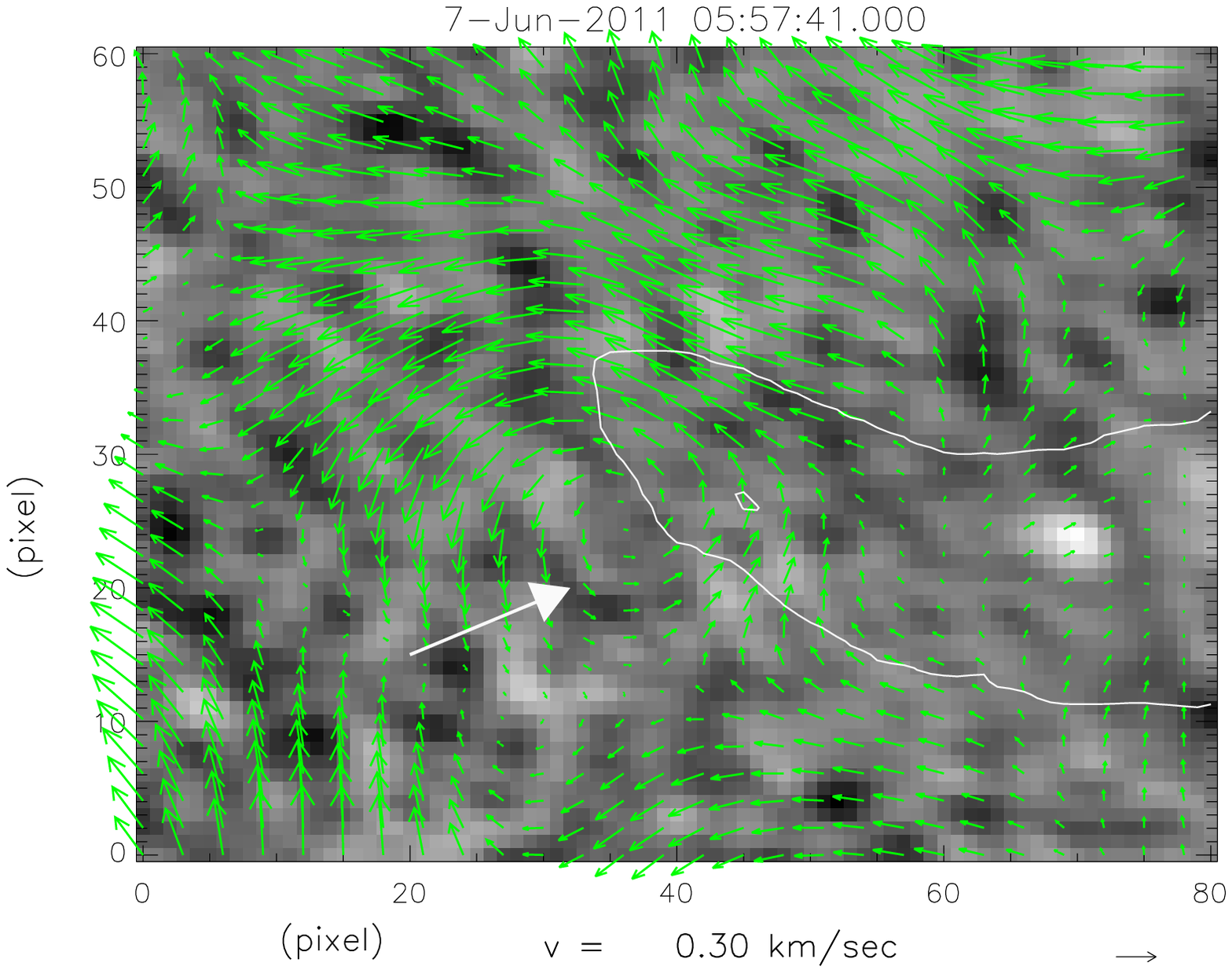}
               \hspace*{-0.06\textwidth}
               \includegraphics[width=0.535\textwidth,clip=]{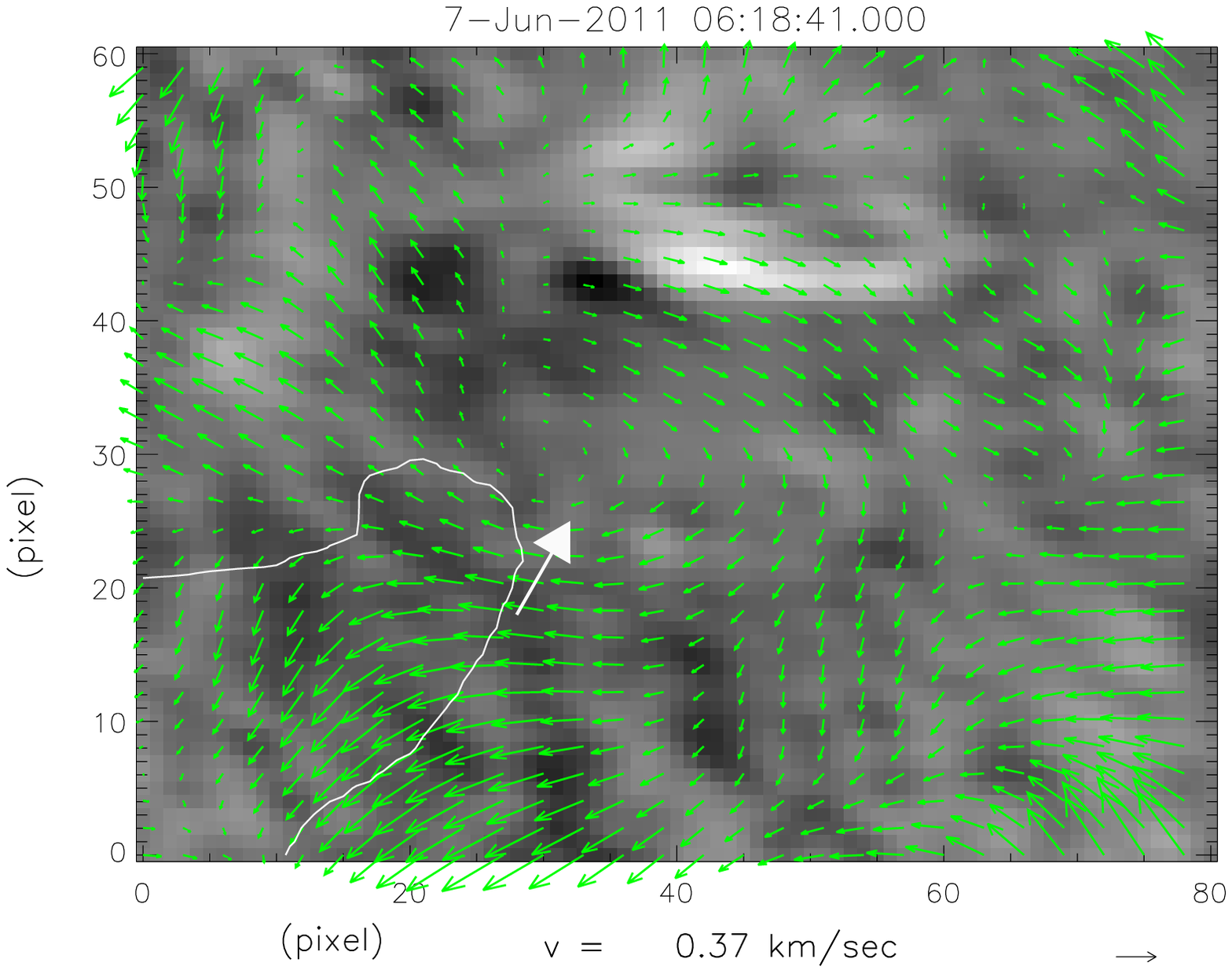}
              }
      \vspace{0.01\textwidth}    
      \centerline{\hspace*{0.015\textwidth}
               \includegraphics[width=0.53\textwidth,clip=]{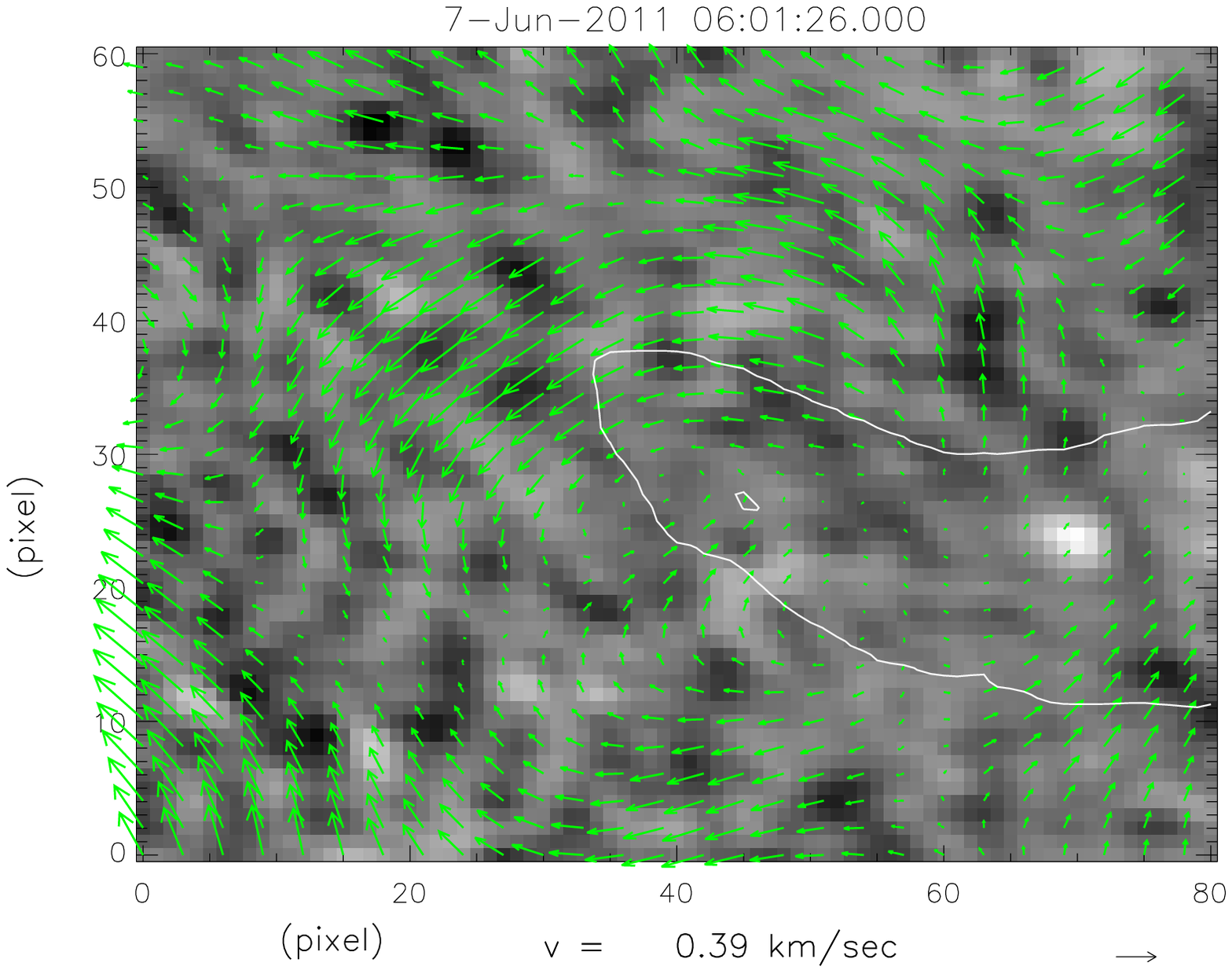}
               \hspace*{-0.06\textwidth}
               \includegraphics[width=0.535\textwidth,clip=]{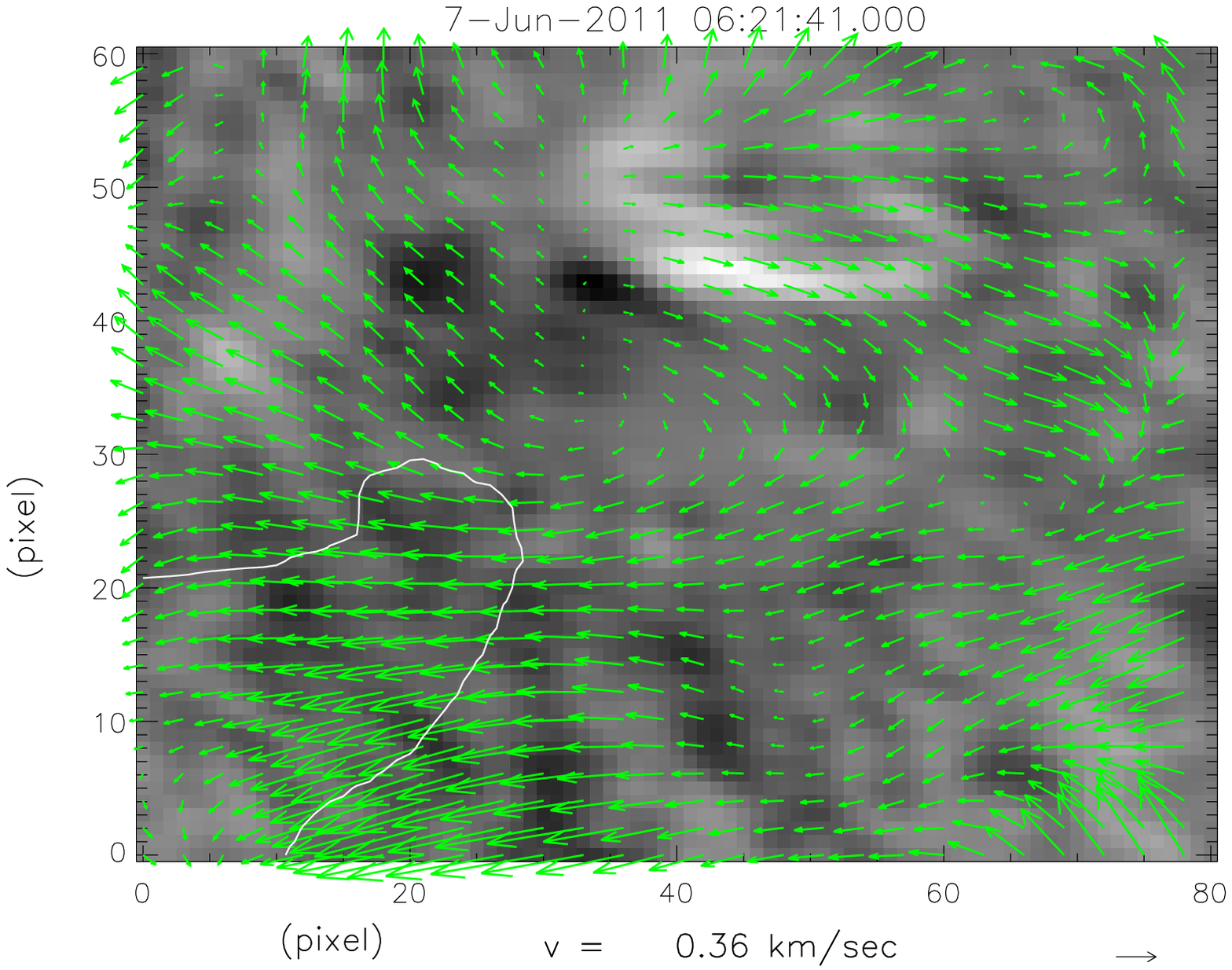}
              }
      \vspace{0.05\textwidth}    
\caption{The temporal evolution of rotational velocity pattern observed for 
event 1. The left column corresponds to location 1 and right column for location 2 of 
Fig.~\ref{fig:2}(a). The filament contour extracted from the
304 \AA~image closest in time is overlaid upon the dopplergram. 
The rotation regions are indicated by a white arrow. The date and time of 
the computed velocity is shown above each map. The size of the arrow below
 each map represents the magnitude of velocity.}
   \label{fig:3}
   \end{figure}

In the boxed regions, during the early stage of the filament eruption the situation
was different. The flow fields in the boxed regions of Fig.~\ref{fig:2} showed rotational 
motions for a few minutes. These velocity flow fields were obtained from time sequence 
of images without averaging the velocity maps.  Figure~\ref{fig:3} shows 
the temporal sequence of flow field in the footpoints of the filaments which erupted on 
07 June 2011 (event 1). We observed counter-clockwise rotation at the eastern footpoint and 
clockwise rotation at the western footpoint. The rotational pattern persisted for 7 and 15 minutes
in the eastern and western footpoints, respectively.

\begin{figure}    
   \centerline{\hspace*{0.015\textwidth}
               \includegraphics[width=0.54\textwidth,clip=]{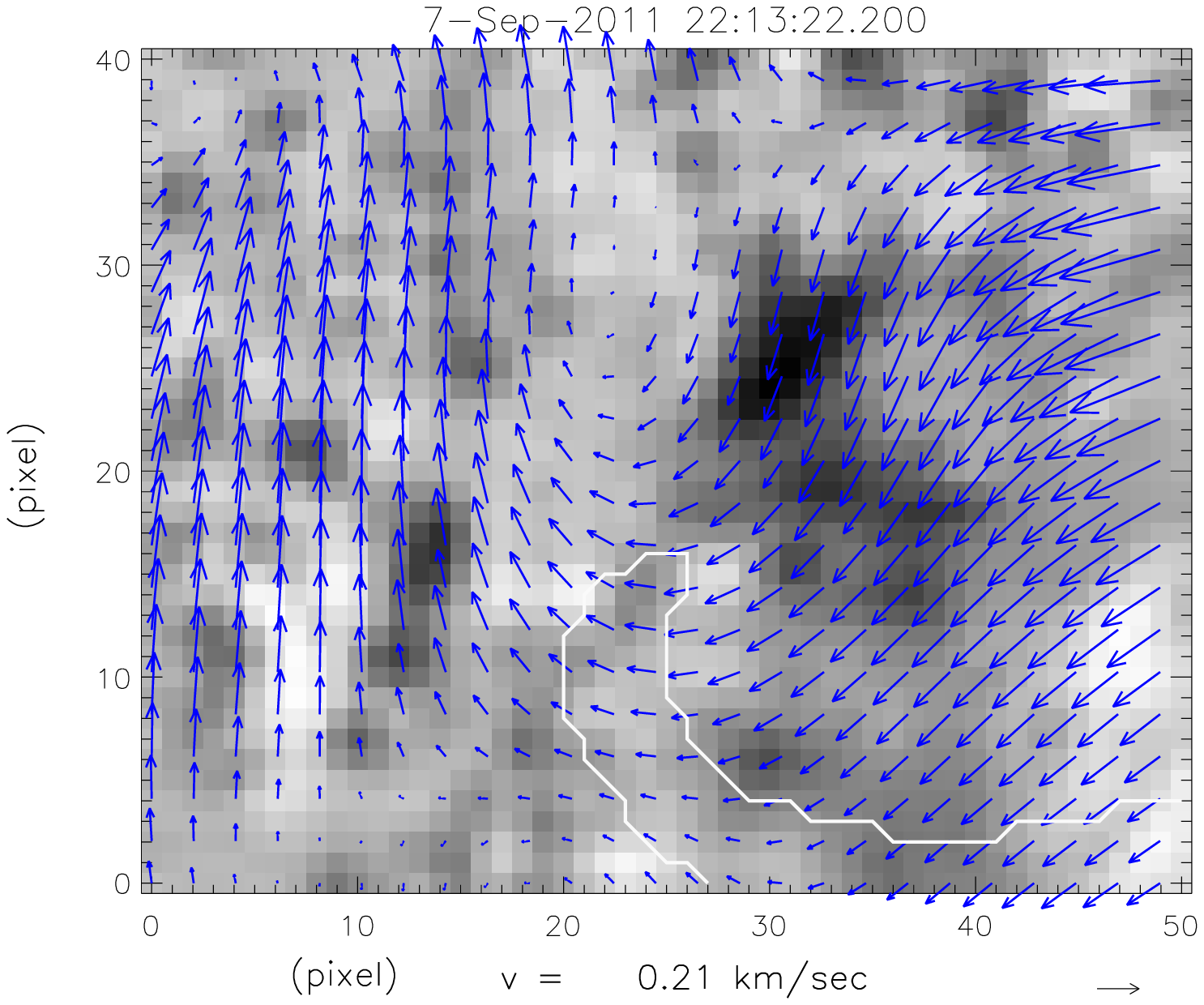}
               \hspace*{-0.08\textwidth}
               \includegraphics[width=0.54\textwidth,clip=]{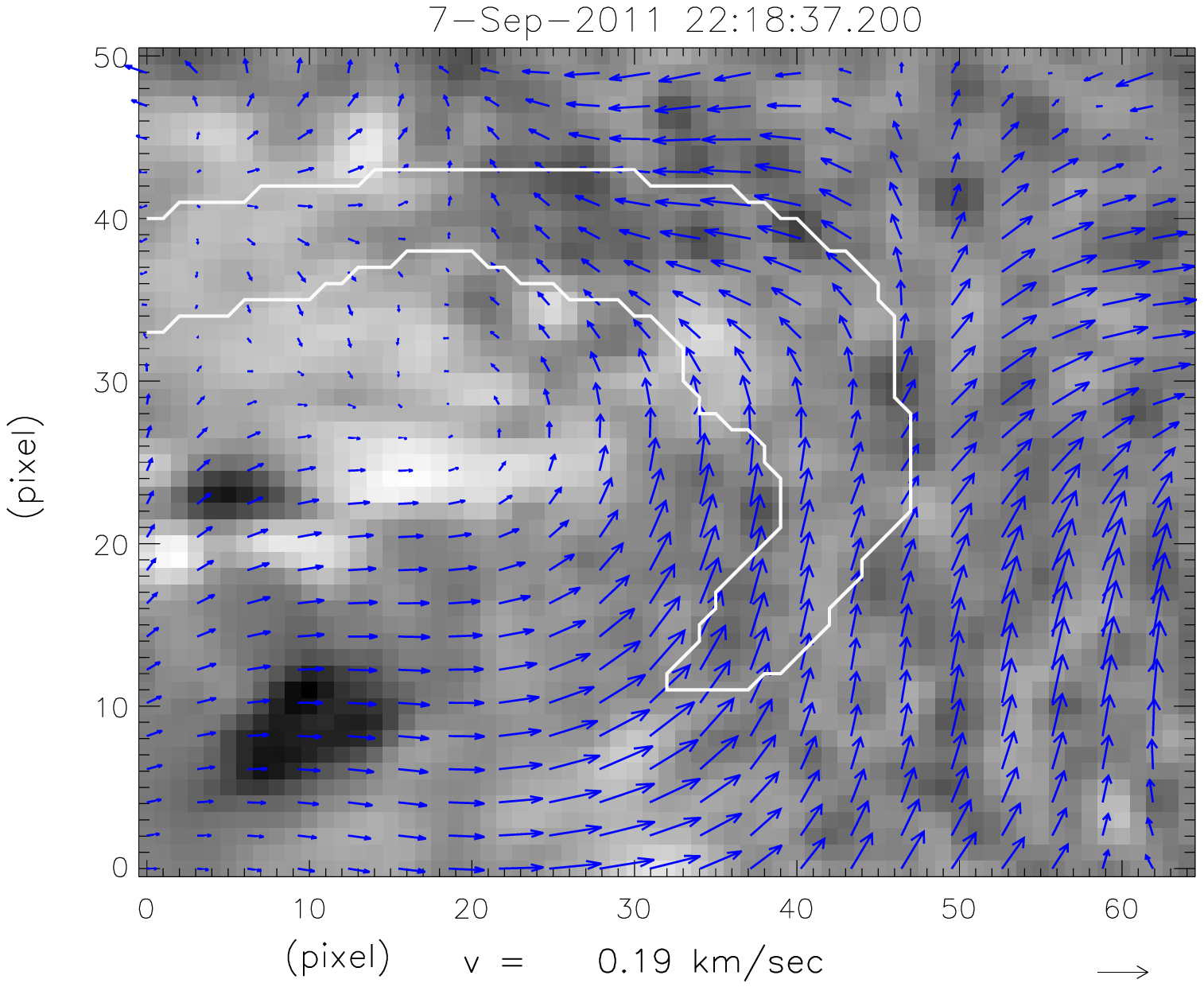}
              }
     \vspace{0.01\textwidth}   
   \centerline{\hspace*{0.015\textwidth}
               \includegraphics[width=0.54\textwidth,clip=]{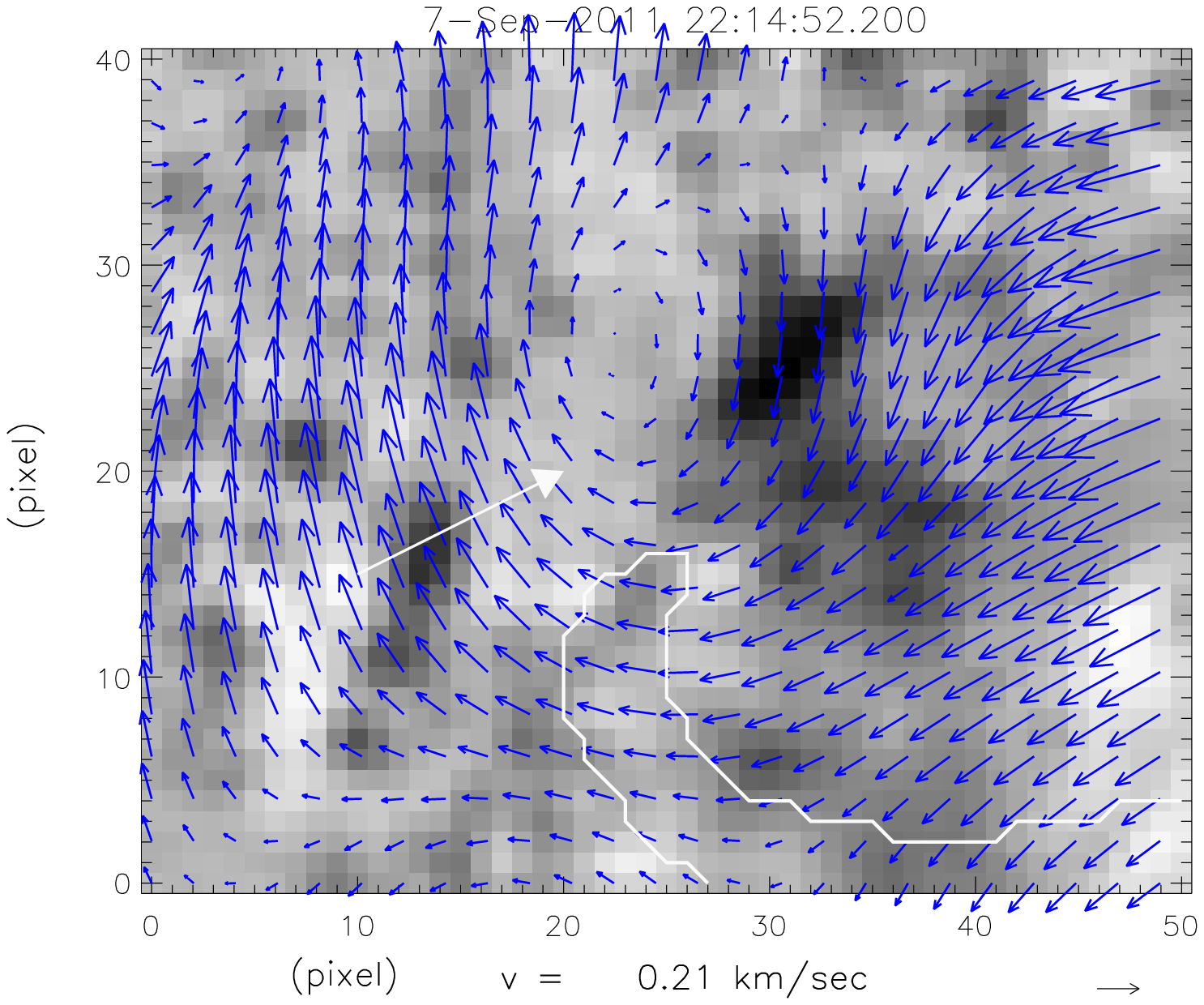}
               \hspace*{-0.08\textwidth}
               \includegraphics[width=0.54\textwidth,clip=]{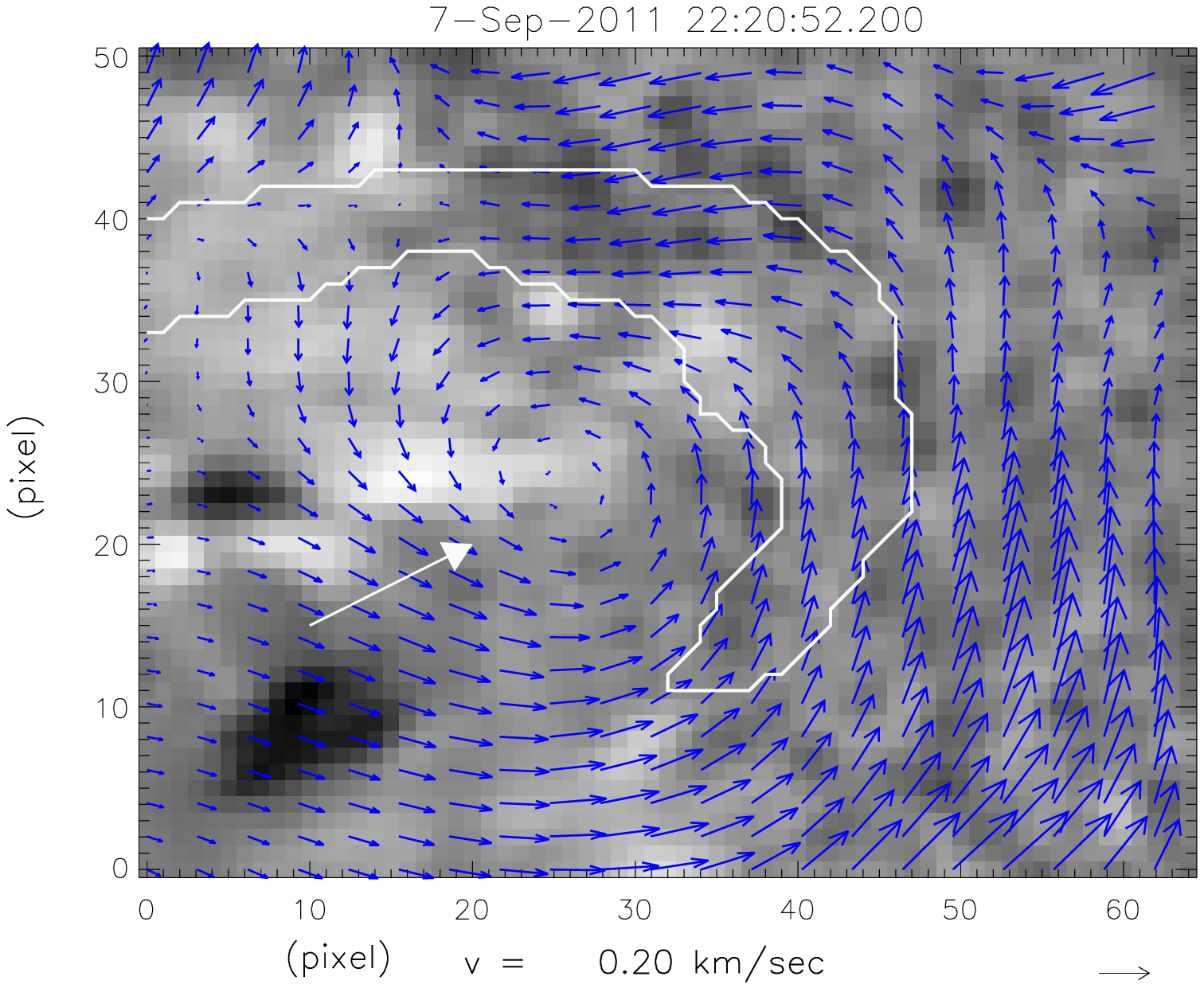}
              }
      \vspace{0.01\textwidth}    
      \centerline{\hspace*{0.015\textwidth}
               \includegraphics[width=0.54\textwidth,clip=]{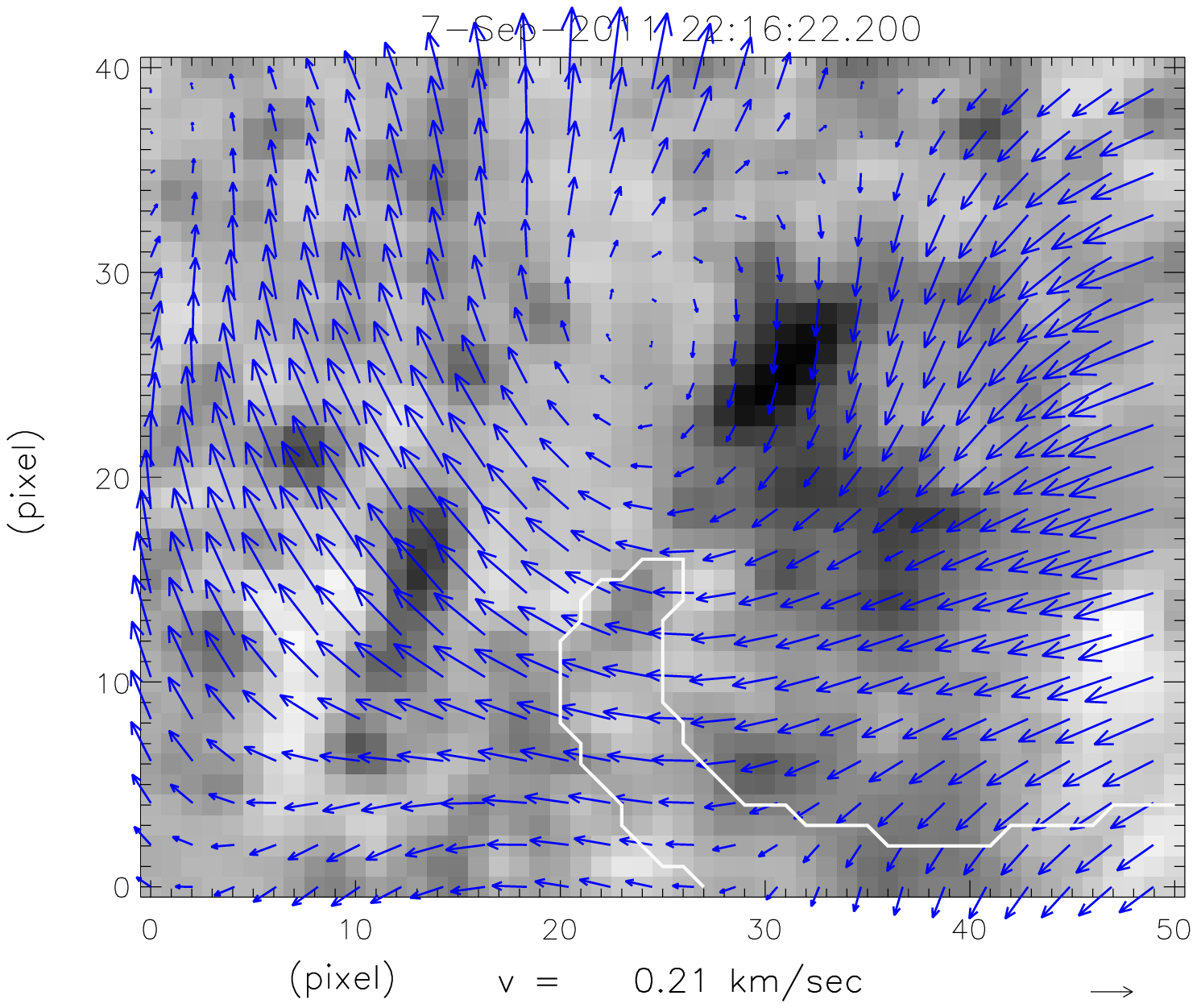}
               \hspace*{-0.08\textwidth}
               \includegraphics[width=0.54\textwidth,clip=]{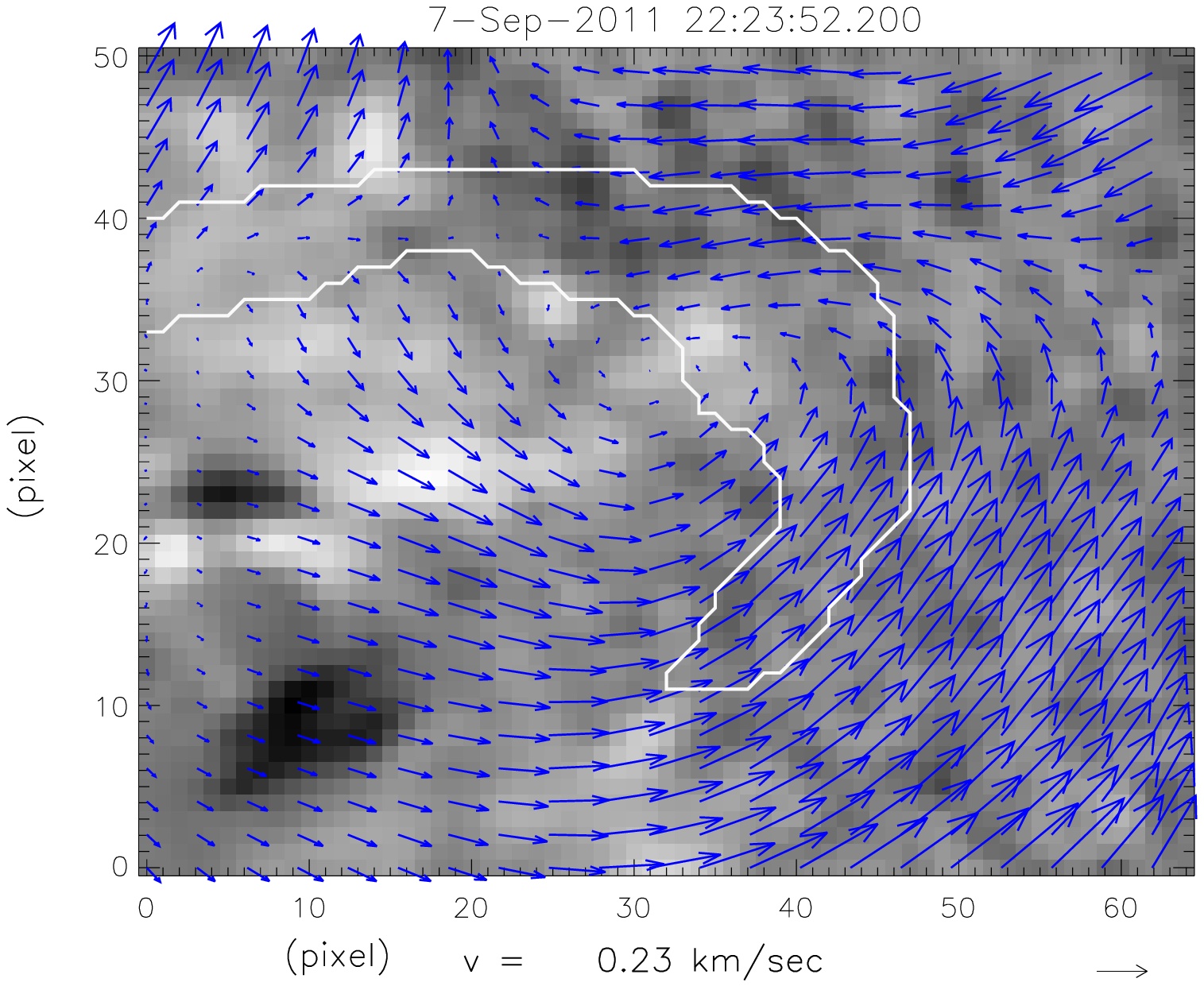}
              }
      \vspace{0.05\textwidth}    
\caption{Same as Fig.~\ref{fig:3}, but for event 2.}
   \label{fig:4}
   \end{figure}

\begin{figure}    
   \centerline{\hspace*{0.015\textwidth}
               \includegraphics[width=0.54\textwidth,clip=]{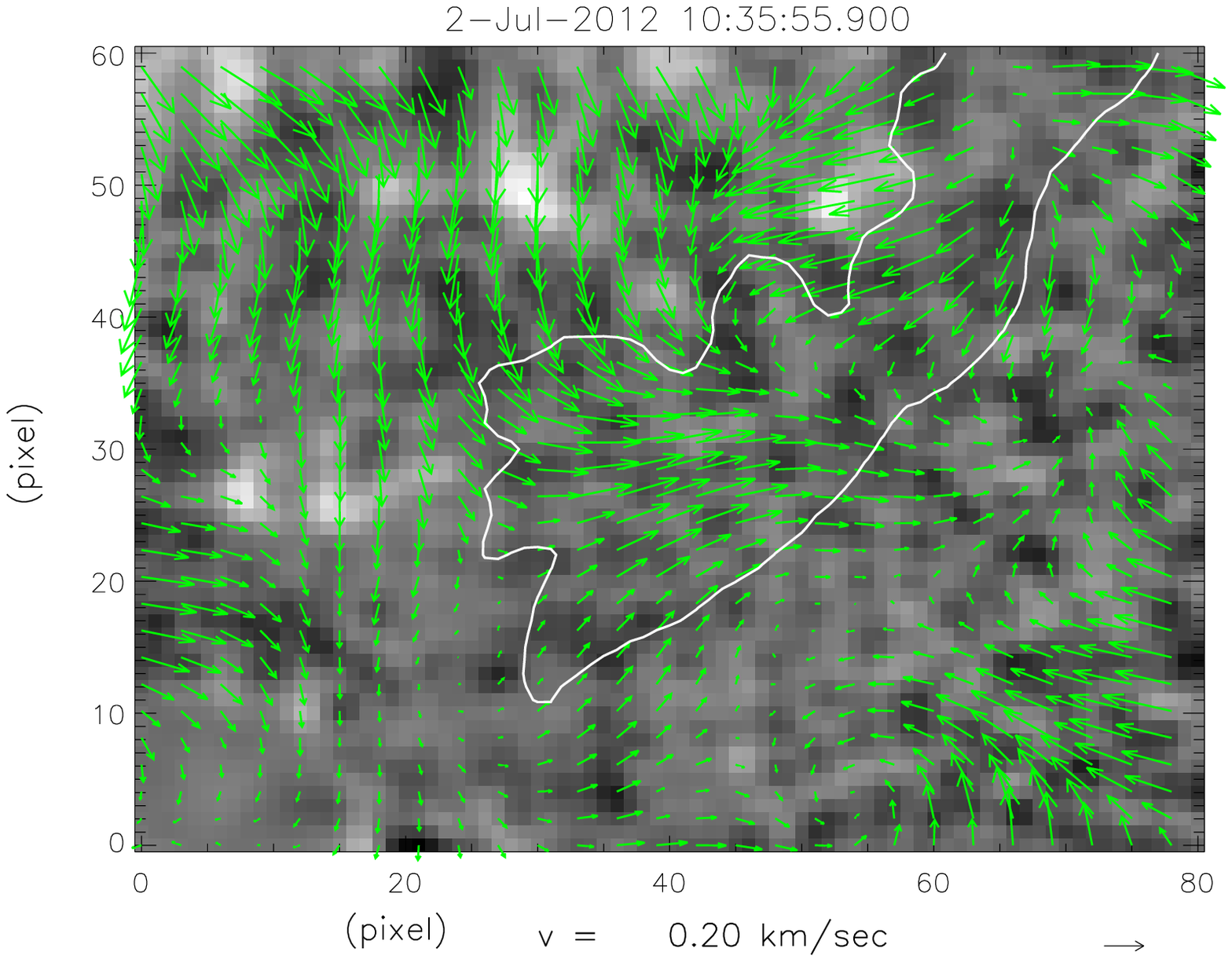}
               \hspace*{-0.08\textwidth}
               \includegraphics[width=0.54\textwidth,clip=]{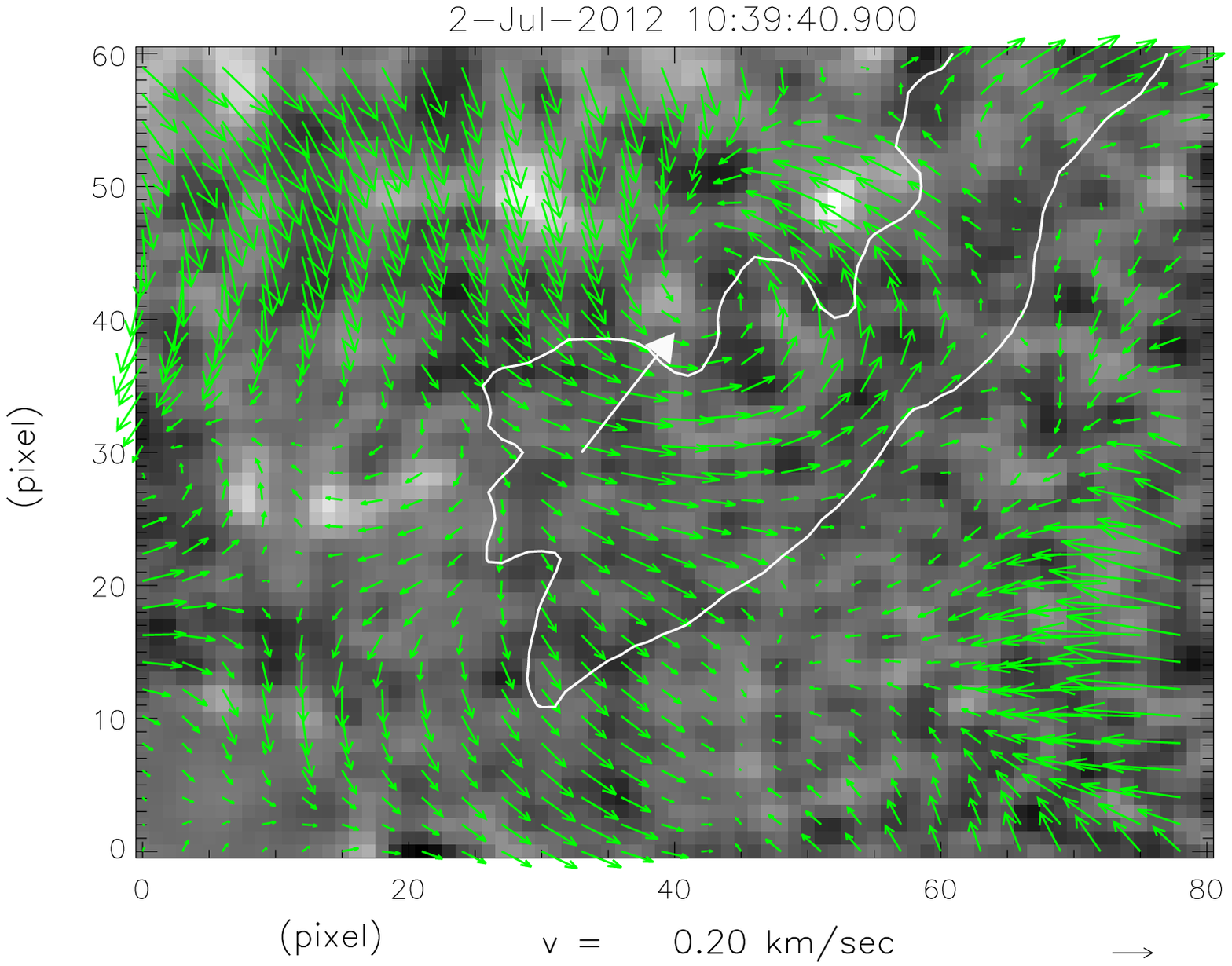}
              }
     \vspace{0.01\textwidth}   
   \centerline{\hspace*{0.015\textwidth}
              \includegraphics[width=0.53\textwidth,clip=]{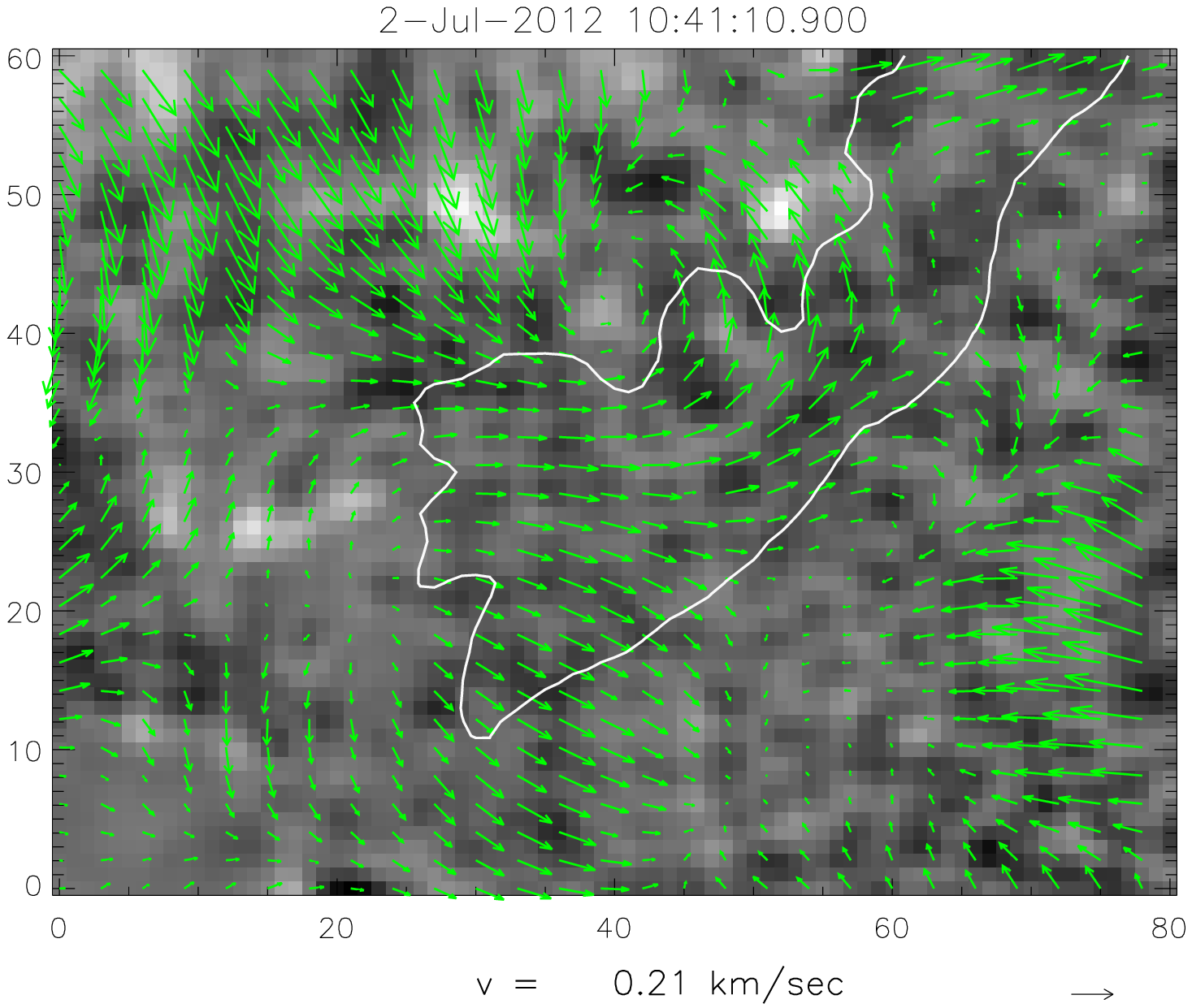}
              \hspace*{-0.065\textwidth}
               \includegraphics[width=0.53\textwidth,clip=]{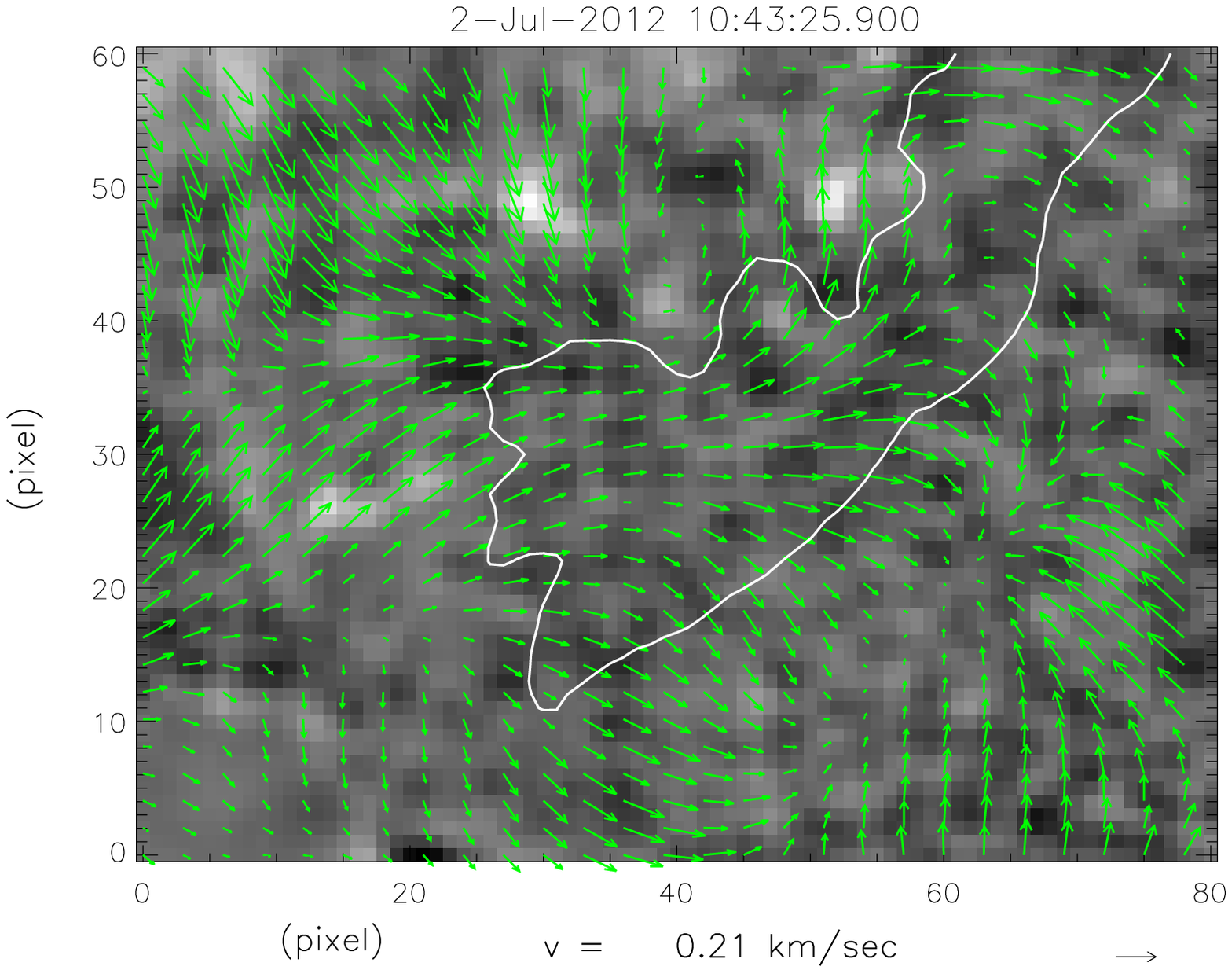}
              }
      \vspace{0.05\textwidth}    
\caption{Same as Fig.~\ref{fig:3}, but for the region 1 of  event 3.}
   \label{fig:5}
   \end{figure}

\begin{figure}    
   \centerline{\hspace*{0.015\textwidth}
               \includegraphics[width=0.54\textwidth,clip=]{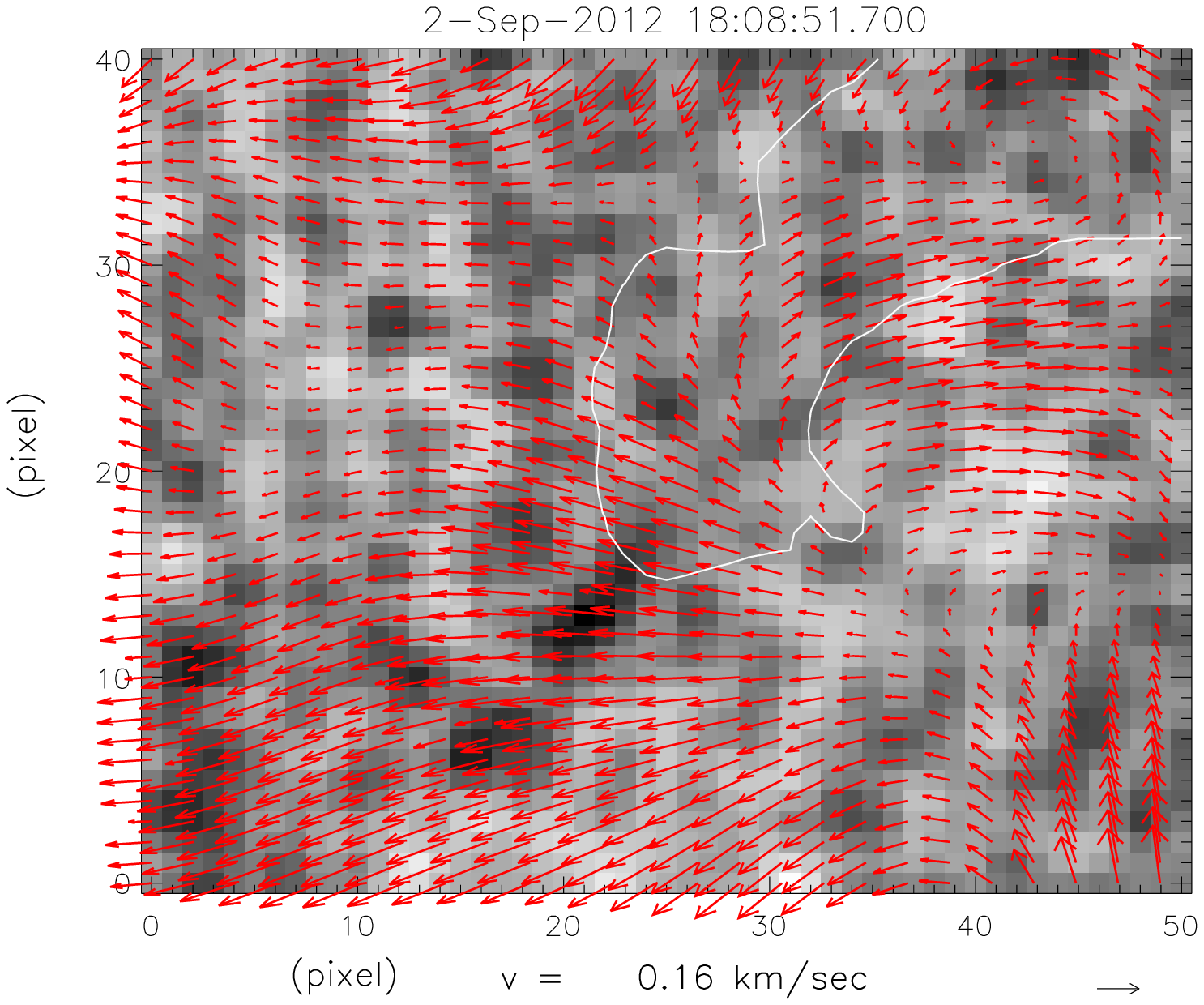}
               \hspace*{-0.08\textwidth}
               \includegraphics[width=0.54\textwidth,clip=]{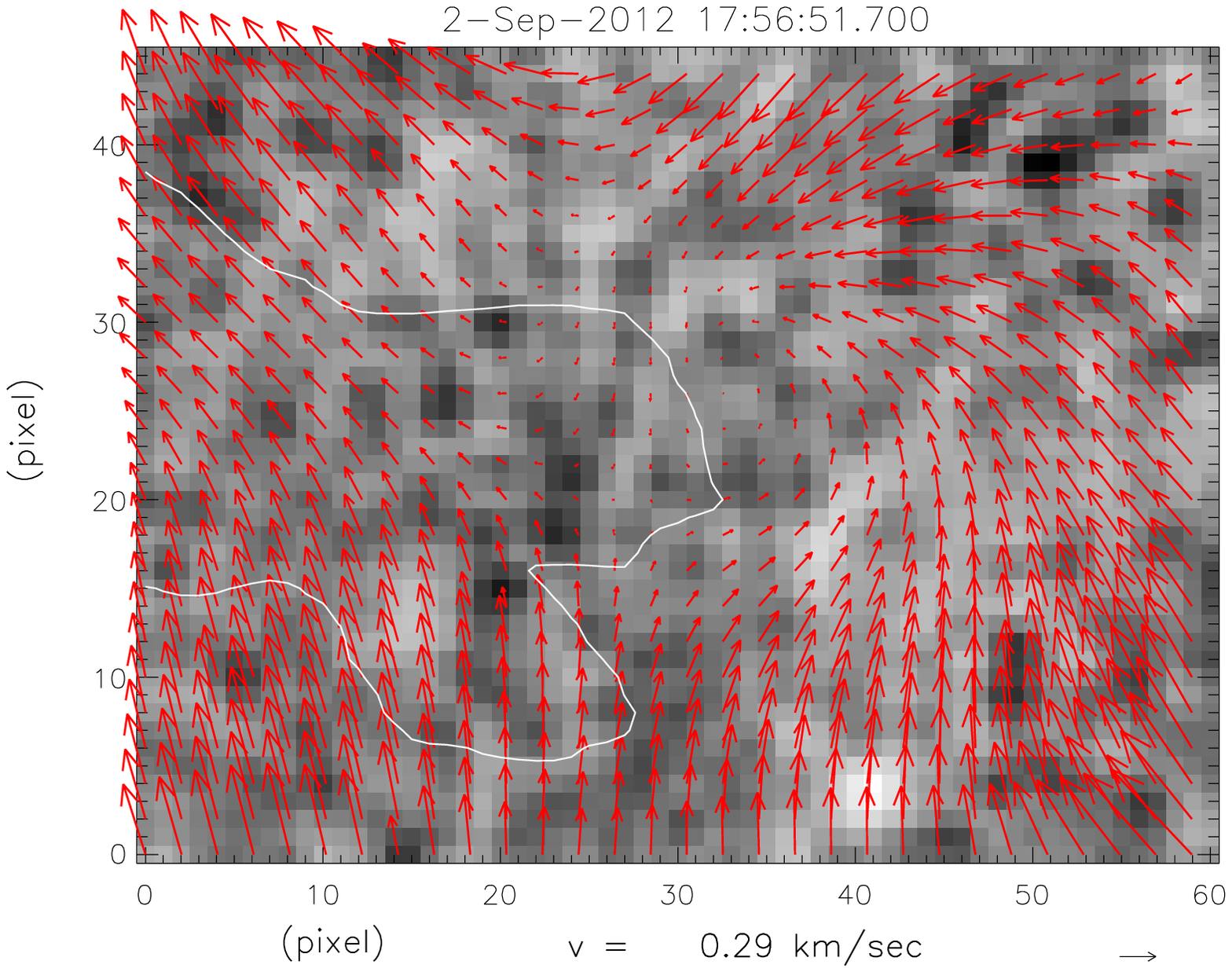}
              }
     \vspace{0.01\textwidth}   
   \centerline{\hspace*{0.015\textwidth}
               \includegraphics[width=0.54\textwidth,clip=]{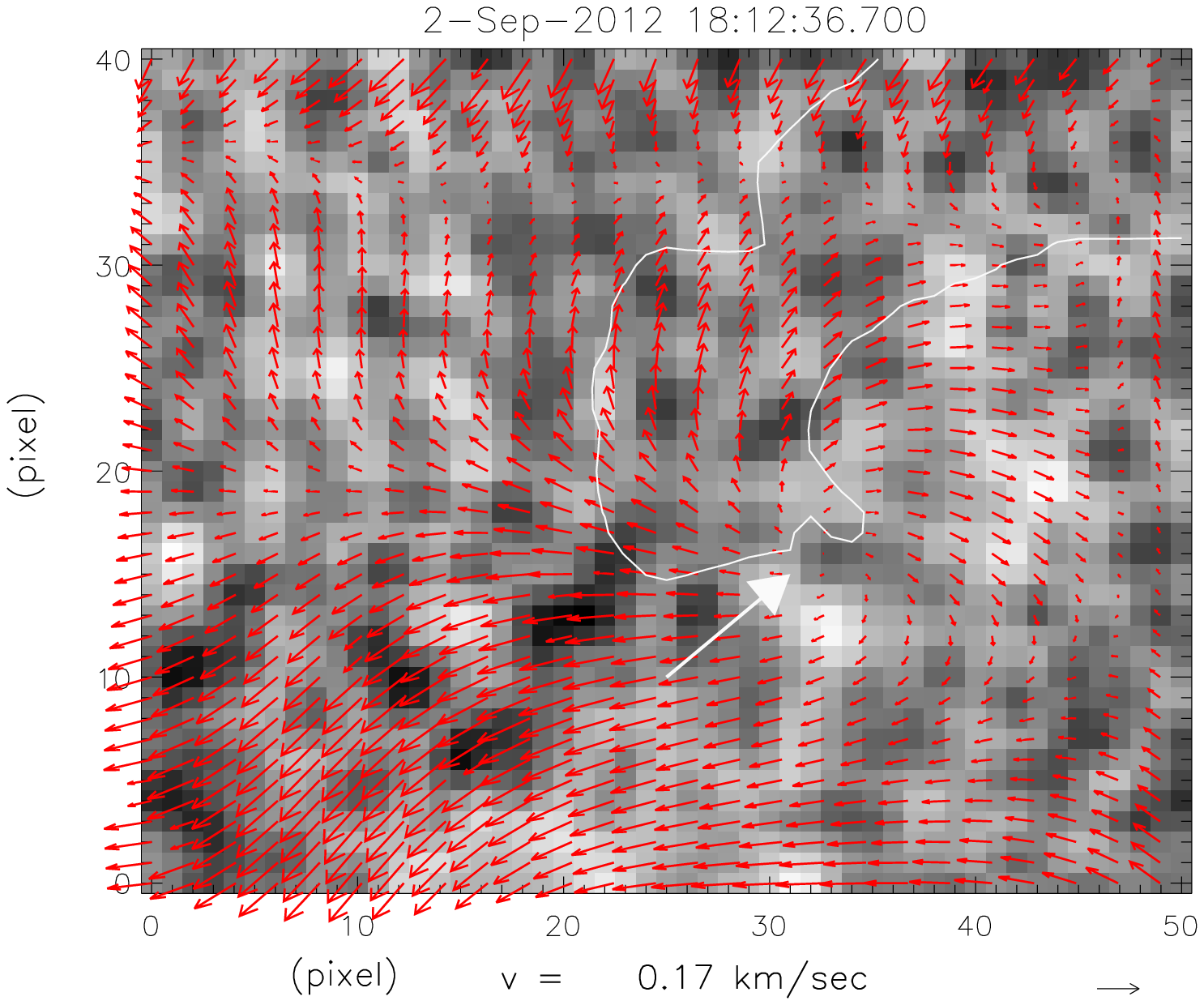}
               \hspace*{-0.08\textwidth}
               \includegraphics[width=0.54\textwidth,clip=]{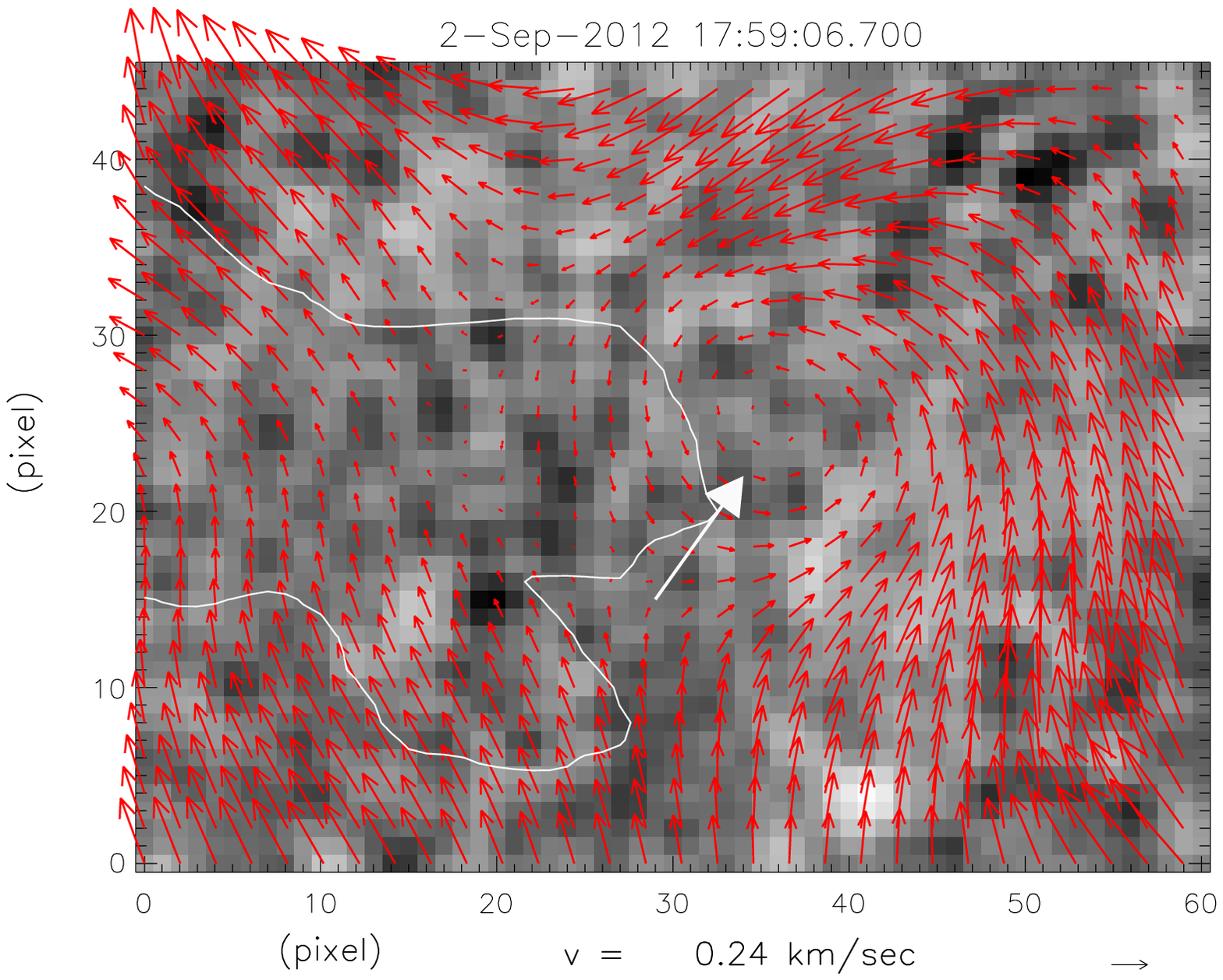}
              }
      \vspace{0.01\textwidth}    
      \centerline{\hspace*{0.015\textwidth}
               \includegraphics[width=0.54\textwidth,clip=]{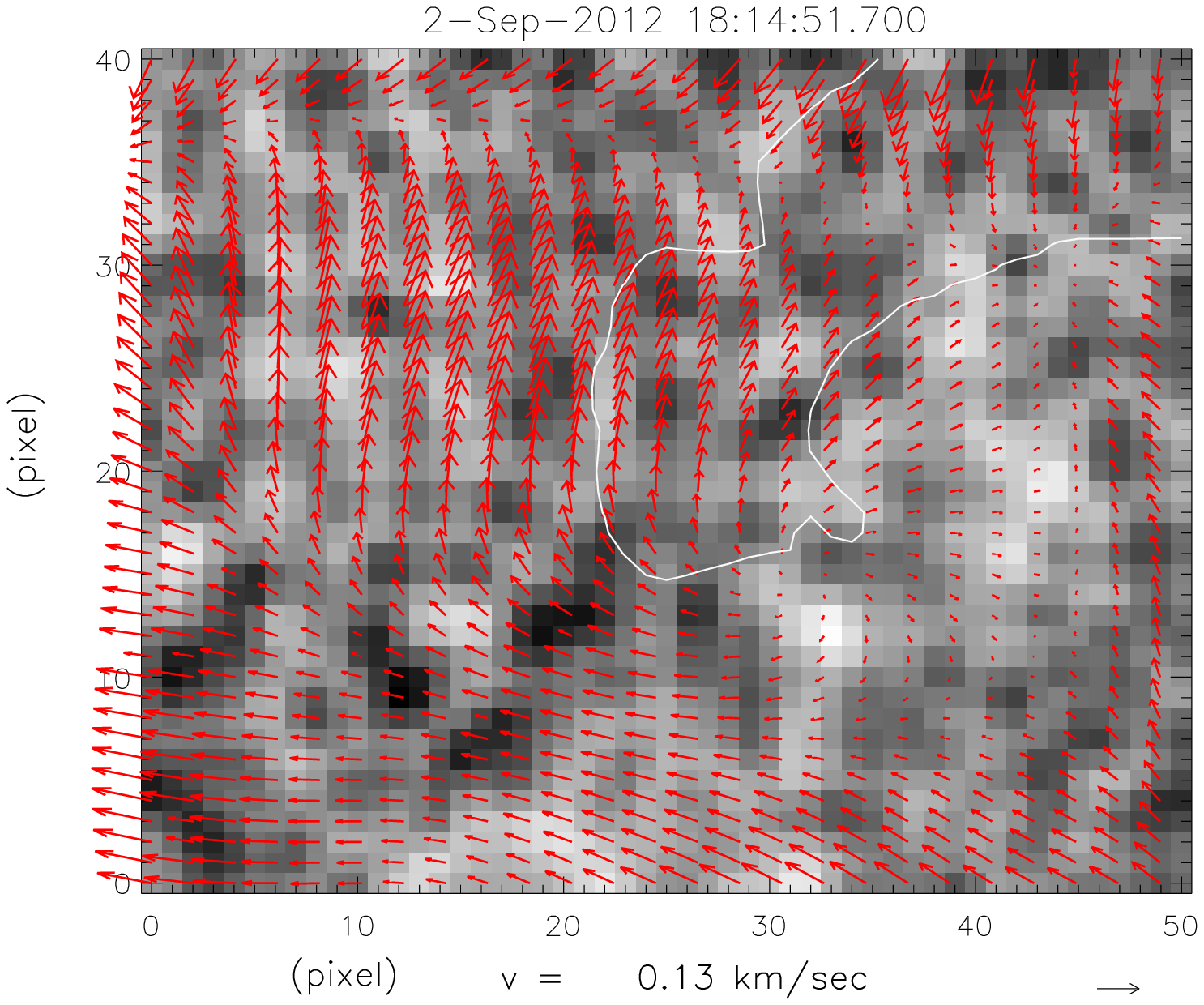}
               \hspace*{-0.08\textwidth}
               \includegraphics[width=0.54\textwidth,clip=]{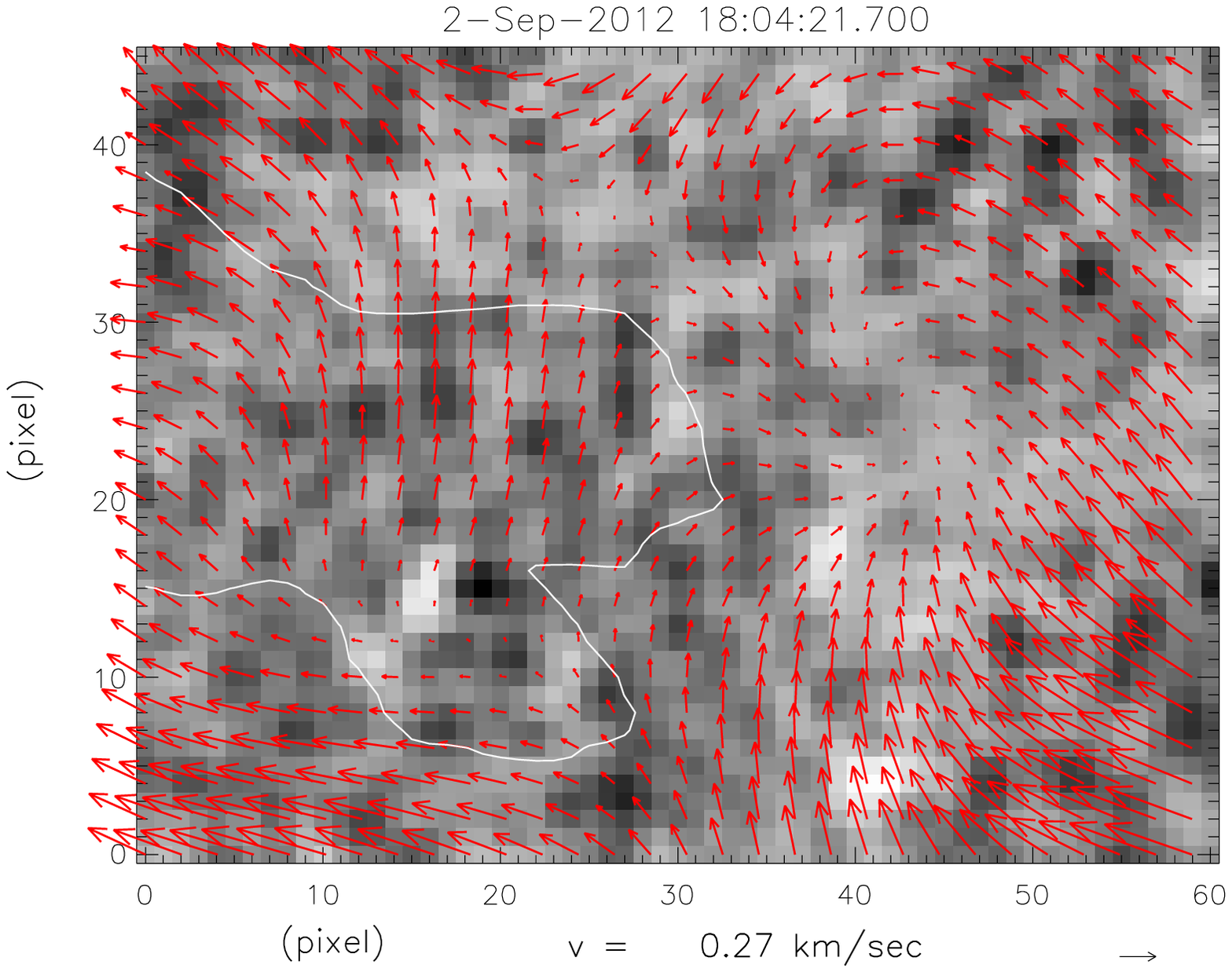}
              }
      \vspace{0.05\textwidth}    
\caption{Same as Fig.~\ref{fig:2}, but for event 4.}
   \label{fig:6}
   \end{figure}

The 07 September 2011 event showed clockwise rotation in the eastern footpoint and 
anti-clockwise rotation in the western footpoint. These rotational motions are shown in Fig.~\ref{fig:4}.
But only in the wastern footpoint of
the erupting filament an anti-clockwise rotational motion was observed on 02 July 2012 event(event 3, Fig.~\ref{fig:5}). 
The western footpoint did not show any rotational motion during the filament eruption in this event. 
This footpoint was located close to the penumbral portion of the sunspot region. 
Similarly, the 02 September 2012 event showed  a clockwise rotation in the eastern footpoint and 
anti-clockwise rotation in the western footpoint of the filament(event 4, Fig.~\ref{fig:6}). In all the events the
filament footpoint was located at the periphery of the rotation center. The rotation lasted 
for 4--20 minutes in each of these events. 

The observed rotational motion is not simultaneous in both ends of the filament. There is
about one to ten minute time difference between the ending of the rotational motion in one
of the footpoints and starting in another footpoint of the filament. Table~\ref{Table:1} provides 
the starting and ending times of the rotational motions seen in the ends of 
the filaments for all the events. The duration of the rotational motions in any one footpoint is about 4--20 minutes.
In each of these events, the fast rise of the filament was observed either during or 
after the rotational motion observed. Table \ref{Table:1} provides the direction of rotation in each of the footpoint.  It 
should be noted here that the direction of rotation is opposite in each filament end points. In two 
events (3 and 10), one end of the filament was located in the sunspot penumbral region where we did 
not find any rotational motion. We did not see any correlation between the speed of the filament eruption and duration of 
rotation. The average rotational speed at the footpoint of the filament is about 0.2~km~s$^{-1}$, or about 6~deg~hr$^{-1}$. 

We have also observed rotational motions at other
locations in the active region during the onset of the filament eruption. But the sizes of 
those rotating regions are significantly smaller compared to the 
reported one and they survived for 2--3 minutes only. One such rotational motion can be seen 
in Fig.~\ref{fig:3} (top-right) at (70$^{\prime\prime}$, 50$^{\prime\prime}$) pixel location. 
The rotational region is small, 10$^{\prime\prime}$ in size. But those seen 
close to the filament ends are of supergranular size. The filaments are considered as the 
lower ends of the flux ropes which are embedded in the cavities of 
helical fields (\opencite{Mackay10}; \opencite{Guo10}). The 193 ~\AA~movies of 
four regions (events no. 1--4) provided on the ftp site\footnotemark[1] show
the erupting filament and the evolving bright flux rope. The general view of the flux rope 
evolution can be obtained through \inlinecite{cheng14}.

\footnotetext[1]{Movies generated from SDO/AIA 193~\AA~images of the filament 
eruptions associated with active region NOAA 11226, 11283, 11515 and 11560 discussed in this paper are available on 
our web site (ftp://ftp.iiap.res.in/sajal/). The movies are named according to the date of observations of the events.}

\section{Summary and Discussion} 
      \label{section4} 
We have analyzed ten filament eruptions from eight different active regions 
at different times during Solar Cycle 24. Nine out of the ten filament eruption events
were followed by a flare. In one event, the filament destabilized a couple of hours before the large 
flare (event no. 7). During the initial stages of filament eruption, in eight events, if one type of 
rotational motion was observed in one end of the filament, the opposite direction of
rotational motion was observed in the other end of the filament. In two events (3 and 10) the rotational
motion was observed only in one end of the filament. The rotational motions  persisted 
for about 4--20 minutes and the rate of rotation was about 6$^{\circ}$~hr$^{-1}$.  
The observed rotational motions in both ends of the filament are not simultaneous. There 
is a few minutes difference between the ending of the rotational motion at one end of 
the filament and starting of the rotational motion in the vicinity of another end of the filament. In all these 
events, the fast rising phase initiated either during the rotational motions or after they have ceased.

The filament destabilization and subsequent eruption may have been initiated due to several reasons such as a 
kink-mode instability in twisted magnetic fields \cite{Sakurai76}, the magnetic reconnection at low level 
\cite{Contarino03}, flux cancelation at the photosphere 
(see, {\it e.g.,} \opencite{Amari03}) etc.
Once the filament starts to erupt, it expands axially towards higher heights. 
As it rises, the local external pressure is smaller compared to the previous 
location of the filament. Because of the pressure difference, the top portion of the filament 
expands radially, thereby inducing a torque imbalance between the 
expanded portion and  undisturbed portion of the filament \cite{Parker74,Jockers78}. 
The consequence of the radial expansion of the filament is a rotational motion in the expanded portion 
of the filament. This can happen by transferring the twist from the undisturbed portion to 
the expanded portion. The effect of the transfer of twist from the undisturbed portion to
the expanded portion can lead to the rotation in the legs of the filaments, which could
extend from the corona to the chromosphere. In this case two footpoints of the filament 
should exhibit oppositely directed motions \cite{Jockers78}. The plasma inside the filament will also rotate 
with the flux rope. However, this cannot cause the observed rotational motions of the plasma in the 
photosphere where the plasma beta is large. 

Alternatively, the bending of the top portion of the erupting filament can induce the rotation
of the legs. \inlinecite{Martin03} and \inlinecite{Panasenco08} have observed a bending motion of the top portion 
of the erupting prominence and subsequently observed an oppositely twisting motion in  two
legs. They name this phenomenon the ``roll effect''. Similarly, \inlinecite{Panasenco13} 
mention that the non-radially erupting prominence first exhibits a bending motion of the top
of the prominence and later an oppositely directed rolling motion propagates into its legs.
In this mechanism the plasma inside the flux rope can rotate with the magnetic field and the rotational motion 
may not reach the photosphere as the torque and force required for the photospheric plasma to 
rotate is probably much larger than this.

The observed photospheric rotational motion near the ends of the filament could possibly be 
resulting from the photospheric or
subphotospheric shearing flows. We observed that the fast rise of the 
filament was initiated during/after the vortical 
motion of the plasma had set in. We have also observed vortical flows offset from the 
ends of the filament. However, their lifetime is too short and their size is about 
10$^{\prime\prime}$, which is much smaller than the ones observed near the filament ends. The spatial correlation
between the observed end points of the filament and the location of the rotational motion,
and the temporal coincidence of the rotational motions with the filament eruption time in ten
events suggest us that the observed rotational motion is not randomly occurring 
 in the vertices of the supergranulation \cite{Innes09}.

When the filament starts to erupt, it expands axially towards higher heights. 
The immediate consequence of the deformation of the 
axial component of the magnetic field in the flux rope is the launch of shear Alfv\`{e}n waves 
by shearing its footpoints (by the Lorentz force)  that in turn carry the 
axial component of the flux into the expanded portion of the flux rope \cite{Manchester04}. 
The increase in the axial flux in the flux rope will increase the magnetic 
pressure. The tendency of the outward directed magnetic pressure will push the flux rope 
slowly to higher heights. Once the top of the flux rope reaches a critical height where the 
“torus instability” criteria are satisfied then the flux rope becomes unstable and eventually it
will erupt \cite{Kliem06}.

Another possibility is that, if there is an emergence of a poloidal flux, it results in a 
rotation of the footpoints. The injection of the poloidal flux 
into the flux rope, will  increase the magnitude of the outward Lorentz self force (hoop force). 
This will push the flux rope slowly to critical height where the external poloidal flux is 
decreasing faster than the decrease of the hoop force. This condition will again lead to the 
“torus instability” and in that condition the flux rope is no longer in equilibrium scenario 
and eventually  it erupts. But the generation and emergence of the poloidal field is very unlikely 
and there is no observational support to this [{\it see also} \opencite{Chen10}].

\inlinecite{Martin03} and \inlinecite{Panasenco11} reported that there is a 
correlation between the direction of the rolling motion 
observed in the erupting filament and the chirality of the filament. The observed directions 
of rotation in both legs of the prominence are opposite to each other.  
In future, we plan to study the rotation in the footpoints of a large number of 
erupting filaments using a spectroscopic technique both in the photosphere and chromosphere 
simultaneously along with their sign of chirality using vector magnetic field measurements. 
This may provide a vital clue to the origin of the rotational motion near the ends of the erupting filaments. 

\begin{acks}
We thank the referee for insightful comments which helped us to improve the content in the
manuscript. The AIA data used here are the courtesy of NASA/SDO and AIA consortium. 
SDO/HMI is a joint effort of many teams and individuals to whom we are 
greatly indebted for providing the data.
\end{acks}

\appendix 

\begin{figure}    
   \centerline{\hspace*{0.005\textwidth}
               \includegraphics[width=0.60\textwidth,clip=]{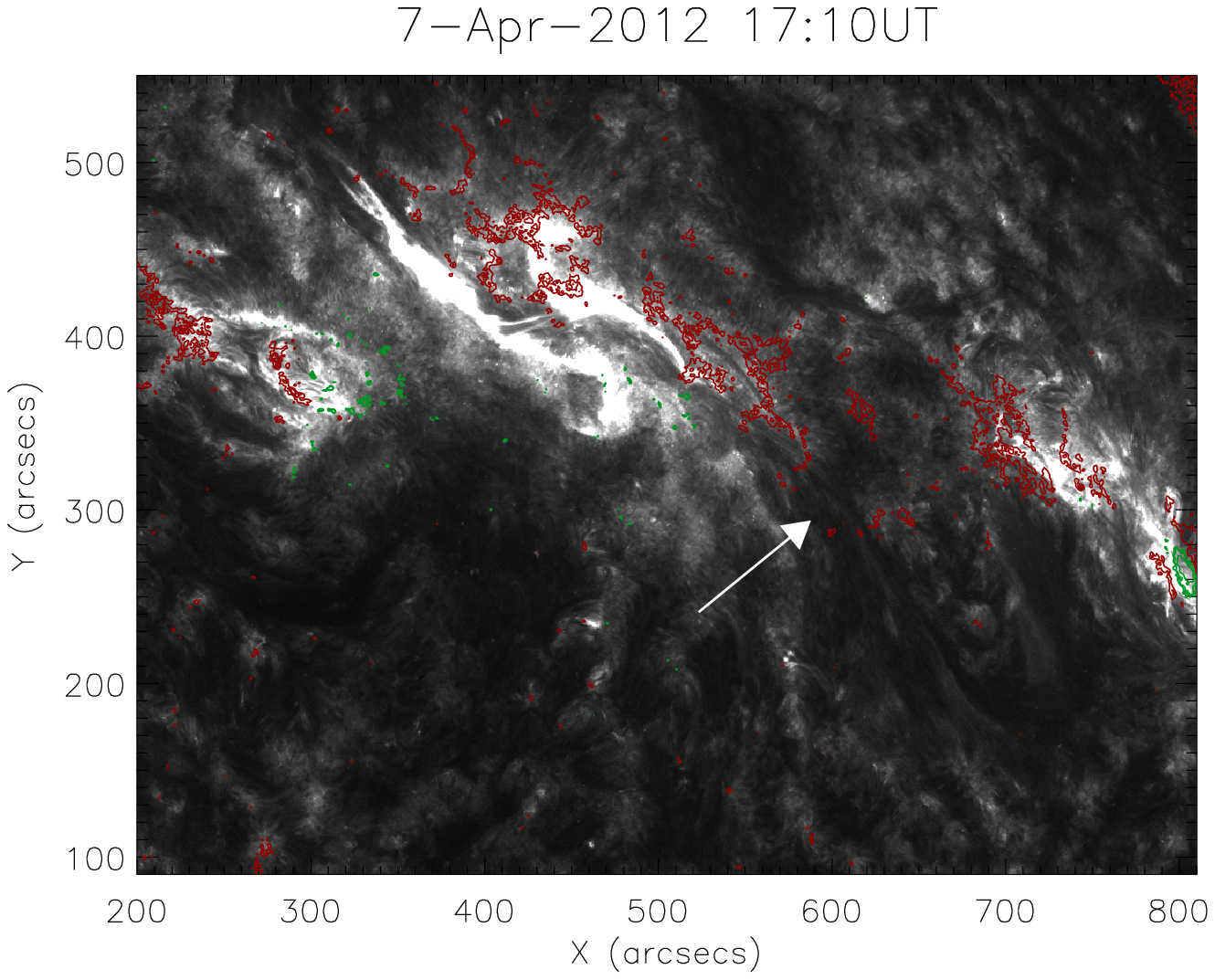}
               \hspace*{-0.11\textwidth}
               \includegraphics[width=0.535\textwidth,clip=]{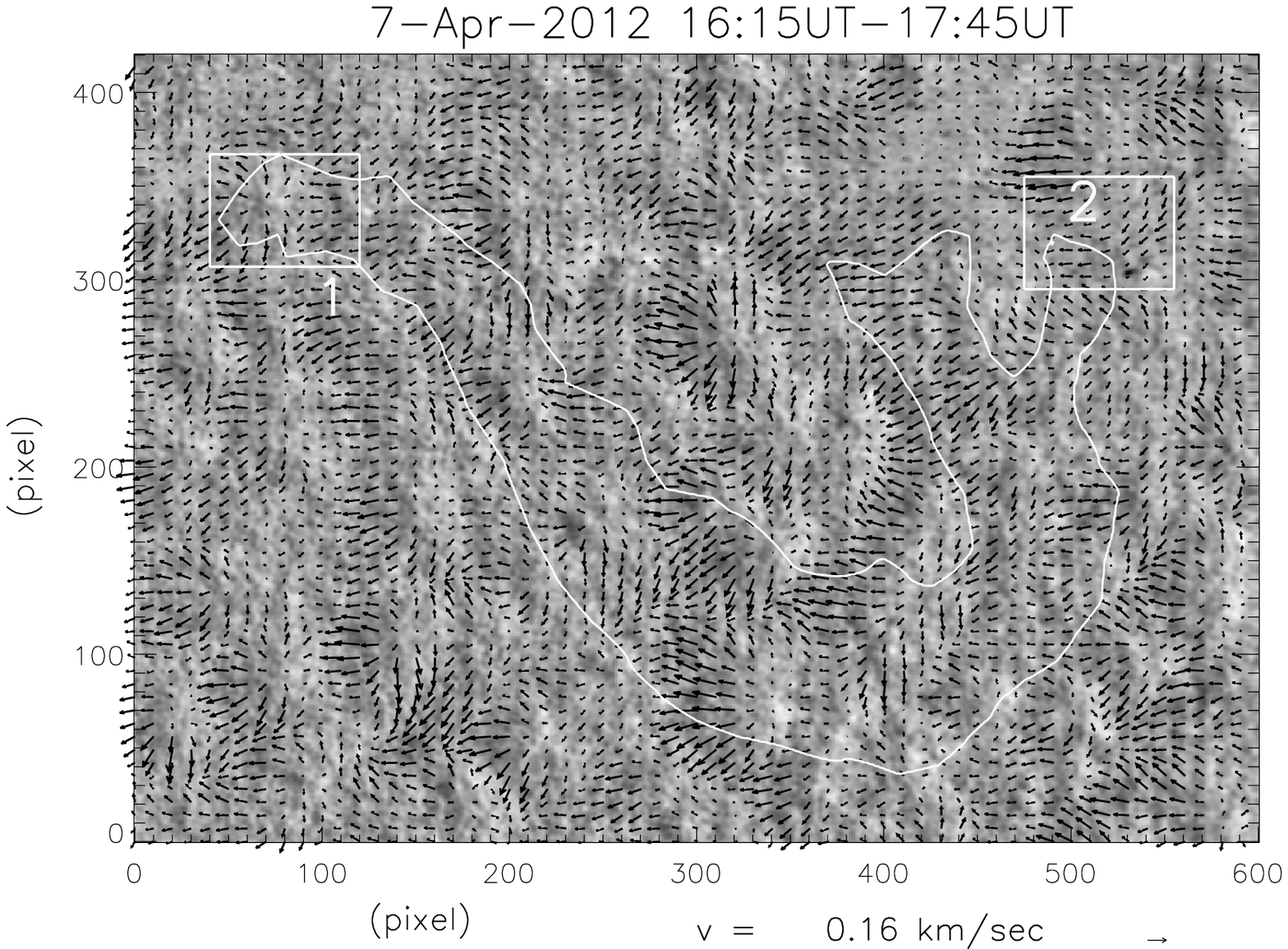}
              }
               \vspace{-0.12\textwidth}   
     
     \centerline{\Large \bf     
    \hspace{0.4 \textwidth}  \color{white}{(a)}
      \hspace{0.43\textwidth}  \color{white}{(b)}
         \hfill}
         \vspace{0.1\textwidth}      
   \centerline{\hspace*{0.012\textwidth}
               \includegraphics[width=0.57\textwidth,clip=]{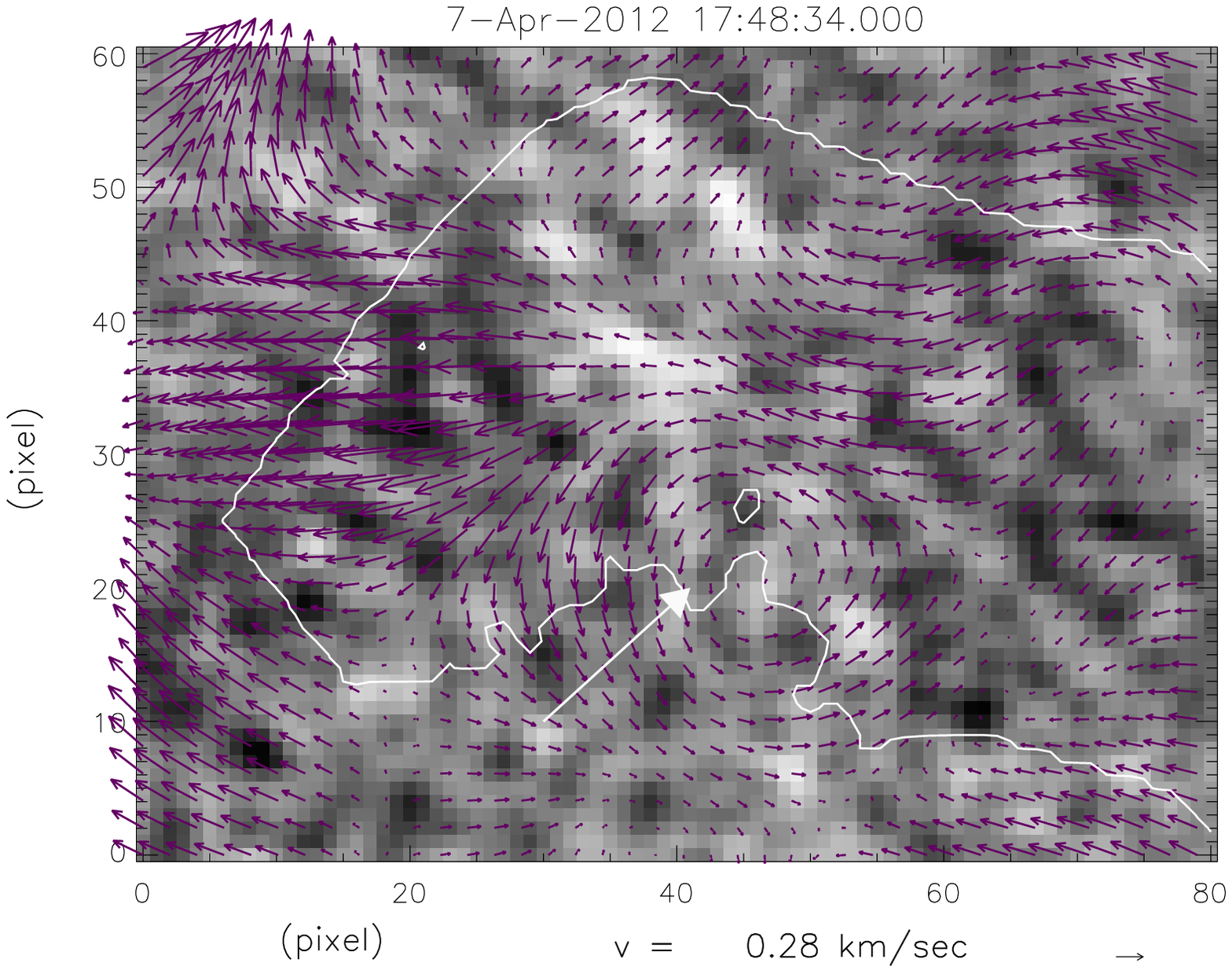}
               \hspace*{-0.06\textwidth}
               \includegraphics[width=0.57\textwidth,clip=]{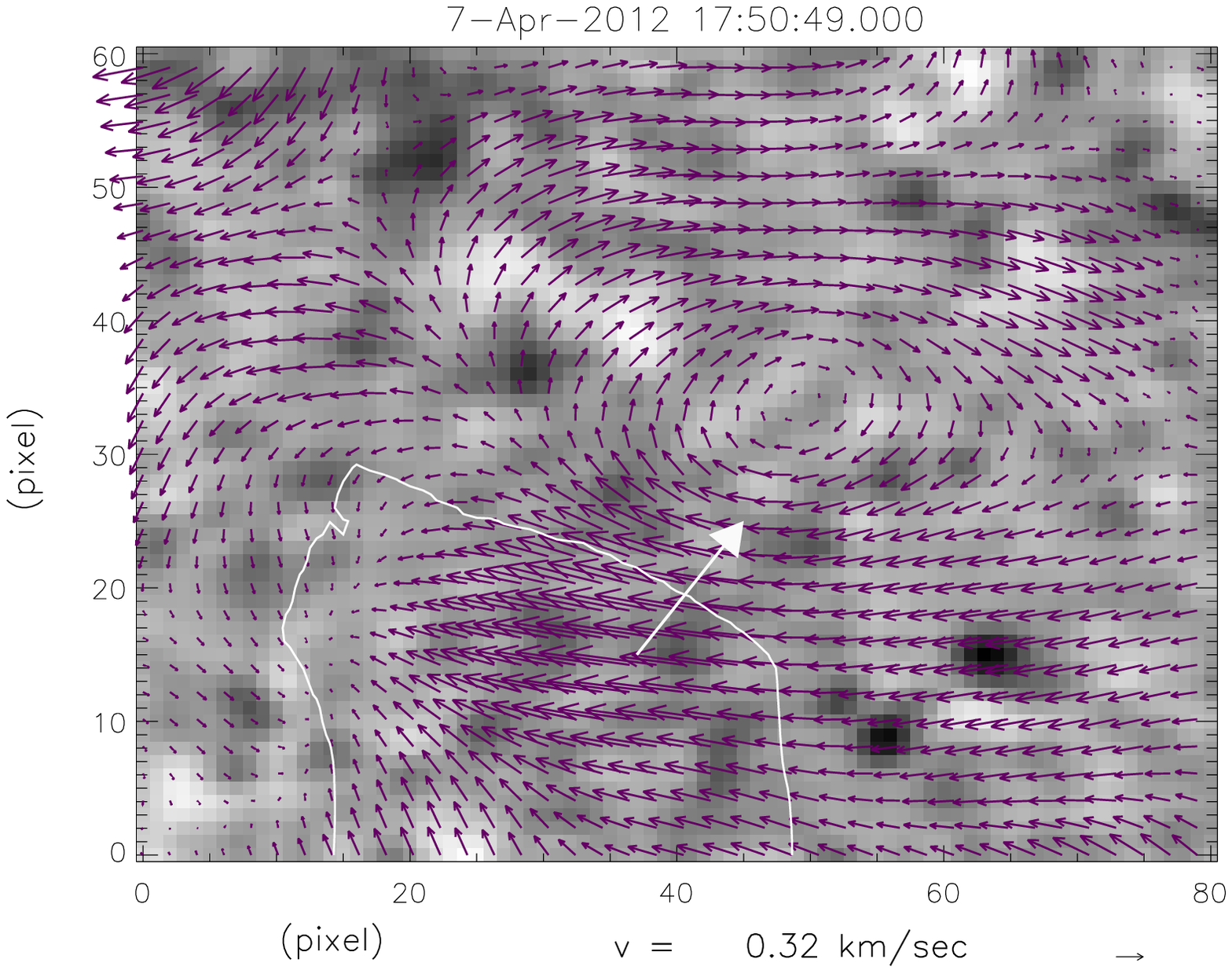}
              }      
        
      \vspace{-0.1\textwidth}   
     
     \centerline{\Large \bf     
    \hspace{0.40 \textwidth}  \color{white}{(c)}
      \hspace{0.45\textwidth}  \color{white}{(d)}
         \hfill}
      \vspace{0.10\textwidth}    
      
\caption{(a) The filament observed in SDO/AIA 304~\AA~channel
for event 5 is indicated by a white arrow.
Magnetic field isocontours from SHO/HMI overlaid on the 304~\AA~image.
The red and green contours represent the 
positive and negative polarities with magnetic field strength values of $\pm$ 150, 400, 600 
and 900~G, respectively. (b) Averaged horizontal velocity vectors shown in arrows are overlaid upon an 
averaged dopplergram. The duration of the averaging time is given above the panel. 
The contour of the filament, extracted 
from the 304~\AA~ image is also shown. 
The boxed regions 1 and 2 show the location of the eastern 
and western ends of the filament respectively.
(c, d) Rotational velocity pattern observed for this
event. (c) image corresponds to location 1 and (d) to location 2 in (b).}

   \label{fig:7}
   \end{figure}

\begin{figure}    
   \centerline{\hspace*{-0.07\textwidth}
               \includegraphics[width=0.58\textwidth,clip=]{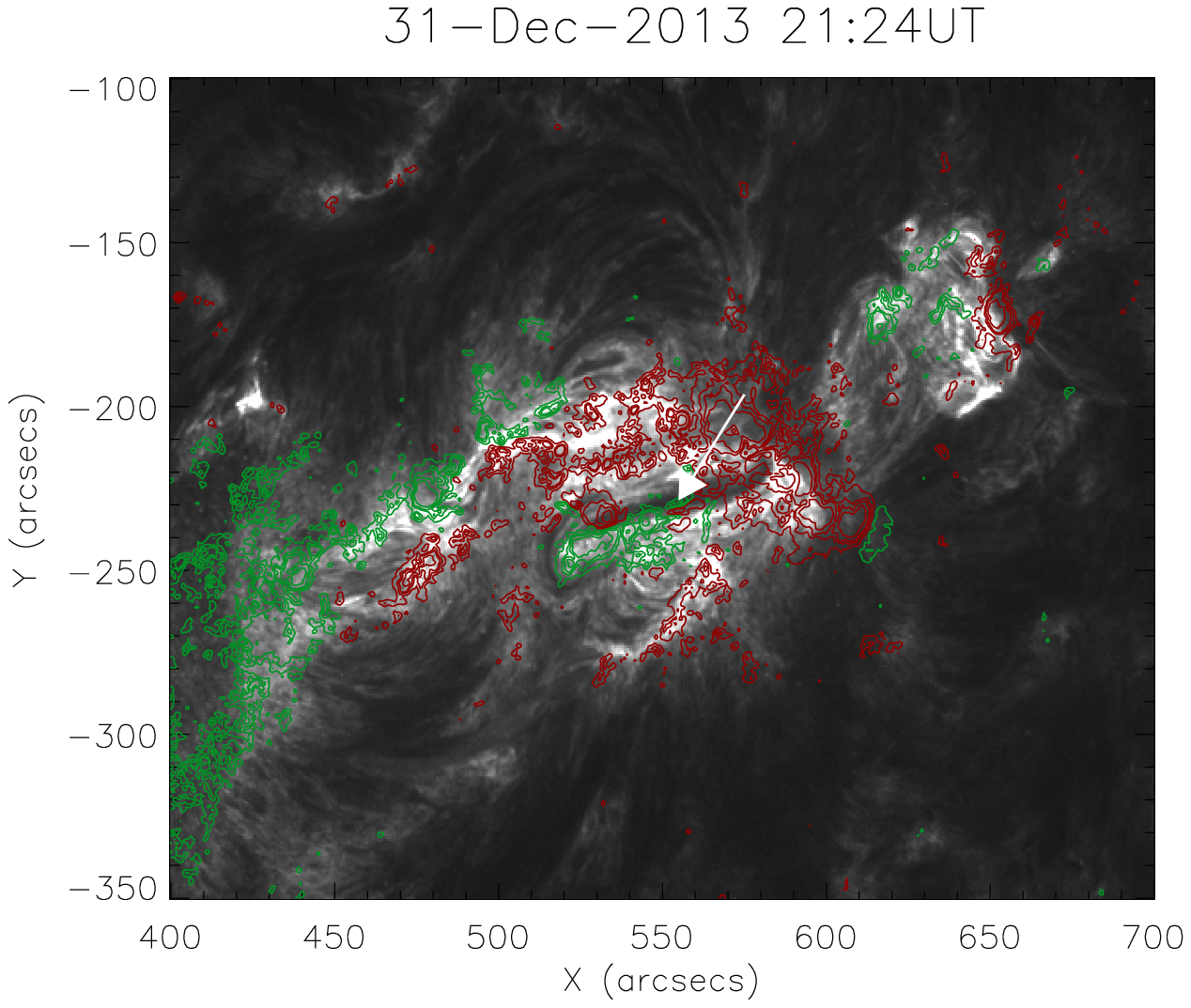}
               \hspace*{-0.13\textwidth}
               \includegraphics[width=0.51\textwidth,clip=]{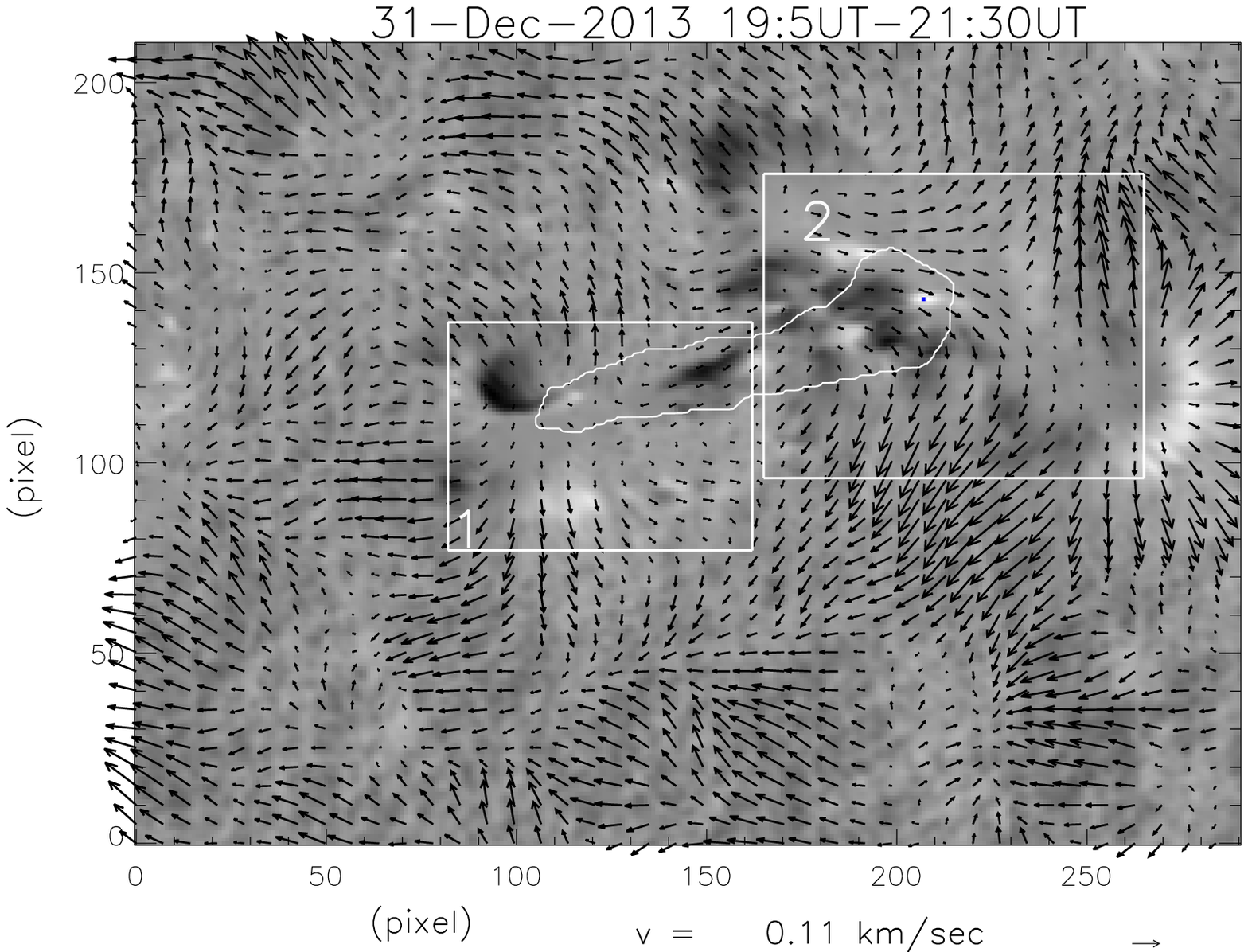}
              }
    
     \vspace{-0.12\textwidth}   
     
     \centerline{\Large \bf     
    \hspace{0.35 \textwidth}  \color{white}{(a)}
      \hspace{0.40\textwidth}  \color{white}{(b)}
         \hfill}
         \vspace{0.08\textwidth}   
   \centerline{\hspace*{-0.045\textwidth}
               \includegraphics[width=0.53\textwidth,clip=]{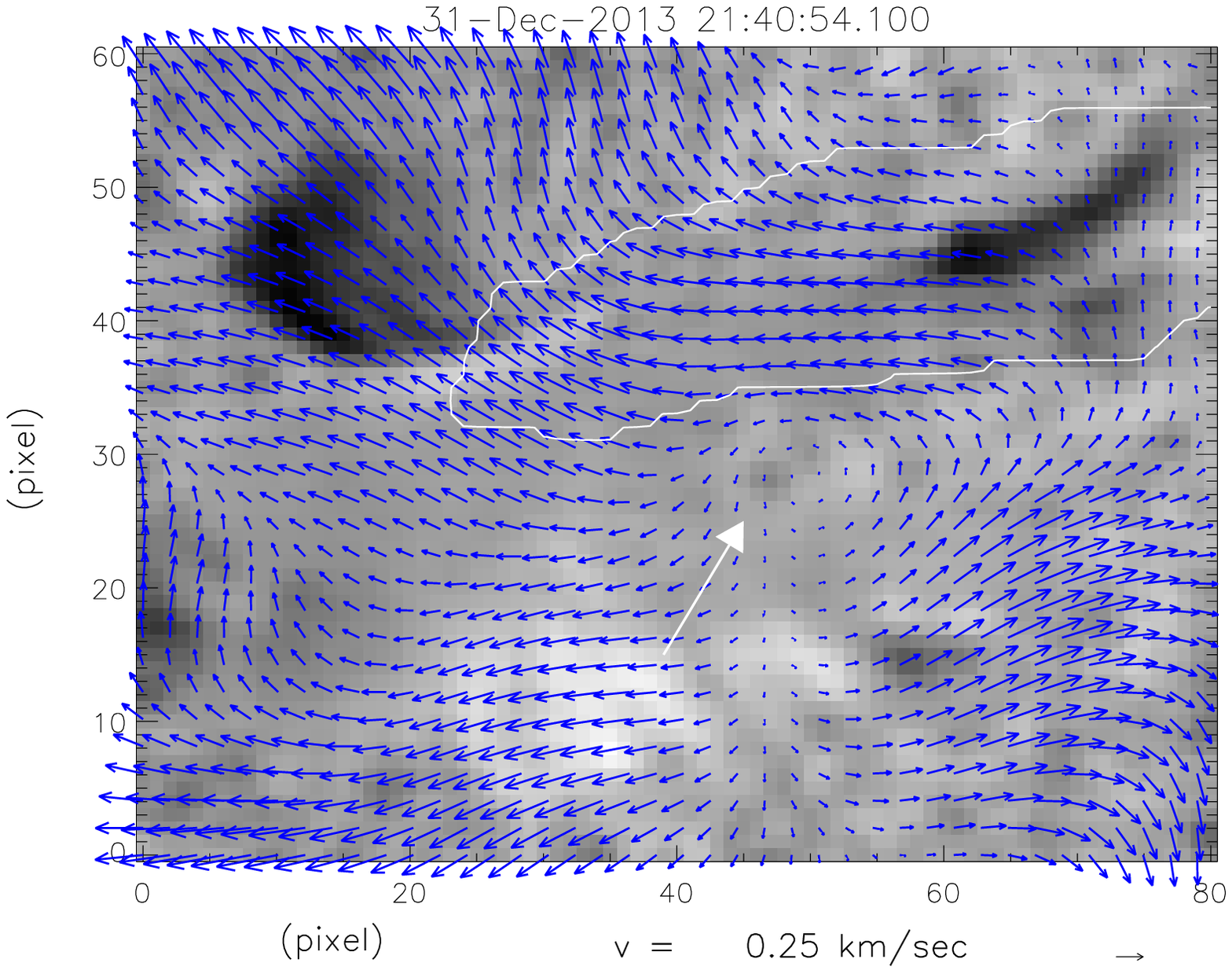}
               \hspace*{-0.06\textwidth}
               \includegraphics[width=0.53\textwidth,clip=]{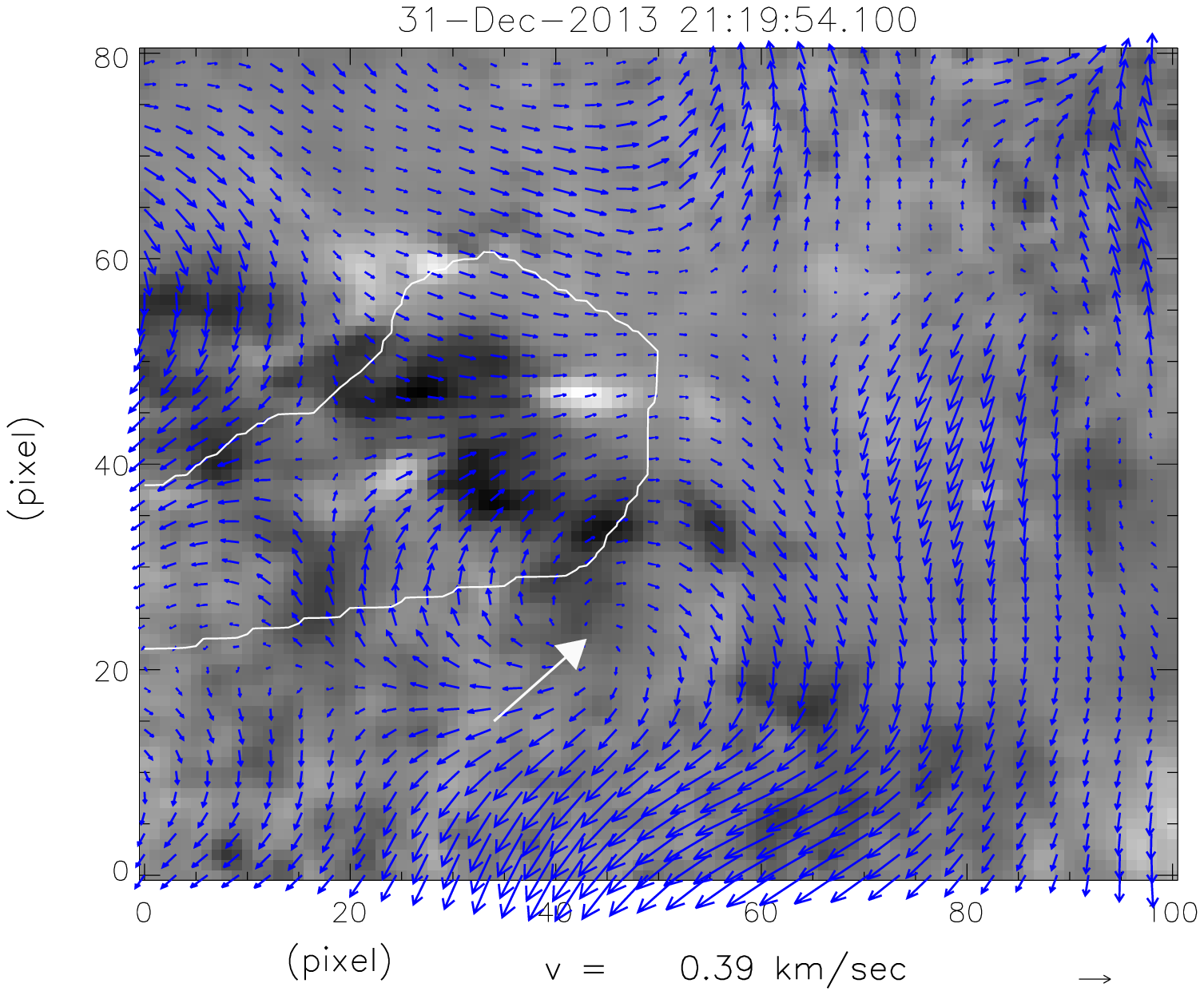}
              }
     
      \vspace{-0.1\textwidth}   
     
     \centerline{\Large \bf     
    \hspace{0.37 \textwidth}  \color{white}{(c)}
      \hspace{0.38\textwidth}  \color{white}{(d)}
         \hfill}
      \vspace{0.10\textwidth}    
         
\caption{Same as Fig.~\ref{fig:7}, but for event 6.}
   \label{fig:8}
   \end{figure}

\begin{figure}    
   \centerline{\hspace*{-0.01\textwidth}
               \includegraphics[width=0.6\textwidth,clip=]{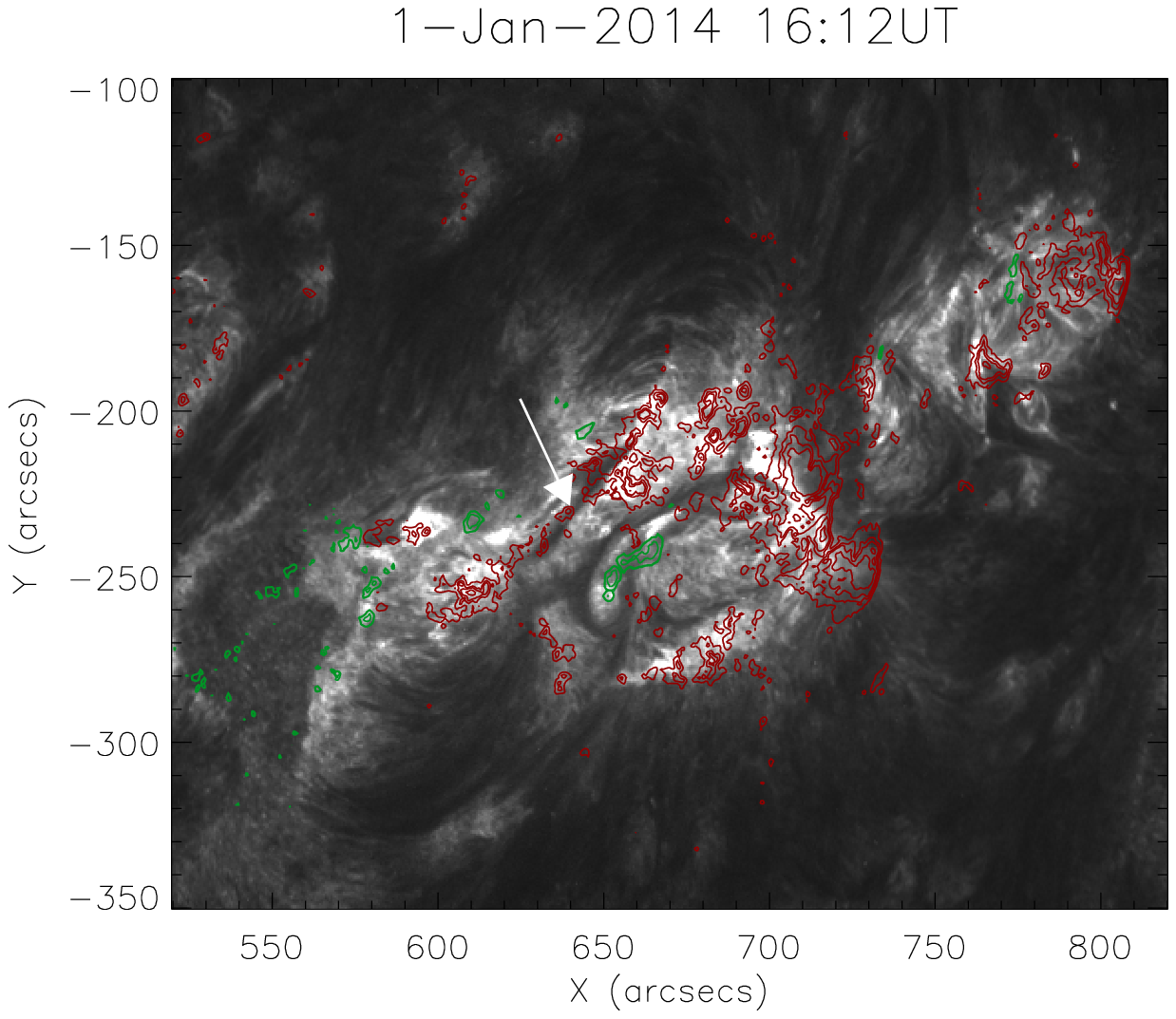}
               \hspace*{-0.08\textwidth}
               \includegraphics[width=0.53\textwidth,clip=]{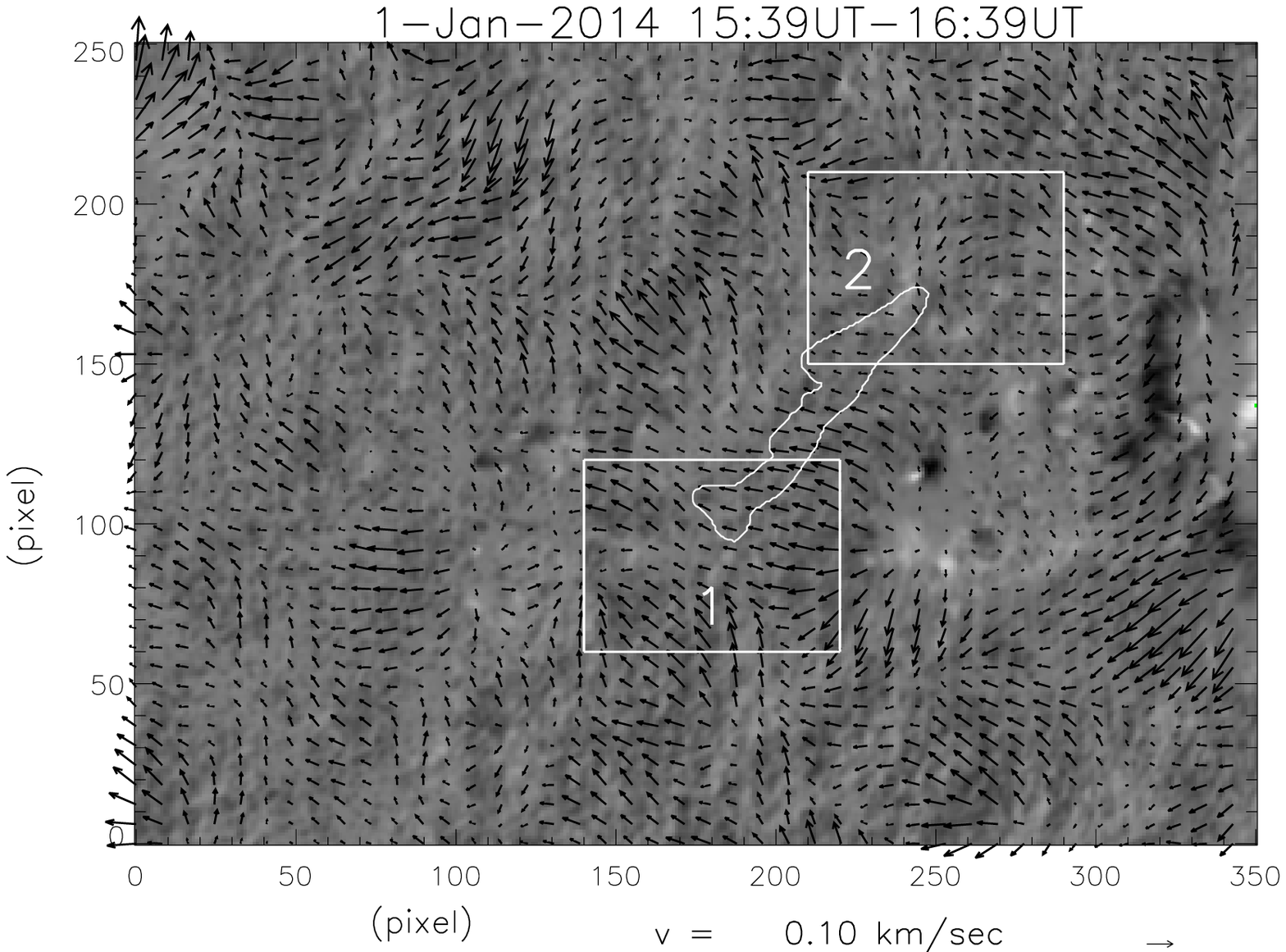}
              }
      
     \vspace{-0.12\textwidth}   
     
     \centerline{\Large \bf     
    \hspace{0.37 \textwidth}  \color{white}{(a)}
      \hspace{0.48\textwidth}  \color{white}{(b)}
         \hfill}
         \vspace{0.08\textwidth}  
   \centerline{\hspace*{0.015\textwidth}
               \includegraphics[width=0.57\textwidth,clip=]{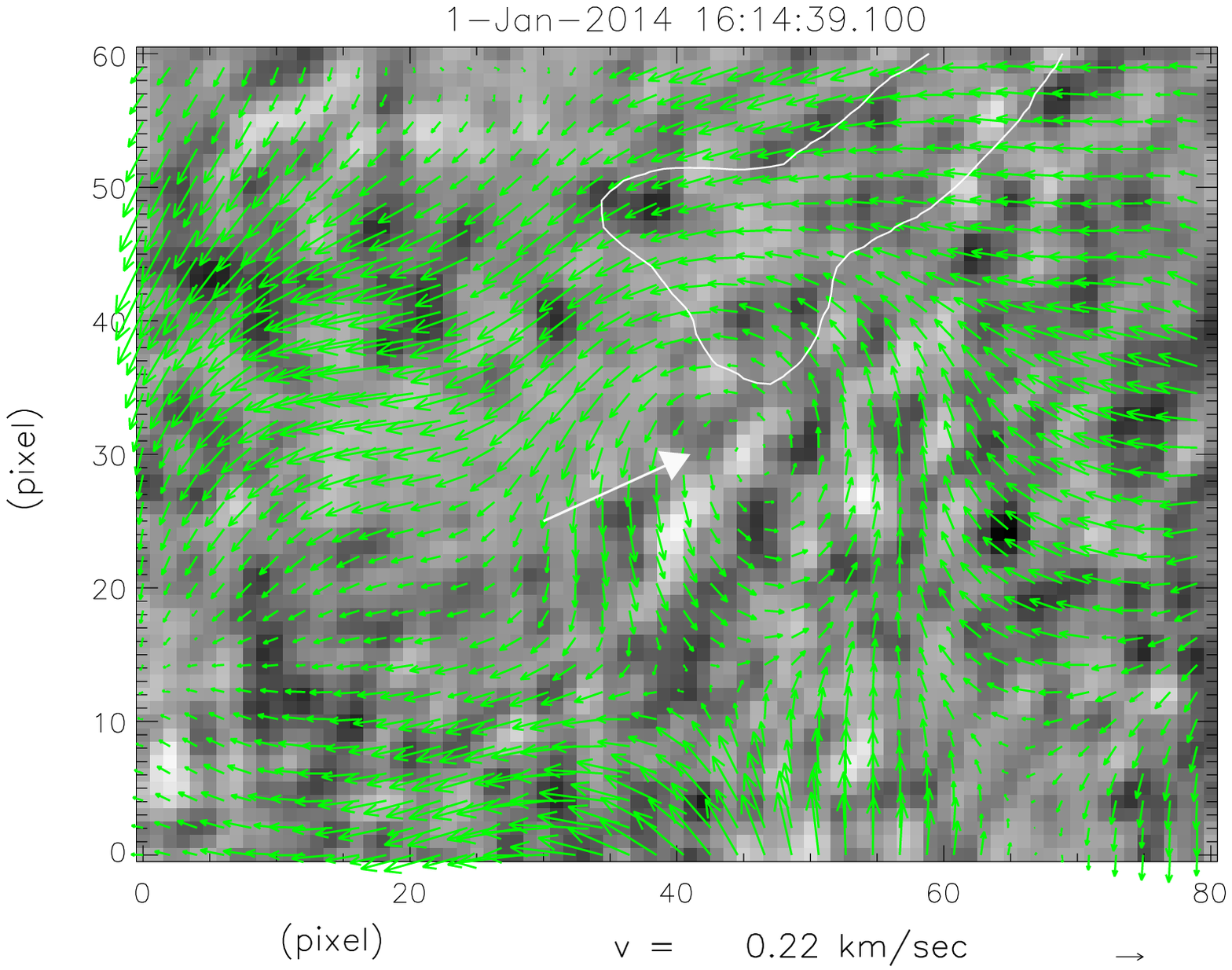}
               \hspace*{-0.06\textwidth}
               \includegraphics[width=0.57\textwidth,clip=]{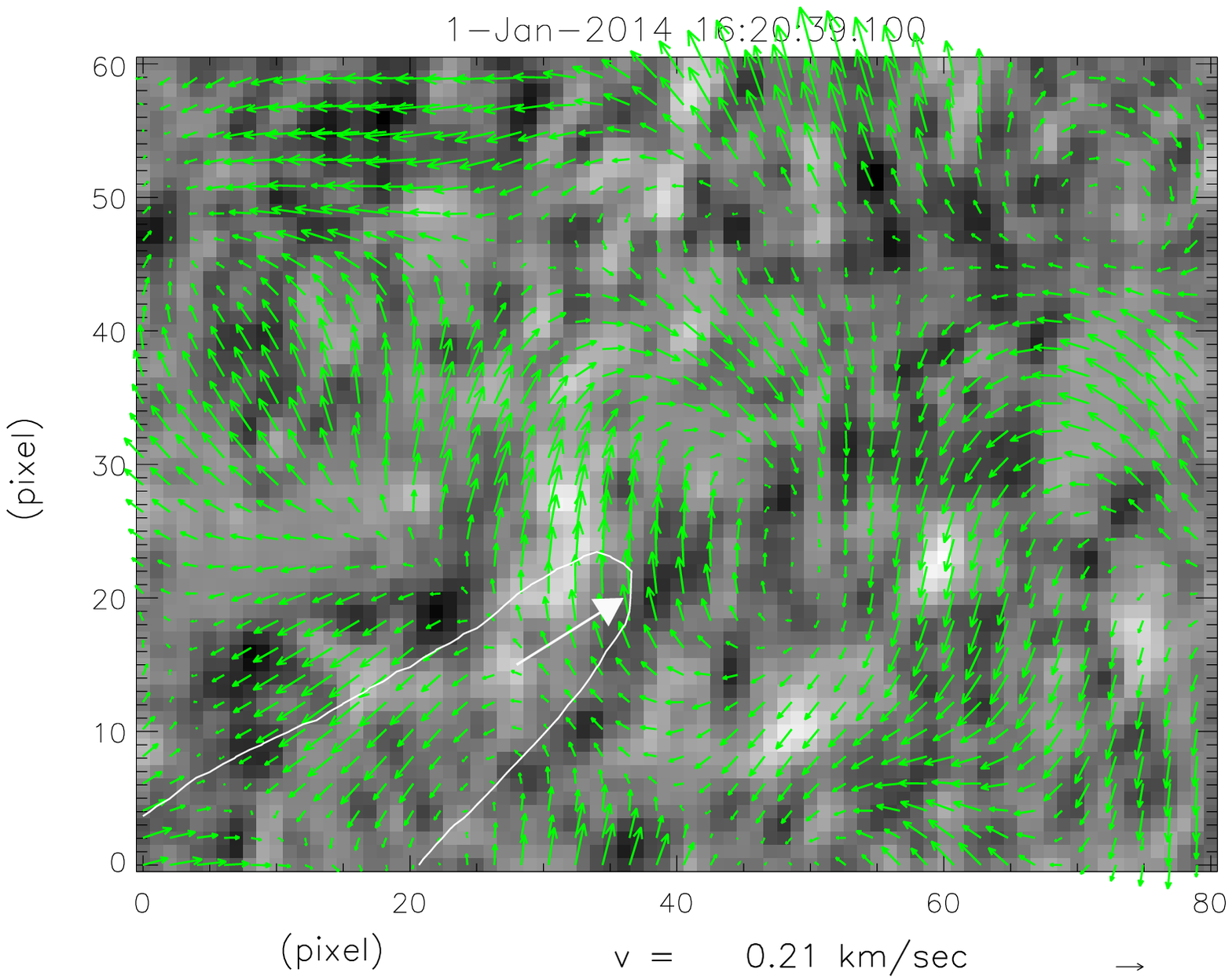}
              }
      
       \vspace{-0.1\textwidth}   
     
     \centerline{\Large \bf     
    \hspace{0.41 \textwidth}  \color{white}{(c)}
      \hspace{0.44\textwidth}  \color{white}{(d)}
         \hfill}
      \vspace{0.10\textwidth}    
\caption{Same as Fig.~\ref{fig:7}, but for event 7.}
   \label{fig:9}
   \end{figure}

\begin{figure}    
   \centerline{\hspace*{-0.08\textwidth}
               \includegraphics[width=0.62\textwidth,clip=]{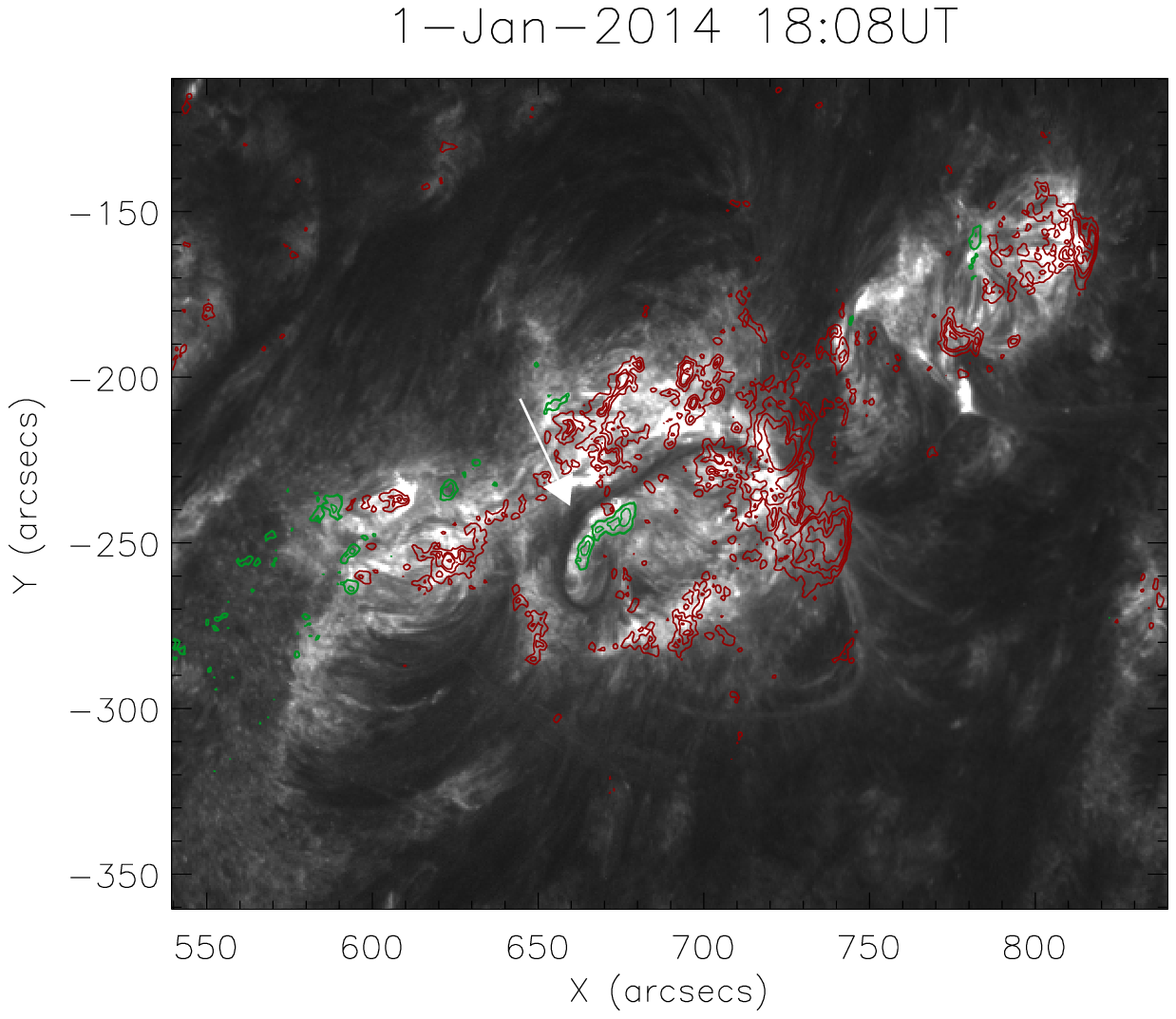}
               \hspace*{-0.12\textwidth}
               \includegraphics[width=0.55\textwidth,clip=]{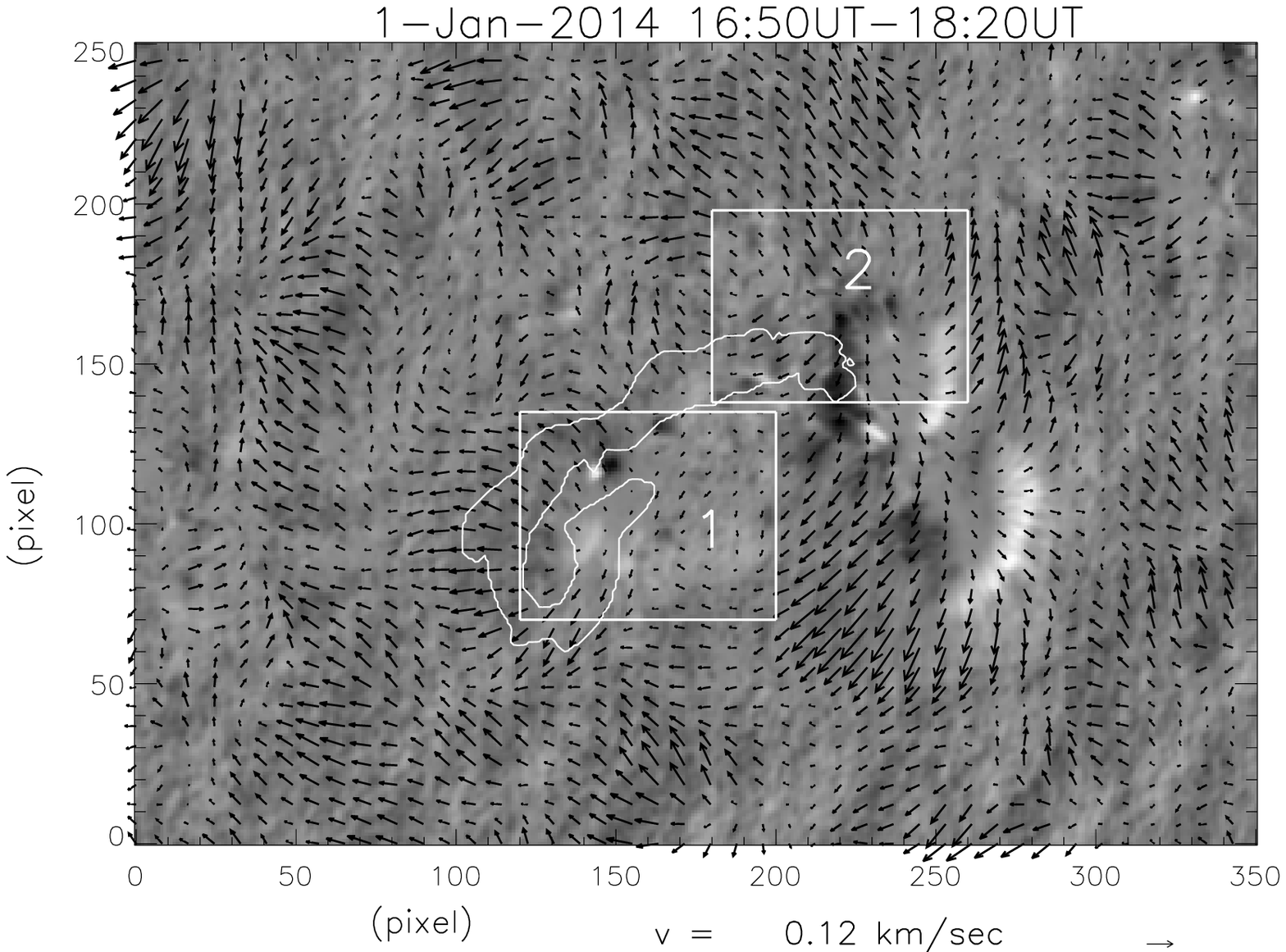}
              }
    
     \vspace{-0.12\textwidth}   
     
     \centerline{\Large \bf     
    \hspace{0.35 \textwidth}  \color{white}{(a)}
      \hspace{0.45\textwidth}  \color{white}{(b)}
         \hfill}
         \vspace{0.08\textwidth}  
   \centerline{\hspace*{-0.01\textwidth}
               \includegraphics[width=0.58\textwidth,clip=]{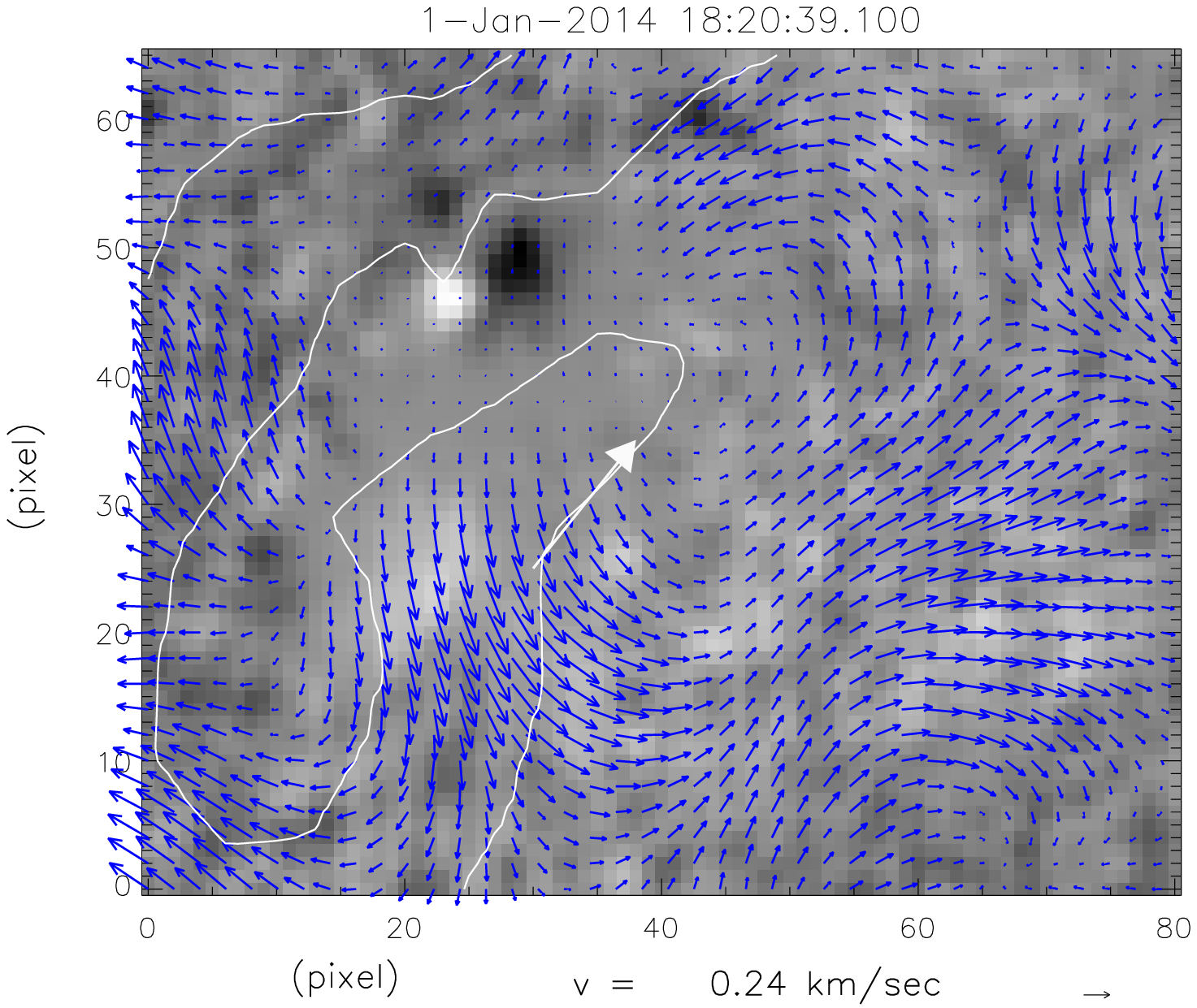}
               \hspace*{-0.11\textwidth}
               \includegraphics[width=0.58\textwidth,clip=]{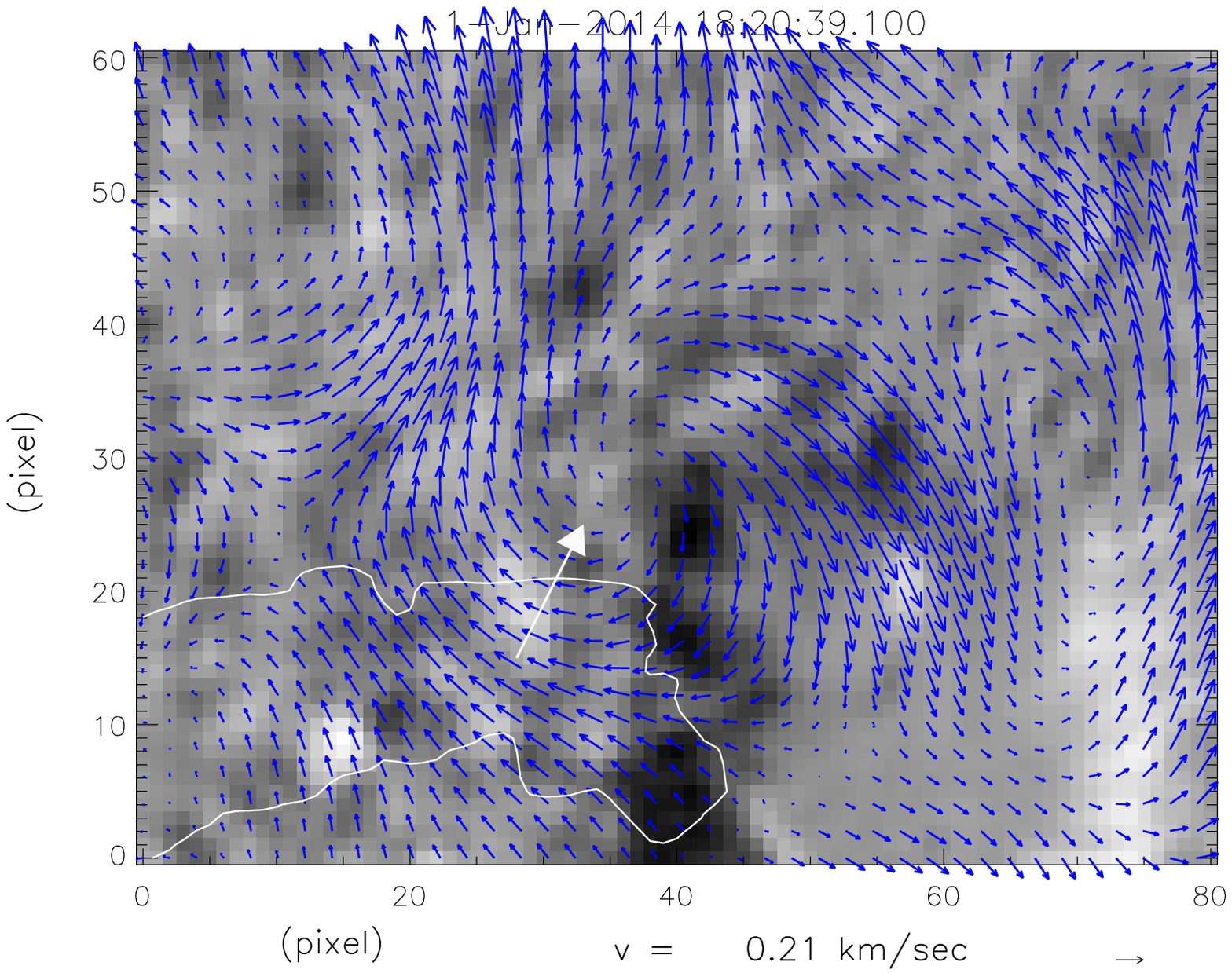}              
              }
      
      \vspace{-0.1\textwidth}   
     
     \centerline{\Large \bf     
    \hspace{0.39 \textwidth}  \color{white}{(c)}
      \hspace{0.42\textwidth}  \color{white}{(d)}
         \hfill}
      \vspace{0.10\textwidth}    
        
\caption{Same as Fig.~\ref{fig:7}, but for event 8.}
   \label{fig:10}
   \end{figure}

\begin{figure}    
   \centerline{\hspace*{0.01\textwidth}
               \includegraphics[width=0.60\textwidth,clip=]{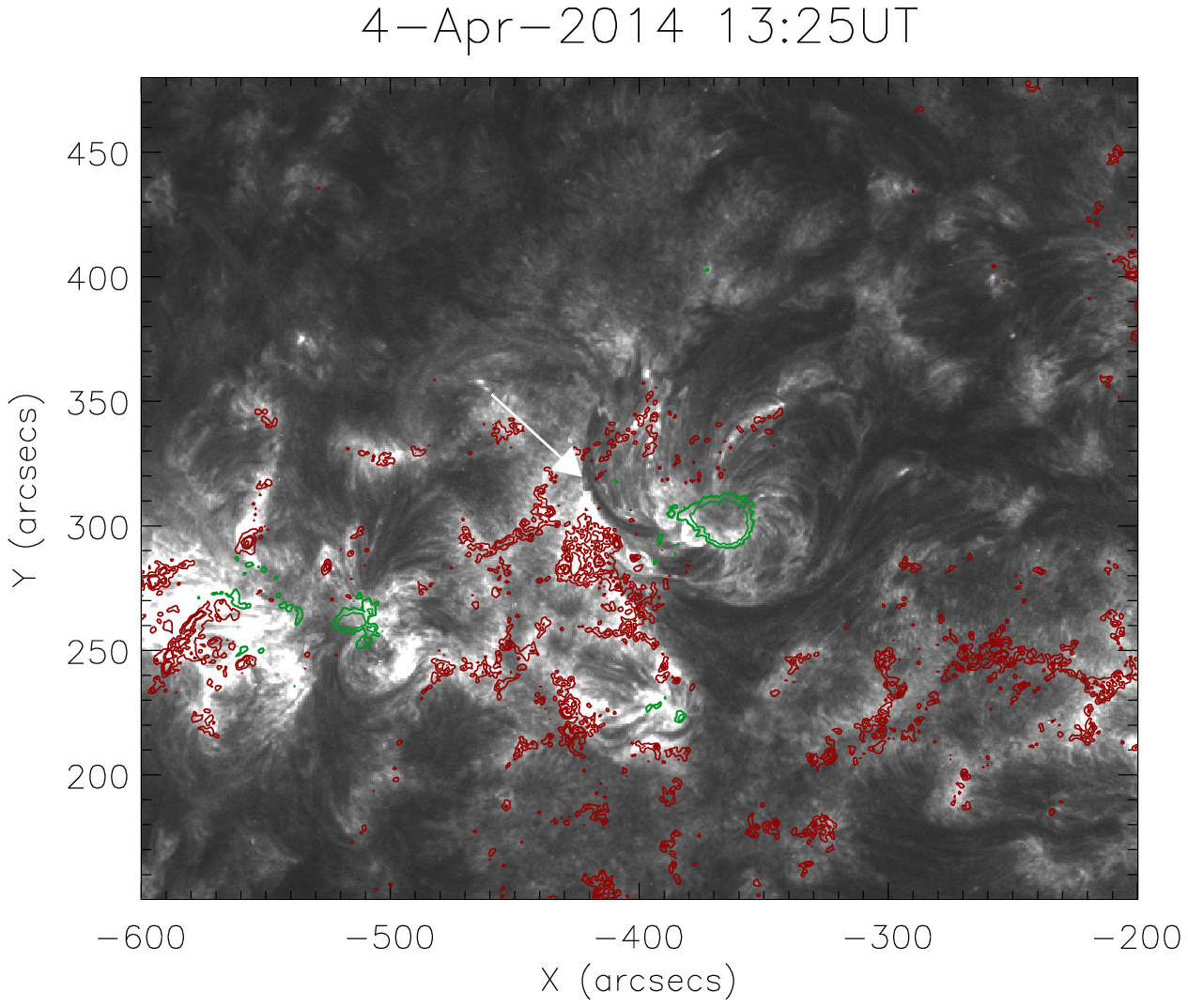}
               \hspace*{-0.12\textwidth}
               \includegraphics[width=0.55\textwidth,clip=]{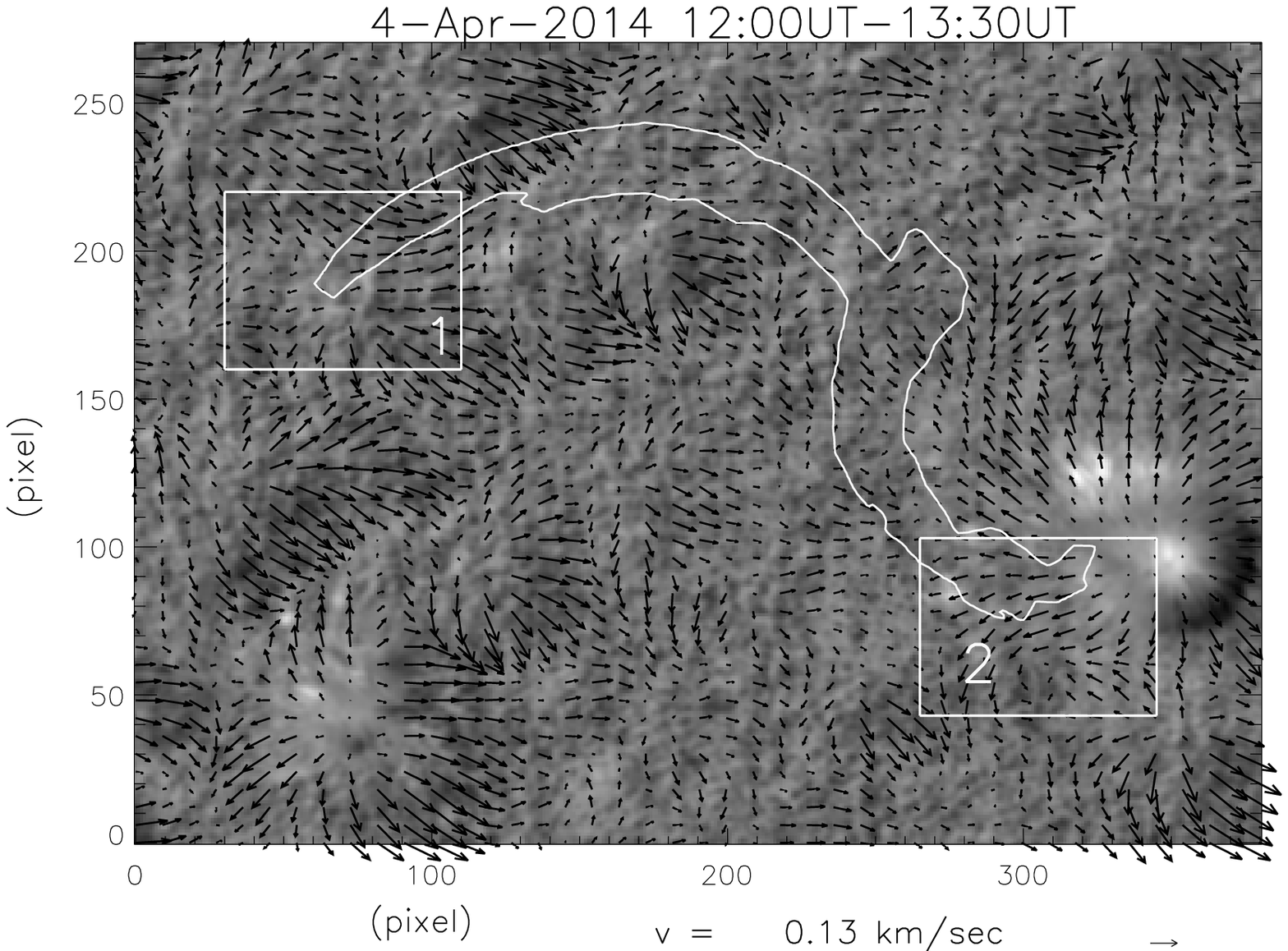}
              }
      
     \vspace{-0.10\textwidth}   
     
     \centerline{\Large \bf     
    \hspace{0.41 \textwidth}  \color{white}{(a)}
      \hspace{0.46\textwidth}  \color{white}{(b)}
         \hfill}
         \vspace{0.07\textwidth}  
   \centerline{\hspace*{0.015\textwidth}
               \includegraphics[width=0.57\textwidth,clip=]{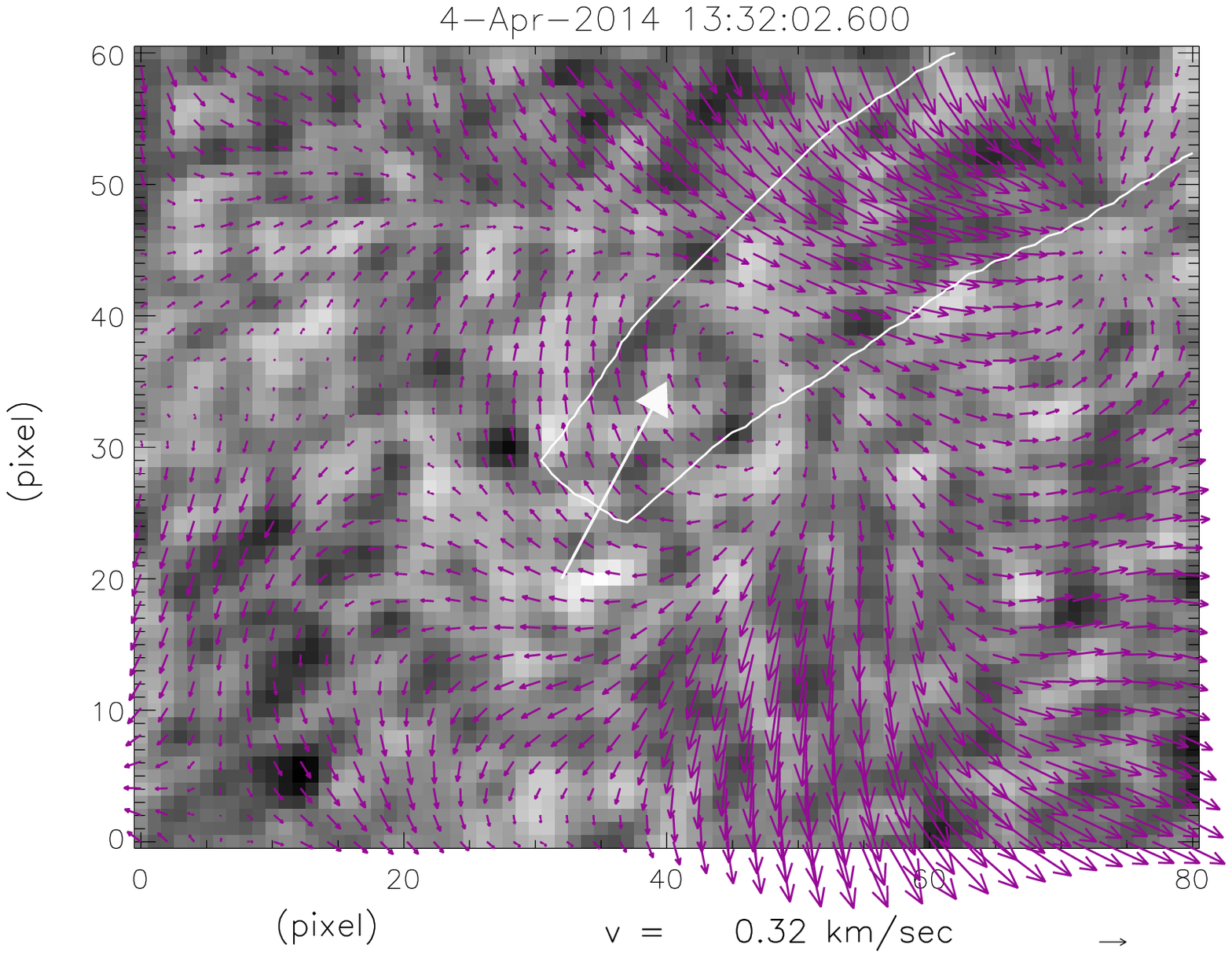}
               \hspace*{-0.06\textwidth}
               \includegraphics[width=0.57\textwidth,clip=]{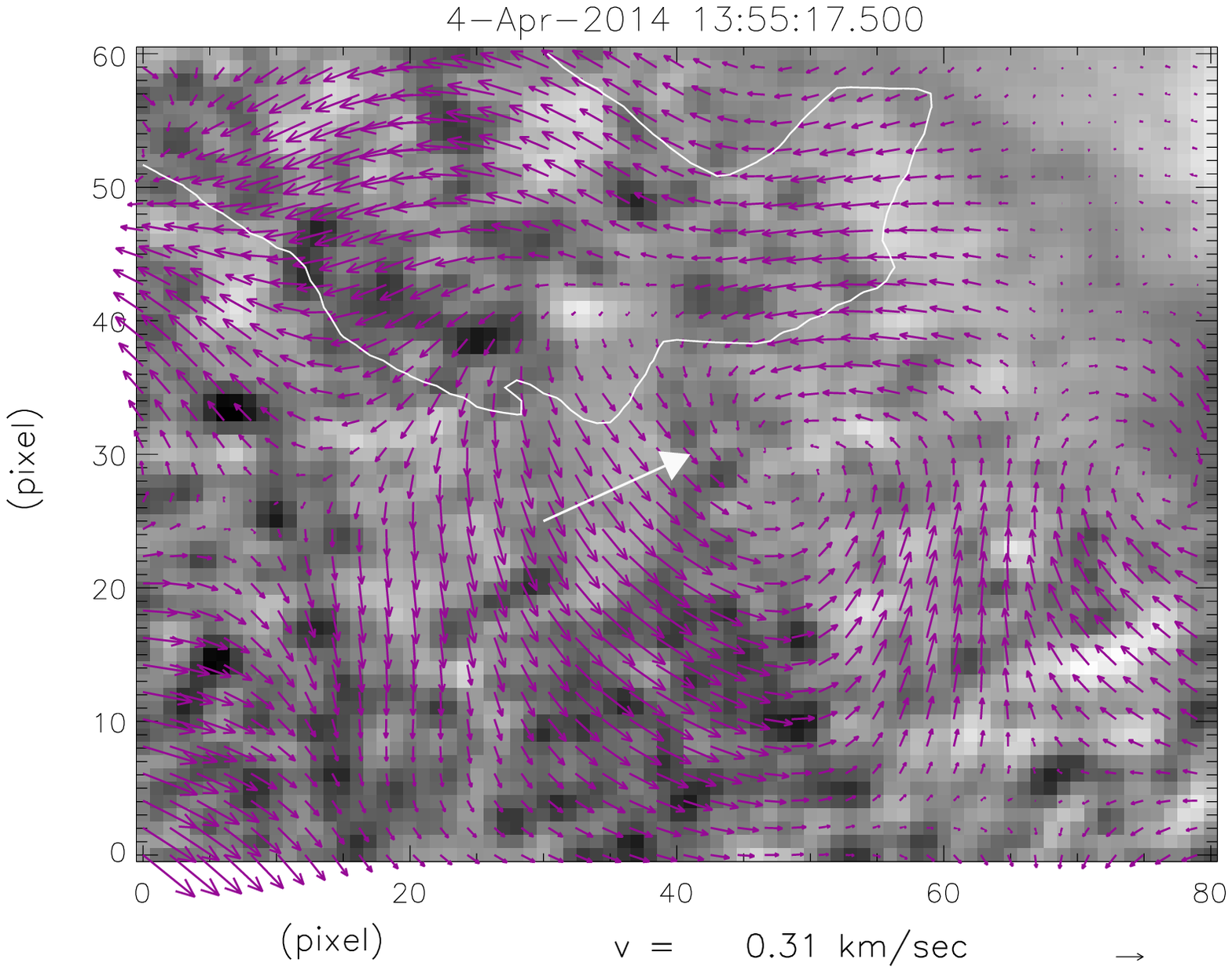}
              }
        
          \vspace{-0.1\textwidth}   
     
     \centerline{\Large \bf     
    \hspace{0.40 \textwidth}  \color{white}{(c)}
      \hspace{0.45\textwidth}  \color{white}{(d)}
         \hfill}
      \vspace{0.10\textwidth}    
\caption{Same as Fig.~\ref{fig:7}, but for event 9.}
   \label{fig:11}
   \end{figure}

\begin{figure}    
   \centerline{\hspace*{0.05\textwidth}
               \includegraphics[width=0.65\textwidth,clip=]{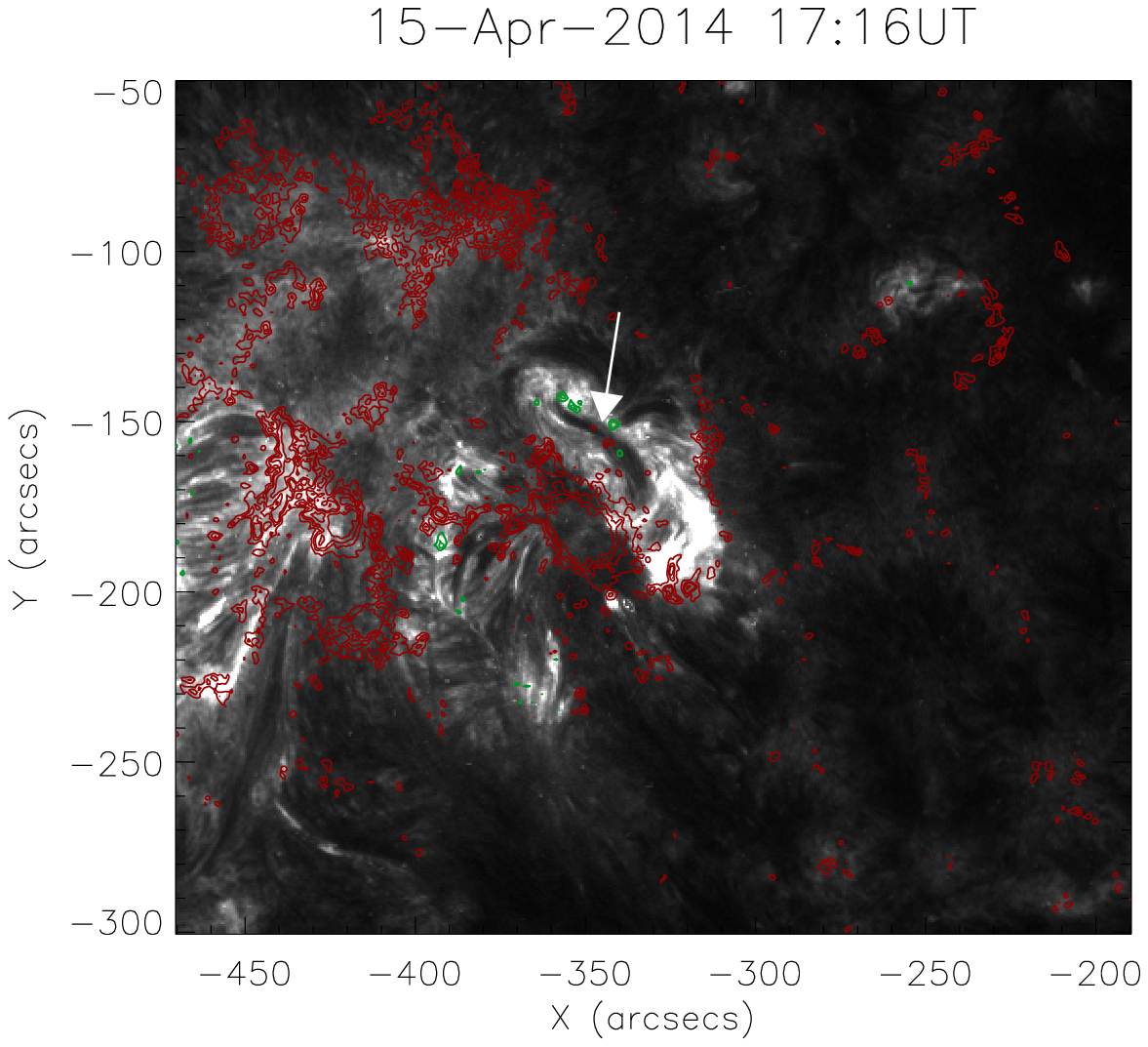}
               \hspace*{-0.16\textwidth}
               \includegraphics[width=0.59\textwidth,clip=]{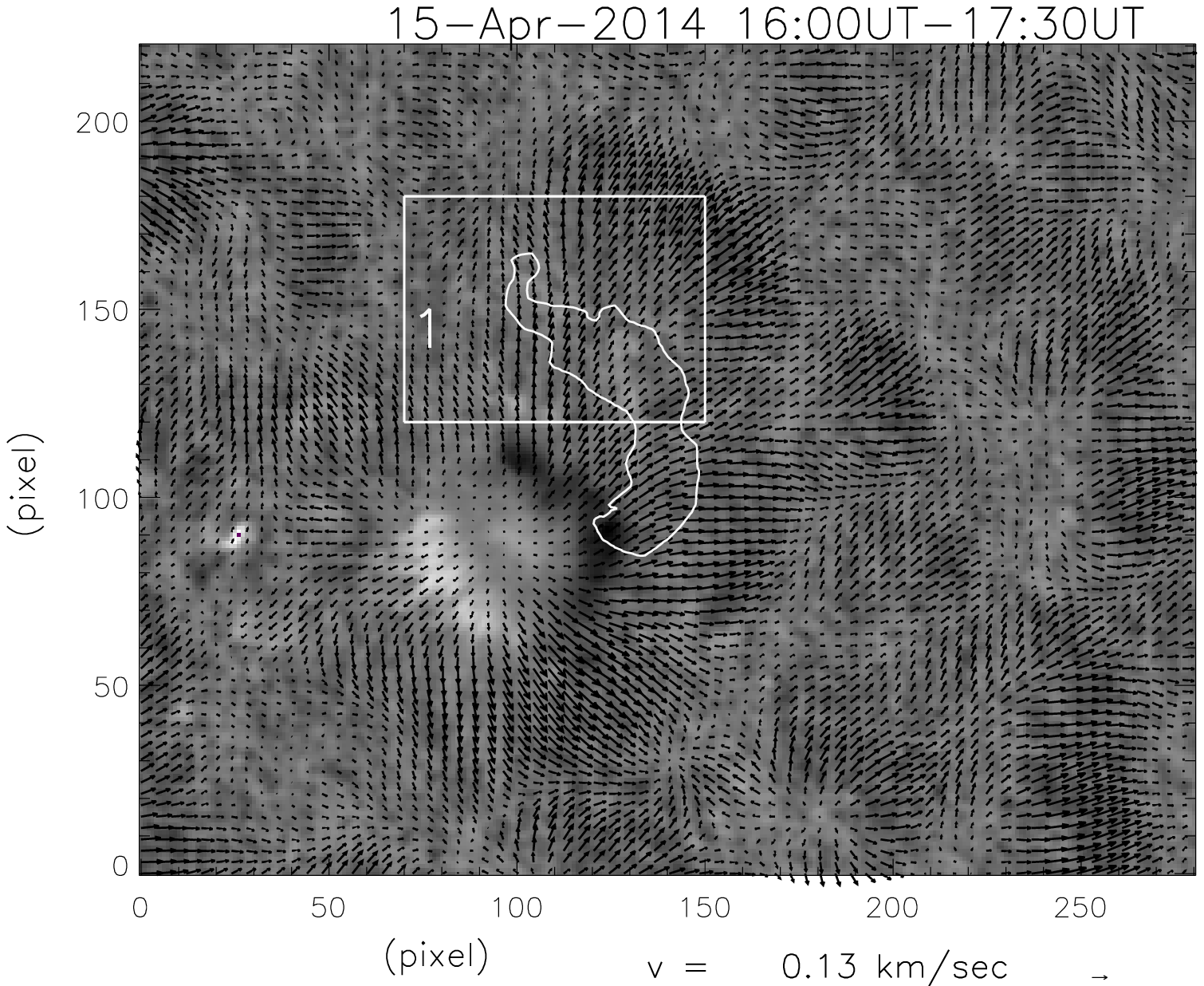}
              }
     \vspace{-0.12\textwidth}   
     
     \centerline{\Large \bf     
    \hspace{0.42 \textwidth}  \color{white}{(a)}
      \hspace{0.44\textwidth}  \color{white}{(b)}
         \hfill}
         \vspace{0.07\textwidth}  
         
   \centerline{\hspace*{0.015\textwidth}
               \includegraphics[width=0.59\textwidth,clip=]{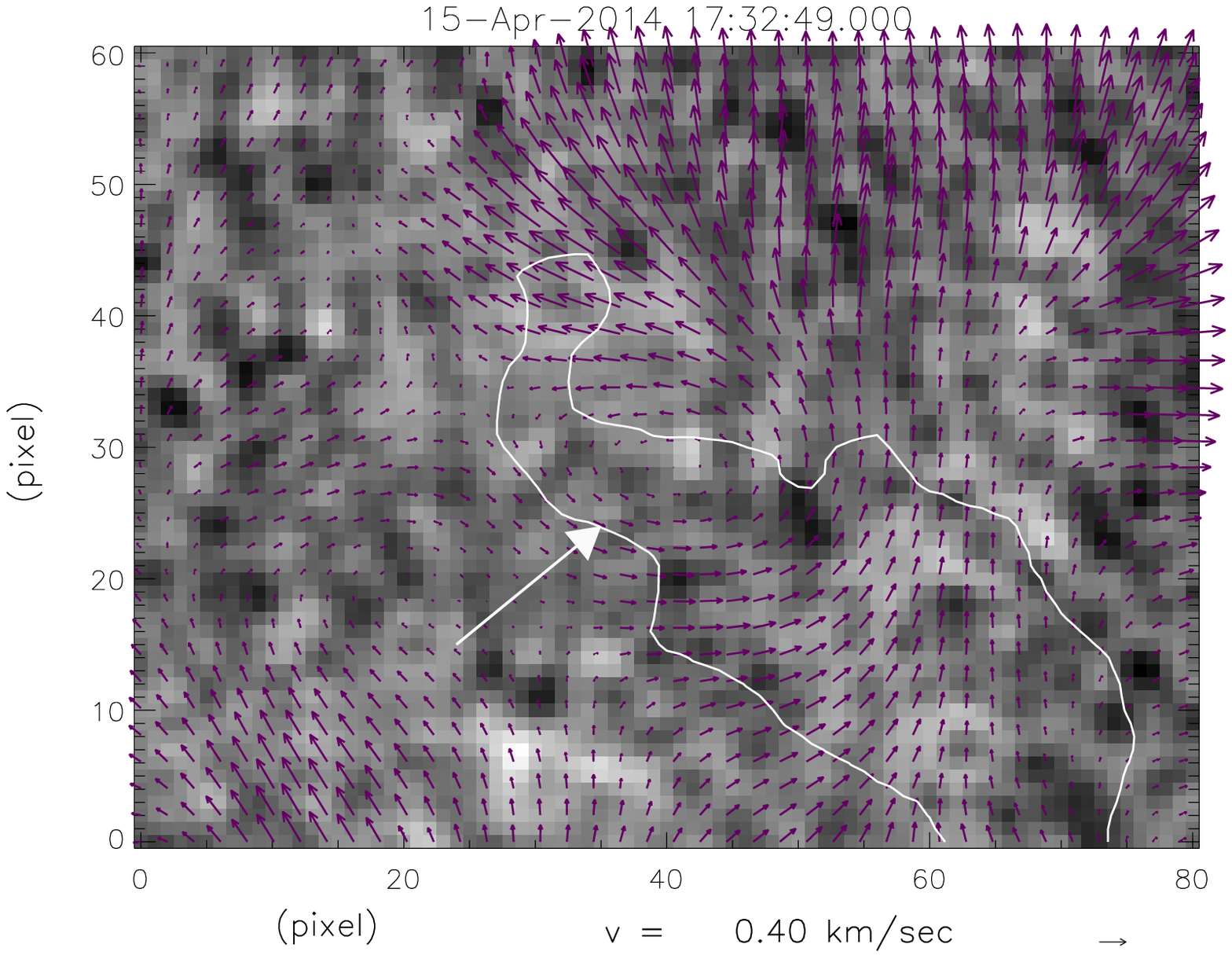}
              }
         \vspace{-0.1\textwidth}   
     
     \centerline{\Large \bf     
    \hspace{0.30 \textwidth}  \color{white}{(c)}
         \hfill}
      \vspace{0.10\textwidth}    
      
\caption{Same as Fig.~\ref{fig:7}, but for event 10. (c) Rotational 
velocity pattern observed in the location 1 for this
event.}
   \label{fig:12}
   \end{figure}

The filament locations in He~II 304~\AA~SDO/AIA channel for events 5--10 are displayed in 
Figs.~\ref{fig:7} to \ref{fig:12}(a), respectively. The position of the filament on 
an averaged dopplergram (greyscale) and averaged horizontal flow map (arrows) is shown
by overlying the contours of the filament extracted from 
304~\AA~images. These are shown in Figs.~\ref{fig:7} to \ref{fig:12}(b), respectively. 
The observed rotational motions at the ends of 
the filament footpoints for all the events are also shown in bottom-left(c) and bottom-right(d) rows in 
Figs.~\ref{fig:7} to \ref{fig:12}. The date and time of the observations, the filament location 
on the Sun, associated active region number, the filament activation time, importance of the associated 
flare, and the velocities of the erupting filaments are given in Table \ref{Table:1}. The 
starting and ending time and the type of the rotational motions seen at both ends of the filaments for all 
events are also given in the same table.

The filament in event 5 is a long one, associated with two active regions. One of its ends is located in  
AR 11451 and the other end is in AR 11450. Before the filament eruption, bidirectional flows
were observed in the filament. There were three filament eruptions observed in AR 11936(events 6--8).
Event 6 was observed on 31 December 2013. The other two events were observed on 
01 January 2014. On 31 December 2013 only one filament existed in the active region 
and before the eruption this filament bifurcated into two halves. The right-side half erupted and the 
left-side one did not erupt. Later, again a filament formed in the same region and the 
bifurcation disappeared. Event 7 also occurred in the same active region, but the filament
was located North of the 31 December 2013 event. This was a failed filament eruption. 
Event 8 also occurred on the same day, a few hours after event 7. The erupting filament
was the same as the one which erupted on 31 December 2013. Unlike the 31st December event, however, here the 
whole filament was activated with no bifurcation. But it was a failed eruption. 
In event 9, AR 12027 was surrounded by 
filaments in all directions. It was not a single filament, in fact it was a group of 3--4 
segmented filaments. On 04 April 2014 the North-East portion of the filament erupted (event 9). 
Before the eruption a large scale mass flow was observed inside the filament.
Event 10 occurred in AR 12035. The filament was hard to see before the 
eruption in the active region. But during the filament activation a cusp shaped filament 
was observed in the active region. This filament erupted completely during a
C7.3~class flare.

\end{article} 

\end{document}